\documentclass[amsmath,amssymb,aps,prd,showpacs,twocolumn,superscriptaddress,letterpaper,longbibliography, floatfix]{revtex4-2}

\usepackage{graphicx}
\usepackage{dcolumn}
\usepackage{subfigure}
\usepackage{orcidlink}
\usepackage{bm}
\usepackage{hyperref}
\usepackage{amsmath, amssymb} 
\usepackage{color}
\usepackage[dvipsnames]{xcolor}
\usepackage{comment} 
\usepackage{hyperref} 
\usepackage{float} 
\usepackage{makecell} 
\usepackage{tikz}
\usepackage{placeins}
\usepackage{environ}
\usepackage{multirow}
\usepackage[normalem]{ulem}

\hypersetup{breaklinks = true, colorlinks = true, citecolor = blue, linkcolor = blue, urlcolor = blue}

\usepackage[textsize=tiny,%
    color=orange!25,%
    linecolor=orange,%
    bordercolor=lightgray]{todonotes}
\usepackage[nameinlink,capitalise]{cleveref}

\begin{document}


\title{Measurements of the electron neutrino-argon differential cross section without pions in the final state in MicroBooNE}



\newcommand{\ANL}{Argonne National Laboratory (ANL), Lemont, IL, 60439, USA}
\newcommand{\Bern}{Universit{\"a}t Bern, Bern CH-3012, Switzerland}
\newcommand{\BNL}{Brookhaven National Laboratory (BNL), Upton, NY, 11973, USA}
\newcommand{\UCSB}{University of California, Santa Barbara, CA, 93106, USA}
\newcommand{\Cambridge}{University of Cambridge, Cambridge CB3 0HE, United Kingdom}
\newcommand{\CIEMAT}{Centro de Investigaciones Energ\'{e}ticas, Medioambientales y Tecnol\'{o}gicas (CIEMAT), Madrid E-28040, Spain}
\newcommand{\Chicago}{University of Chicago, Chicago, IL, 60637, USA}
\newcommand{\Cincinnati}{University of Cincinnati, Cincinnati, OH, 45221, USA}
\newcommand{\CSU}{Colorado State University, Fort Collins, CO, 80523, USA}
\newcommand{\Columbia}{Columbia University, New York, NY, 10027, USA}
\newcommand{\Edinburgh}{University of Edinburgh, Edinburgh EH9 3FD, United Kingdom}
\newcommand{\FNAL}{Fermi National Accelerator Laboratory (FNAL), Batavia, IL 60510, USA}
\newcommand{\Granada}{Universidad de Granada, Granada E-18071, Spain}
\newcommand{\IIT}{Illinois Institute of Technology (IIT), Chicago, IL 60616, USA}
\newcommand{\ICL}{Imperial College London, London SW7 2AZ, United Kingdom}
\newcommand{\Indiana}{Indiana University, Bloomington, IN 47405, USA}
\newcommand{\Kansas}{The University of Kansas, Lawrence, KS, 66045, USA}
\newcommand{\KSU}{Kansas State University (KSU), Manhattan, KS, 66506, USA}
\newcommand{\Lancaster}{Lancaster University, Lancaster LA1 4YW, United Kingdom}
\newcommand{\LANL}{Los Alamos National Laboratory (LANL), Los Alamos, NM, 87545, USA}
\newcommand{\Louisiana}{Louisiana State University, Baton Rouge, LA, 70803, USA}
\newcommand{\Manchester}{The University of Manchester, Manchester M13 9PL, United Kingdom}
\newcommand{\MIT}{Massachusetts Institute of Technology (MIT), Cambridge, MA, 02139, USA}
\newcommand{\Michigan}{University of Michigan, Ann Arbor, MI, 48109, USA}
\newcommand{\MSU}{Michigan State University, East Lansing, MI 48824, USA}
\newcommand{\Minnesota}{University of Minnesota, Minneapolis, MN, 55455, USA}
\newcommand{\Nankai}{Nankai University, Nankai District, Tianjin 300071, China}
\newcommand{\NMSU}{New Mexico State University (NMSU), Las Cruces, NM, 88003, USA}
\newcommand{\Oxford}{University of Oxford, Oxford OX1 3RH, United Kingdom}
\newcommand{\Pitt}{University of Pittsburgh, Pittsburgh, PA, 15260, USA}
\newcommand{\QMUL}{Queen Mary University of London, London E1 4NS, United Kingdom}
\newcommand{\Rutgers}{Rutgers University, Piscataway, NJ, 08854, USA}
\newcommand{\SLAC}{SLAC National Accelerator Laboratory, Menlo Park, CA, 94025, USA}
\newcommand{\SDSMT}{South Dakota School of Mines and Technology (SDSMT), Rapid City, SD, 57701, USA}
\newcommand{\Maine}{University of Southern Maine, Portland, ME, 04104, USA}
\newcommand{\TelAviv}{Tel Aviv University, Tel Aviv, Israel, 69978}
\newcommand{\UTA}{University of Texas, Arlington, TX, 76019, USA}
\newcommand{\Tufts}{Tufts University, Medford, MA, 02155, USA}
\newcommand{\VTech}{Center for Neutrino Physics, Virginia Tech, Blacksburg, VA, 24061, USA}
\newcommand{\Warwick}{University of Warwick, Coventry CV4 7AL, United Kingdom}
\newcommand{\NotreDame}{University of Notre Dame, Notre Dame, IN, 46556, USA}

\affiliation{\ANL}
\affiliation{\Bern}
\affiliation{\BNL}
\affiliation{\UCSB}
\affiliation{\Cambridge}
\affiliation{\CIEMAT}
\affiliation{\Chicago}
\affiliation{\Cincinnati}
\affiliation{\CSU}
\affiliation{\Columbia}
\affiliation{\Edinburgh}
\affiliation{\FNAL}
\affiliation{\Granada}
\affiliation{\IIT}
\affiliation{\ICL}
\affiliation{\Indiana}
\affiliation{\Kansas}
\affiliation{\KSU}
\affiliation{\Lancaster}
\affiliation{\LANL}
\affiliation{\Louisiana}
\affiliation{\Manchester}
\affiliation{\MIT}
\affiliation{\Michigan}
\affiliation{\MSU}
\affiliation{\Minnesota}
\affiliation{\Nankai}
\affiliation{\NMSU}
\affiliation{\Oxford}
\affiliation{\Pitt}
\affiliation{\QMUL}
\affiliation{\Rutgers}
\affiliation{\SLAC}
\affiliation{\SDSMT}
\affiliation{\Maine}
\affiliation{\TelAviv}
\affiliation{\UTA}
\affiliation{\Tufts}
\affiliation{\VTech}
\affiliation{\Warwick}
\affiliation{\NotreDame}

\author{P.~Abratenko\,\orcidlink{0000-0001-6945-5941}}\affiliation{\Tufts} 
\author{D.~Andrade~Aldana\,\orcidlink{0009-0008-3143-3374}} \affiliation{\IIT}
\author{L.~Arellano\,\orcidlink{0000-0002-1093-1824}} \affiliation{\Manchester}
\author{J.~Asaadi\,\orcidlink{0000-0001-6915-5279}}   \affiliation{\UTA}
\author{A.~Ashkenazi\,\orcidlink{0000-0002-1995-3851}}   \affiliation{\TelAviv}
\author{S.~Balasubramanian} \affiliation{\FNAL}
\author{B.~Baller\,\orcidlink{0000-0001-8731-9281}}  \affiliation{\FNAL}

\author{A.~Barnard\,\orcidlink{0000-0001-6117-1768}} \affiliation{\Oxford}

\author{G.~Barr\,\orcidlink{0000-0002-9763-1882}} \affiliation{\Oxford}
\author{D.~Barrow\,\orcidlink{0000-0001-5844-709X}} \affiliation{\Oxford}
% ...

\collaboration{The MicroBooNE Collaboration}
\thanks{microboone\_info@fnal.gov}\noaffiliation


\begin{abstract}

We present a new measurement of the electron neutrino charged current cross section on argon without pions in the final state.  This measurement uses the full MicroBooNE booster neutrino beam dataset of $1.3\times 10^{21}$ protons on target collected at Fermi National Accelerator Laboratory.  Events are considered both with and without protons above the kinetic energy visibility threshold.
Differential cross sections are extracted in proton and electron kinematics, including energy and angle relative to the neutrino beam direction. The relationship between the hadronic and leptonic systems is explored through the angle between the proton and electron directions.
The resulting cross sections are compared to a variety of available generator predictions using different models of neutrino interactions. We find good agreement with most models in lepton kinematics and some discrepancies in the hadronic system modeling, particularly in proton angle.
\end{abstract}

\maketitle

\section{Introduction}
\subsection{Motivation}
There are still many open questions about neutrinos~\cite{ATHAR2022103947} that current and future experiments are trying to answer. These include understanding and precisely measuring neutrino oscillations~\cite{ MAK62,GRI69,Giganti:2017fhf}, describing neutrino interactions~\cite{NuSTEC:2017hzk},
determining the neutrino mass ordering~\cite{massordering}, mass mechanism~\cite{massmodelsreview}~\cite{nunature},
and explaining previously observed anomalies~\cite{LSND,MiniBooNE} through standard model or beyond the standard model (BSM) processes. Part of the next generation Deep Underground Neutrino Experiment (DUNE)~\cite{DUNE_FD} physics program is to precisely measure neutrino oscillations. This will include measurements of the mixing angles that determine the relationship between the neutrino mass and flavor states, including if one of these parameters, $\theta_{23}$, is maximal, the mass ordering, and if there is charge-parity violation in the neutrino sector~\cite{LowexpDUNEsensi,LBLsensiDUNE}.  An important signature for these measurements is electron neutrino appearance in a muon neutrino beam, which means that a detailed understanding of electron neutrino interactions is critical.  Electron neutrinos are also a potential signature for BSM physics based on previously observed anomalies.  DUNE will leverage the excellent reconstruction capabilities and low detection thresholds of liquid argon time projection chamber (LArTPC) detectors~\cite{DUNE_ND,DUNE_FD}, which makes measurements of electron neutrino interactions on an argon target particularly important.\\

In this paper, we present an electron neutrino cross section measurement on argon without pions in the final state using data from MicroBooNE.  The MicroBooNE experiment is a LArTPC that collected data at Fermi National Accelerator Laboratory (FNAL) between 2015 and 2020 from two beams: the Booster Neutrino Beam~\cite{MiniBooNEFlux} (BNB) and the Neutrinos at the Main Injector~\cite{NuMIbeam} (NuMI) beamline. It provides a dataset of neutrino interactions on argon that covers part of the DUNE energy spectrum~\cite{DUNEuBfluxComp}.\\

Despite the critical role electron neutrinos play in understanding open questions in neutrino physics there are few measurements of this kind in the literature. This is because electron neutrinos form only a small component of a neutrino beam when it is produced. Most previous electron neutrino measurements are on lighter nuclear targets than argon \cite{T2Knuexsec1,T2Knuexsec2,T2Knuexsec3,NOvanueXsec,COHERENTnuexsec,recentxsecT2K,MinervaExcnuexsec}. Those on argon, both inclusive~\cite{argoneutnuexsec,UBnueInc21,UBnueInc22} and exclusive~\cite{UBnueExcpip}, typically combine electron neutrinos and anti-neutrinos into their signal definition. 
To date, there are two exclusive argon measurements that only use electron neutrinos in their signal definitions. Each of these use only a portion of the MicroBooNE data, one with $\sim30\%$ from NuMI~\cite{nueCCNuMI} and one with $\sim50\%$ from BNB~\cite{MicroBooNEXsec}. 

The measurement presented here 
builds on the BNB measurement to use the full MicroBooNE dataset and include further exploration of the relationship between the leptonic and hadronic final state systems. Like the earlier BNB measurement~\cite{MicroBooNEXsec} this measurement explores electron neutrinos with and without visible protons, and across a visibility threshold of 50 MeV. 
In addition, it is complementary to the NuMI measurement as the neutrino flux is different.
There are similar MicroBooNE measurements of muon neutrinos with visible protons~\cite{MicroBooNE:2025tcm, Stevenpaper} and across the proton visibility threshold \cite{BensAna,BensAnaPRL} of 35 MeV. \\ 

Electron neutrino measurements are also motivated by the fact that most theoretical models in neutrino interaction generators have been tuned primarily to muon neutrinos as there are more available measurements and datasets. However, it has been demonstrated~\cite{Day:2012gb,elecvsMuon2,EvsMuMartini,Ankowski:2017yvm} that the mass difference of the final lepton can have non-trivial effects on the interaction kinematics depending on the allowed phase space in the models, especially at low energy and in the forward region. Electron neutrino measurements can inform the implementation of such effects in the generators and help correct the tuning to better represent data in both flavors.

\subsection{The beamline and detector}
The MicroBooNE detector \cite{Microboonedetector} is located at FNAL on-axis relative to the BNB and 468.5 m after the neutrino beam source. The BNB delivers an 8 GeV proton beam that collides with a beryllium target to produce a neutrino beam with energy peaked at about 700 MeV. It is mostly a muon neutrino beam with an electron neutrino contribution of only 0.51\% \cite{MiniBooNEFlux}. This small intrinsic electron neutrino contribution is used in this measurement. The electron anti-neutrino component is very small, corresponding to 0.05\% of the beam content. This allows us to make a neutrino-only measurement. MicroBooNE has collected $13.09\times10^{20}$ protons on target (POT) in the BNB over the course of its operation.  This represents a factor of 1.9 increase in data statistics relative to MicroBooNE's earlier measurement with BNB data \cite{MicroBooNEXsec}. \\

The MicroBooNE detector is a 85 metric ton LArTPC. Its dimensions are 2.56~m (drift direction) $\times$ 2.32~m (vertical) $\times$ 10.36~m (beam direction). An electric field of 273~V/cm is applied between the anode and cathode. The anode is composed of three wire planes with 3~mm wire spacing. Two of these planes are induction planes placed at an angle of $\pm$ 60° from the vertical and the third is a vertically oriented collection plane. When a charged particle is produced by a neutrino interaction on argon in the TPC, it will ionize the argon along its trajectory. The ionization electrons drift towards the anode where their detection on the three planes allows a 3D reconstruction of the particle track or electromagnetic shower. Photomultiplier tubes located behind the wire planes detect the argon scintillation signal produced by a neutrino interaction with timing resolution on the order of nanoseconds. This makes it possible to select events in time with the beam bunches. For comparison, the maximum electron drift time in MicroBooNE is 2.3~ms \cite{Drifttime2}. MicroBooNE is a surface detector, so after the first year of operation, a cosmic ray tagger (CRT) \cite{CRT_UB} was installed and commissioned. It consists of plastic scintillator strips placed on each side of the TPC and is used to veto cosmic rays.\\

\section{Methods and inputs}
\subsection{Simulation and reconstruction}
\label{sec:F&I}
Three major elements of the analysis that require modeling are the flux, the cross sections, and the detector response.  These allow us to produce a predicted number of signal events in the experiment. 
The flux is estimated using tools from the MiniBooNE experiment \cite{MiniBooNEFlux,UBBNBFlux} adapted to the MicroBooNE location 
since the same beam is used by both experiments.\\

The neutrino interaction model uses a custom MicroBooNE tune~\cite{TuneUB} of the G18\_10a\_02\_11a model in GENIE v3.0.6~\cite{GENIE,Geniev3}.
This model uses a local Fermi Gas (LFG) nucleon momentum distribution within the argon target \cite{LFG}, and the Valencia charged current (CC) Quasi-Elastic (QE)  \cite{CCNieves} and CC meson exchange current (MEC) \cite{2p2hNieves} models. 
It also uses the Kuzmin-Lyubushkin-Naumov Berger-Sehgal (BS) resonant (RES) model \cite{4MomMiniBooNE,BergerSehgal2007,LepPOL,QuarkRES}, the BS coherent \cite{BSCOH} scattering model, and the Bodek-Yang (BY) deep inelastic scattering (DIS) model~\cite{BYDIS} with the hadronization derived from PYTHIA~\cite{PYTHIA}. The tune is a fit of T2K CC0$\pi$ data on carbon to CCQE and CC2p2h parameters of importance to MicroBooNE and for which theoretical uncertainties are significant. There are four: the CCQE axial mass, the strength of the random phase approximation (RPA) effect, and the shape and normalization of the CCMEC contribution in the total interaction cross-section. 
The A scaling from carbon to argon is performed through GENIE, with Final State Interactions (FSI), binding energy, and medium dependence scaling applied as described in~\cite{TuneUB}.
This produces a model that agrees with the MicroBooNE muon neutrino data using only external data~\cite{TuneUB}. The simulated particles from this model are propagated through the detector using GEANT4 \cite{GEANT4}. This simulation and the reconstruction described in the next section are performed through the LArSoft framework \cite{LArSoft}.\\ 

As MicroBooNE is a surface detector that relies on the relatively slow drift of electrons, it is essential to correctly predict external sources of charged particles that are inevitable background to neutrino measurements. This background is mainly muons produced by cosmic-rays and it is evaluated using data taken while the beam was off and overlaying it with the neutrino interaction prediction from GENIE.
This provides a reliable, detector-specific prediction of background that directly incorporates noise and other detector effects without relying on modeling. \\

This analysis uses the pattern recognition toolkit Pandora \cite{Pandora} for reconstruction.  Pandora first eliminates events that cross the detector vertically as those are most likely cosmic rays. It then clusters the remaining charge into regions of interest. The neutrino candidate is the region of interest that provides the best time coincidence with the beam window. Particles that originate from the neutrino interaction are reconstructed as either showers, primarily electrons, photons, and neutral pions, or tracks, primarily protons, muons, and charged pions.  
The Pandora output is also used to construct additional information about the event.
For example, electrons and photons are separated by the conversion distance between the vertex and the starting position of the shower, as well as by the energy deposited per unit distance (d$E$/d$x$) at the start of the shower. Photons manifest in the detector after pair production as an $e^+e^-$ pair, which deposit twice as much energy as electrons.  Tools like these that rely on the topology of the selected particles are used to separate remaining cosmic muons from particles produced in the neutrino interaction and showers from tracks.  The identification strategy is the same as has been described in \cite{oldPELEE,PELEEnew,MicroBooNEXsec}.
 
\subsection{Signal definition and selections}

\begin{figure*}
    \subfigure[]{\includegraphics[width=8cm]{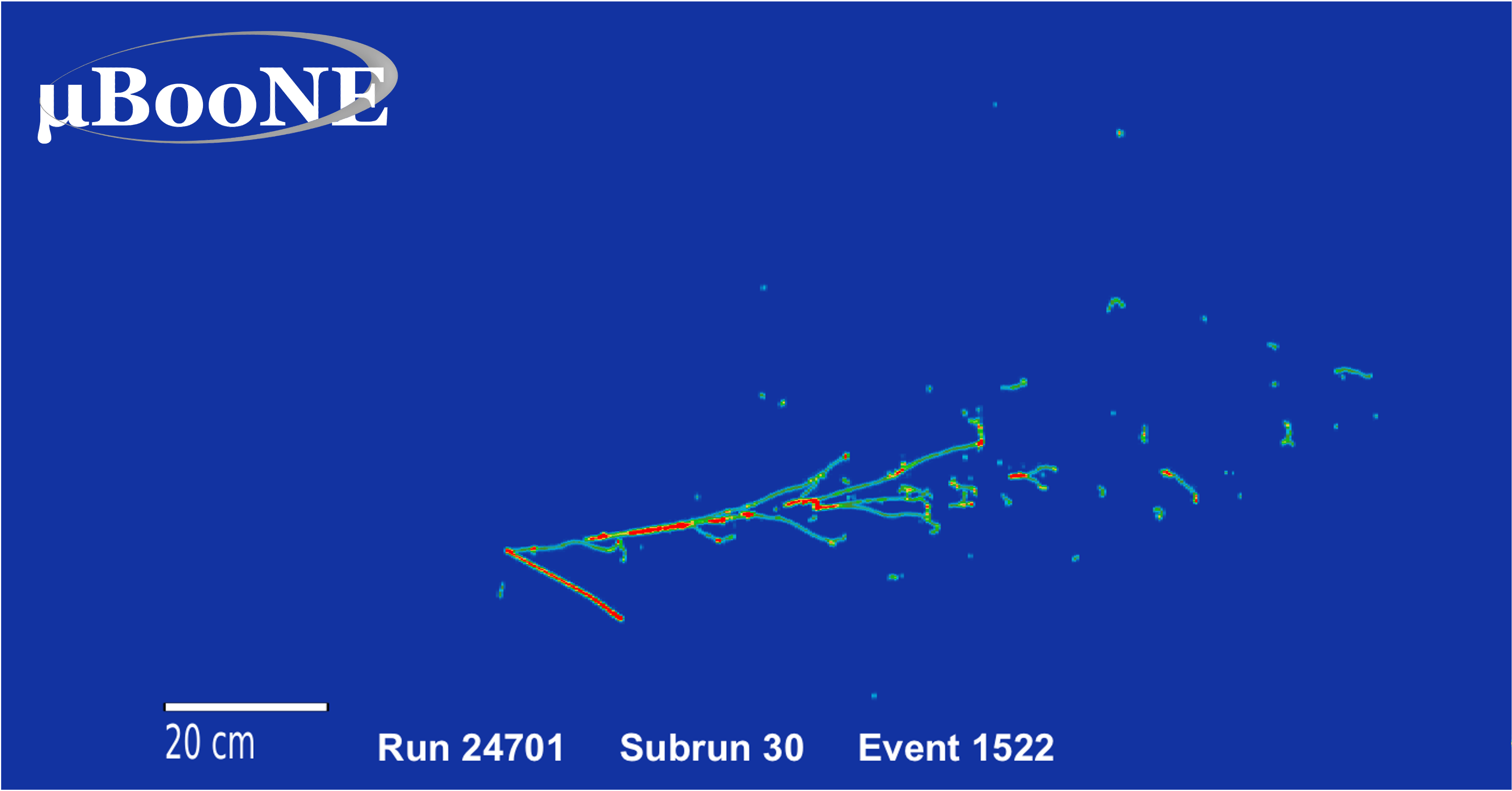} \label{fig:EVTDisNp}} \hspace{2em}
    \subfigure[]{\includegraphics[width=8cm]{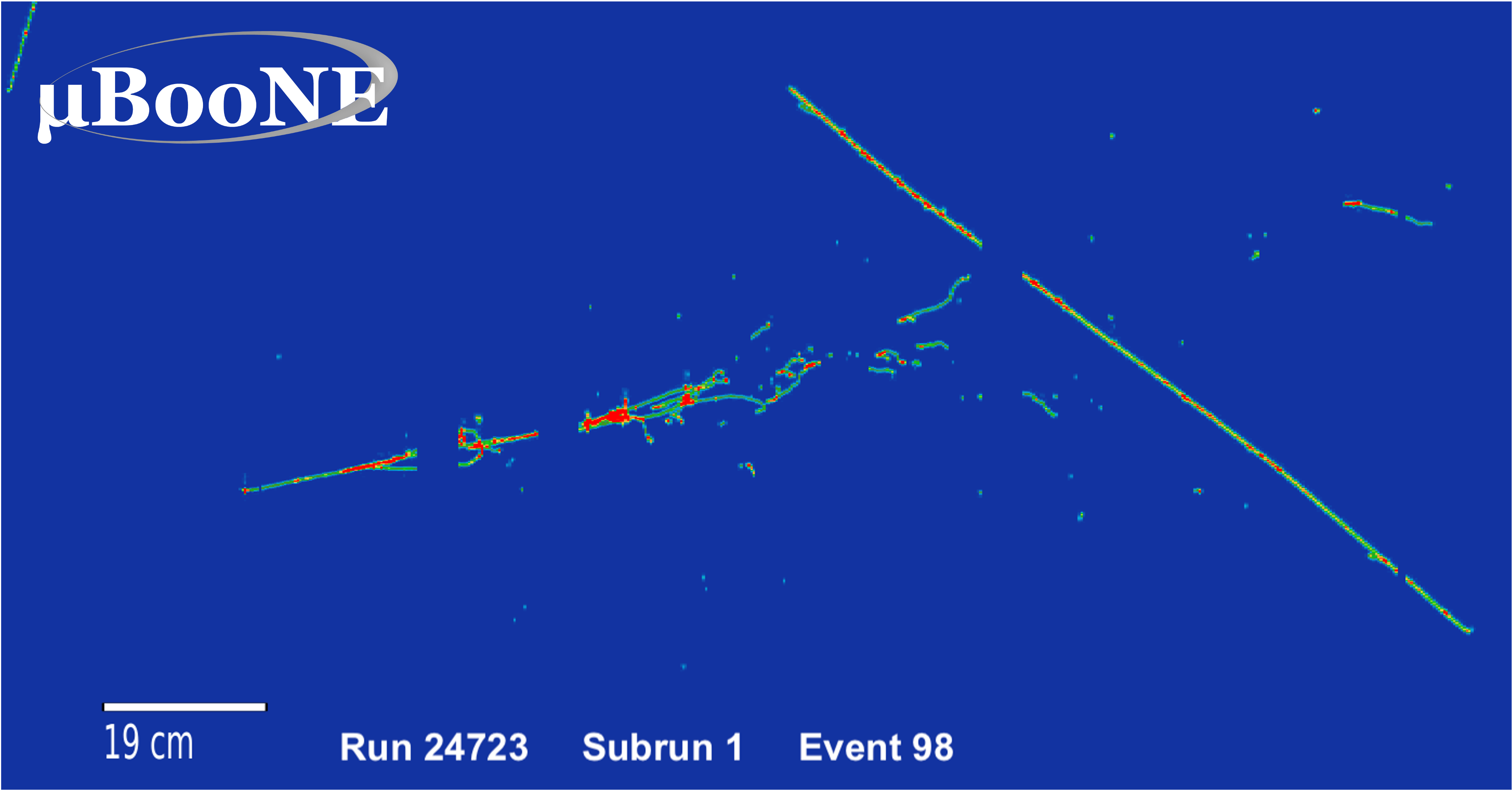}\label{fig:EVTDis0p}}
\caption{Event displays on the collection plane for two of the selected data events: (a) 1eNp0$\pi$ candidate event and (b) 1e0p0$\pi$ candidate event. The x-axis corresponds to wire number and the y-axis corresponds to time. The color scale indicates the amount of deposited charge.}\label{fig:EVTDis}
\end{figure*}

This analysis presents a differential cross section measurement of charged current electron neutrinos on argon without neutral or visible charged pions in the final state. We have two sets of events, the first with at least one visible proton (1eNp0$\pi$) and the other without any visible protons (1e0p0$\pi$) in the final state. Example events in each topology are shown in Fig.~\ref{fig:EVTDis}.
The 1eNp0$\pi$ channel offers a clear topology to reliably identify electron neutrinos, while the 1e0p0$\pi$ channel is more sensitive to differences in interaction models
and complements the 1eNp0$\pi$ measurement to explore data across the phase space of proton kinetic energy (KE). Moreover, these event categories match the ones used for MicroBooNE electron neutrino low energy excess searches, providing additional ability to interpret this dataset~\cite{oldPELEE,PELEEnew}. \\

The true signal definition for this measurement requires an electron neutrino charged current event in the fiducial volume with $KE_e >$ 30 MeV, $KE_{\pi^{\pm}} <$ 40 MeV, and no neutral pions.  
Protons above 50~MeV are considered visible in both the true signal definition and in the selection. 
Protons with kinetic energy above this threshold can be well reconstructed, with an energy resolution under 9\% based on Monte-Carlo (MC) predictions~\cite{oldPELEE}.
This is also the proton kinetic energy threshold where the 1eNp0$\pi$ and 1e0p0$\pi$ selection efficiencies overlap, which indicates the threshold at which these orthogonal selections are able to reliably identify protons~\cite{MicroBooNEXsec}. 
In order to select regions with good signal purity, we apply the following criteria on the interaction phase space in the selection and the true signal definition. In the 1e0p0$\pi$ channel, the electron energy $E_e$ is required to be more than 0.5~GeV and the cosine of the electron angle with respect to the beam direction $\cos\theta_e$ is required to be more than 0.6 as shown in Appendix~\ref{app:cuts}. In the 1eNp0$\pi$ channel, the cosine of the opening angle between the leading proton and the electron is required to be more than -0.9.\\

We report the differential cross section in five kinematic quantities: the electron energy, leading proton kinetic energy, electron and proton angles with respect to the beam direction, and the opening angle between the leading proton and the electron. The 1e0p0$\pi$ channel is presented in the first bin of the leading proton kinetic energy measurement, below 50~MeV. We also provide a measurement across the proton visibility threshold, one bin for 1e0p0$\pi$ events with invisible protons (below 50 MeV) and one bin for all 1eNp0$\pi$ events with visible protons (above 50 MeV). This measurement is analogous to the proton kinetic energy measurement but presented in only two bins.\\

A similar measurement with a partial MicroBooNE dataset, about half of the BNB data, has been published previously \cite{MicroBooNEXsec}. The same selection of events is used in this work, which is adapted from the selections used in \cite{oldPELEE,PELEEnew} in order to increase the electron neutrino efficiency. The same boosted decision trees (BDTs) and training are used here and are described in \cite{oldPELEE}. 
This measurement also uses the information from the CRT to reject cosmic rays in the 1e0p0$\pi$ channel. This is a new addition relative to the earlier iteration of this analysis~\cite{MicroBooNEXsec}, and matches the CRT implementation in \cite{PELEEnew}.
We obtain a purity of 57\% (67\%) and a signal efficiency of 14\% (19\%) for the 1e0p0$\pi$ (1eNp0$\pi$) channel.\\

\begin{figure*}

    \subfigure[]    
    {\includegraphics[width=8cm]{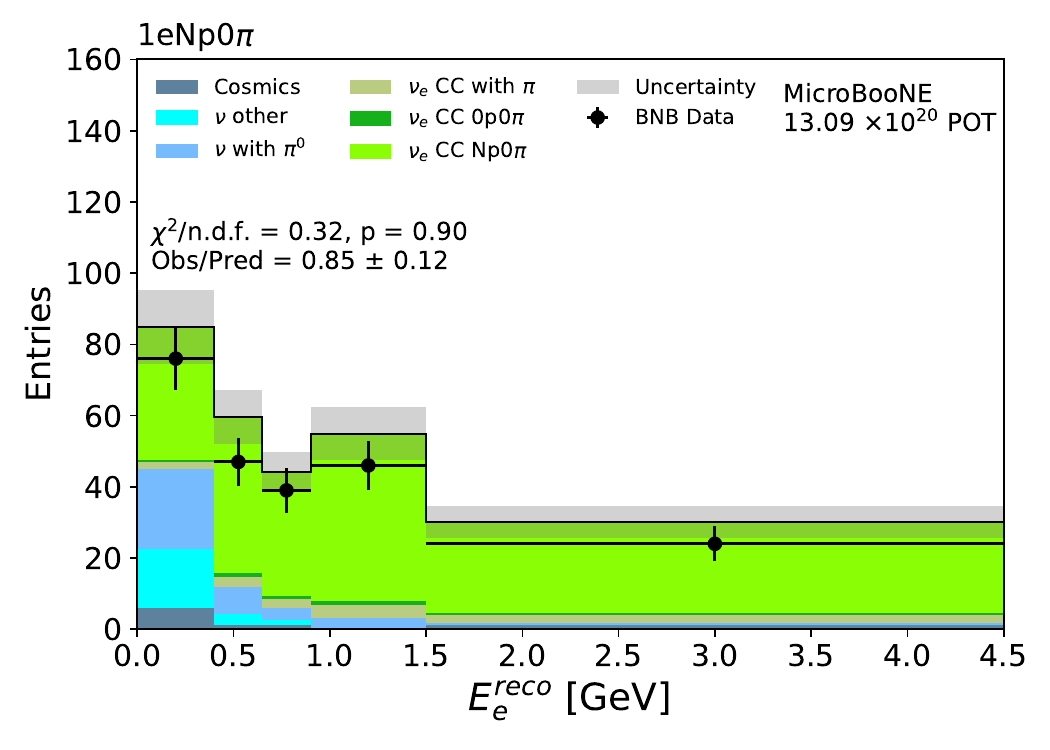}}
    \subfigure[]{\includegraphics[width=8cm]{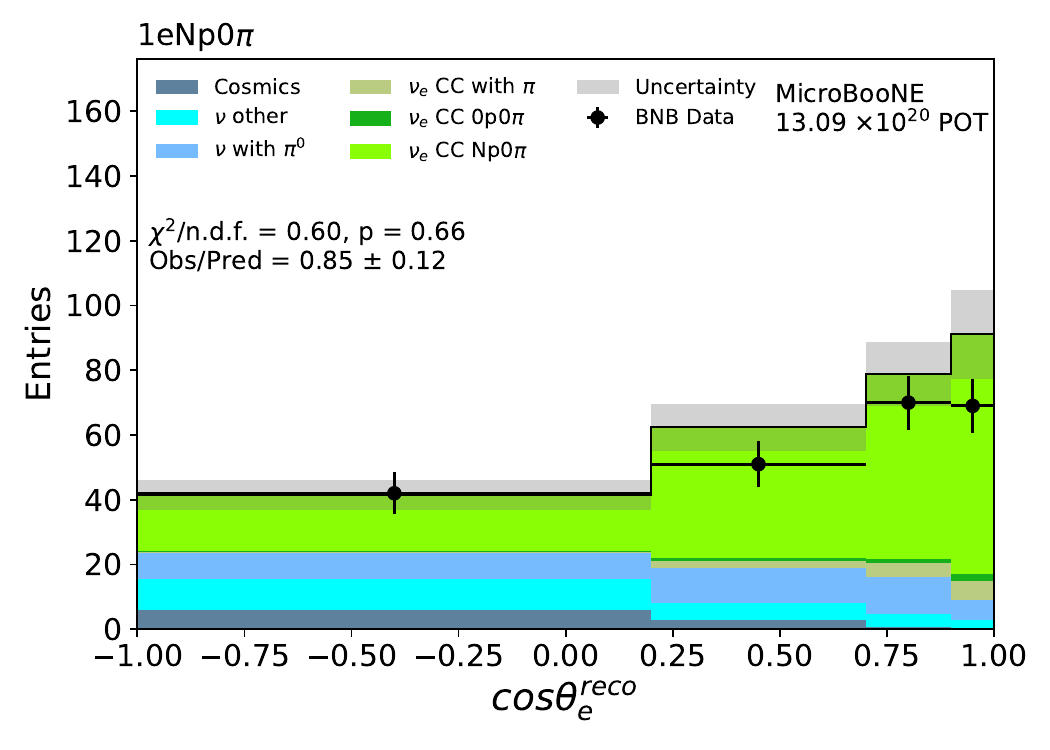}}
    \subfigure[]{\includegraphics[width=8cm]{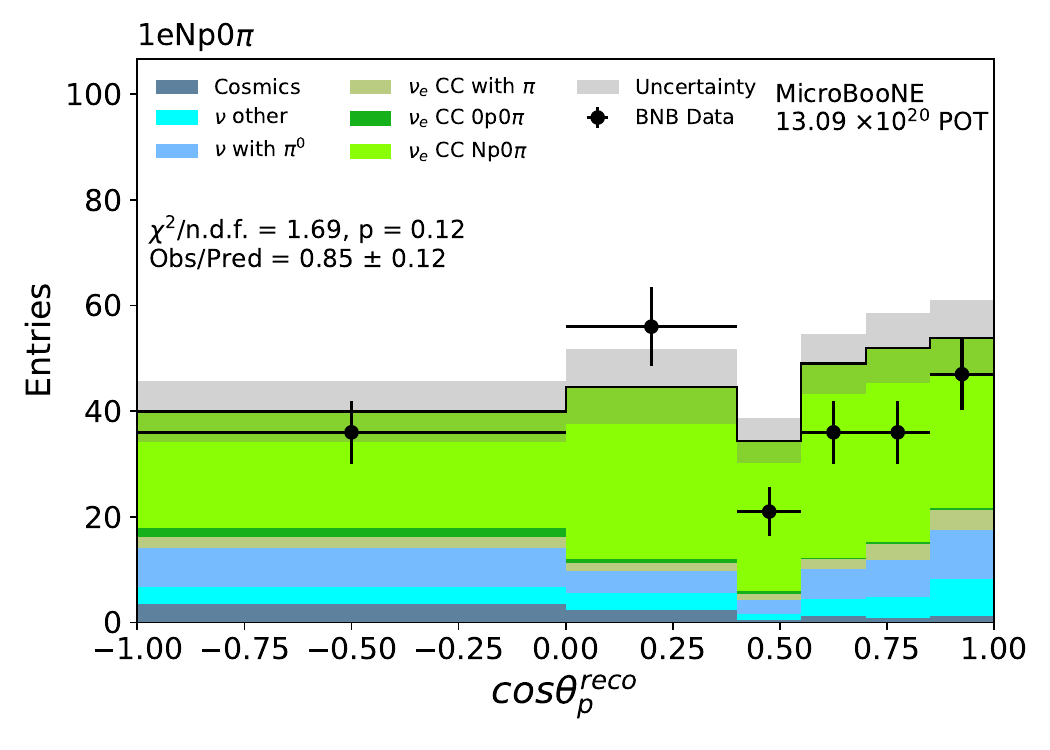}}
    \subfigure[]{\includegraphics[width=8cm]{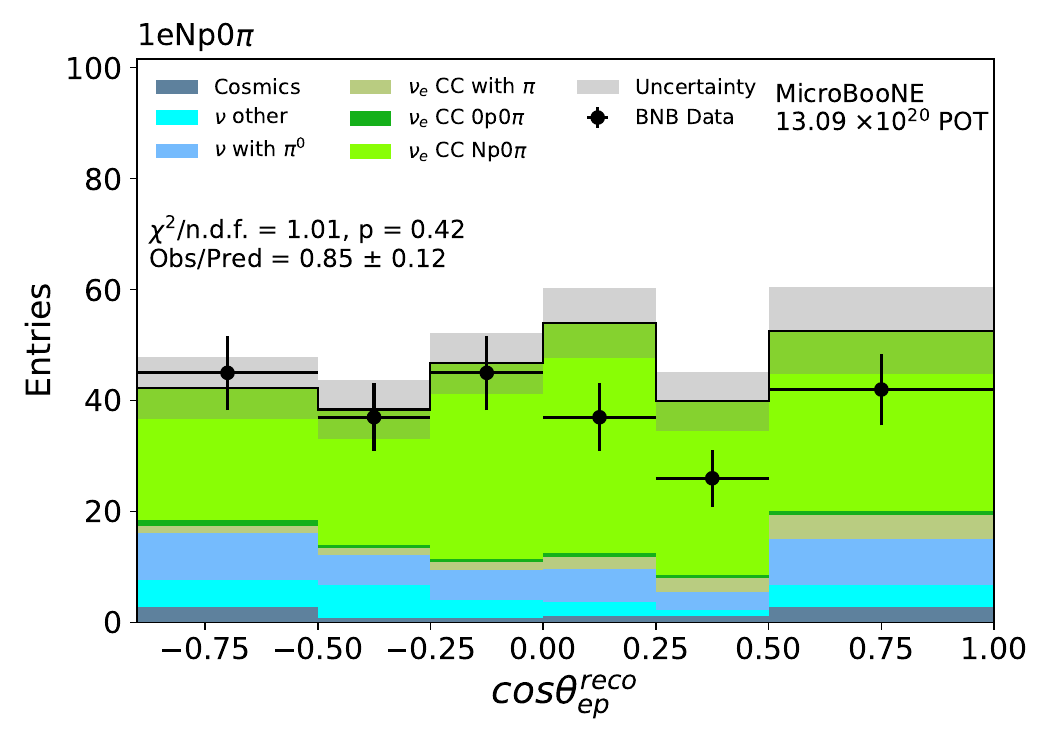}}
\caption{Predicted distribution and data in variables using the 1eNp0$\pi$ channel: (a) electron energy, (b) cosine of electron angle, (c) proton angle, (d) cosine of the electron-proton opening angle. } \label{fig:DistribRes}
\end{figure*}

\begin{figure*}
    \subfigure[]{\includegraphics[width=8cm]{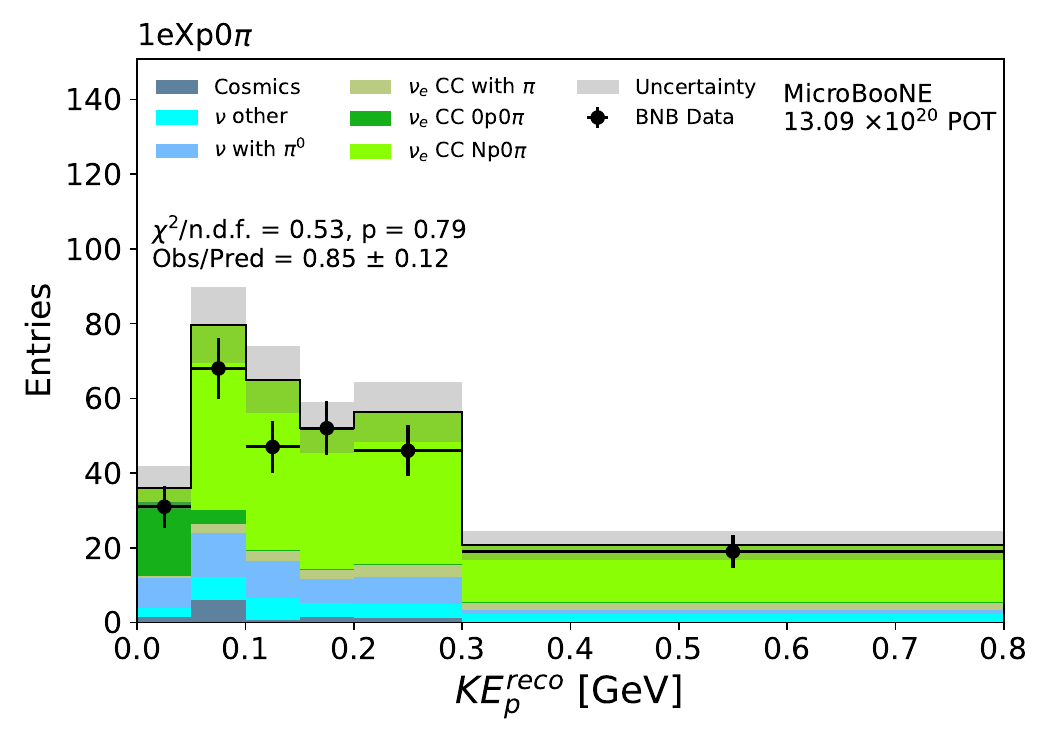}}
    \subfigure[]{\includegraphics[width=8cm]{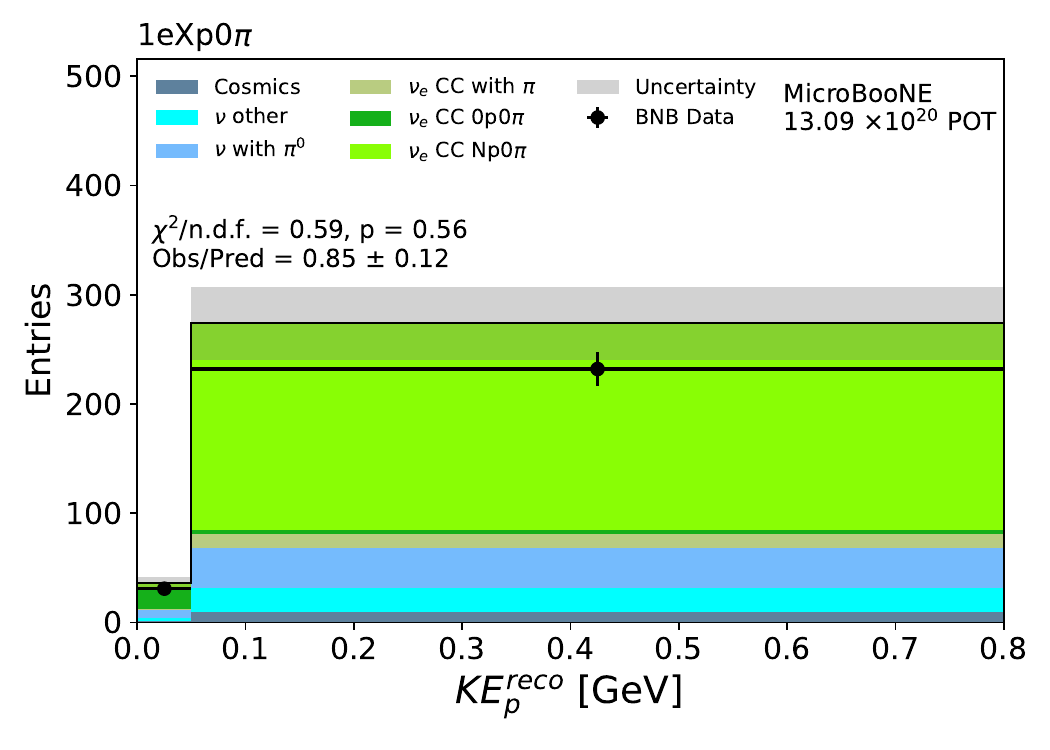}}
\caption{Predicted distribution and data for proton kinetic energy for the 1eXp0$\pi$ channel (1eNp0$\pi$ + 1e0p0$\pi$). (a) shows this distribution with multiple bins for 1eNp0$\pi$ events, and (b) the same distribution with one bin per channel.} \label{fig:DistribResEp}
\end{figure*}

After the event selection, the majority of events originate from QE interactions both for the 1eNp0$\pi$ (47\%) and 1e0p0$\pi$ (53\%) channels. The main subdominant contributions are predicted to come from MEC at 21\% and RES interactions at 27\% for 1eNp0$\pi$. Those proportions are similar for 1e0p0$\pi$ which has 12\% of MEC interactions and 20\% of RES interactions. Other contributions come from DIS and electron scattering. The primary backgrounds come from events with a neutral pion in the final state ($\nu$ with $\pi^0$) representing 40 to 52\% of the background for the 1eNp0$\pi$ and 1e0p0$\pi$ channels respectively. For 1eNp0$\pi$ events, the second main background is muon neutrinos without neutral pions (21\%), while for 1e0p0$\pi$ events it is true 1eNp0$\pi$ events where the proton is missed by the reconstruction (26\%). In~\cref{fig:DistribRes,fig:DistribResEp}, the former muon-like events are combined with other minor neutrino interaction backgrounds in a ``$\nu$ other'' category.

\subsection{Systematic uncertainties}
\label{sec:systs}
Several sources of systematic uncertainties are taken into account in this analysis. \\

Flux uncertainties due to beamline hadron production and geometry are computed by reweighting predicted event rates as in~\cite{UbNCDelta,Ubincnumu}. They are found to be around 6\% across all analysis bins.\\ 

Detector systematics are evaluated for a specific set of variations using dedicated MC samples \cite{detsysts}.  These variations include effects such as ion recombination in the argon and space-charge effects, as well as variations in the light yield and wire responses. 
The distributions, in analysis binning, from these variation samples are smoothed by convolving the event count per bin with a pseudo-Gaussian filter. This minimizes the impact of MC statistical effects from the variation samples in the uncertainty computation. This procedure is described in \cite{PELEEnew}. The fractional detector uncertainties are around 5\% in most bins and can be up to 12\%. \\

Uncertainties on the interaction model are included by reweighting our base model (``MicroBooNE tune'') under different assumptions, as prescribed in~\cite{TuneUB}. Their impact is estimated on the background event rate and on the efficiency and smearing of all events in true space, but not on the number of signal events as this is a quantity of interest in the reported cross section measurement. Uncertainties on the re-interaction of final state particles that exit the nucleus are also taken into account by reweighting their propagation through the detector~\cite{ReINT}. This uncertainty is sub-dominant, except at high proton energy, reaching 8\%, where the probability of proton re-interaction is higher.
\\

A flat uncertainty of 1\% is attributed to the number of target nucleons and a flat 2\% uncertainty, based on direct flux measurements~\cite{MiniBooNEFlux}, is added to account for uncertainty on the POT counting. A MC statistical uncertainty that covers the size of the samples used to build the predicted distributions is also included.\\

In total, the fractional systematic uncertainties are between 9 and 15\% depending on the analysis bin whereas the data statistical uncertainties are between 6 and 22\%. All analysis bins are statistically limited except the lowest electron energy bin and the 1eNp0$\pi$ proton energy bin in the two-bin configuration. 
The systematic uncertainty breakdown for each variable in the measurement binning is included in Appendix~\ref{app:Systs}. Covariance matrices encoding both the systematic and statistical uncertainties are used to propagate uncertainties to the final cross section result and are available in Appendix~\ref{app:matrix}.

\subsection{Cross section extraction}
\label{sec:unfold}
To extract a differential cross section $\frac{d\sigma}{dx}$ with respect to an analysis observable $x$ in true space, we use the following formula:
\begin{equation}
\label{eq:XsecCalc}
<\frac{d\sigma}{dx}>_i~=~\frac{\Sigma_j R_{ij}^{-1}(N_j-B_j)}{N_{\text{target}}\times \phi \times \Delta x_i} .
\end{equation}
The index $i$ denotes a true bin and the index $j$ a reconstructed bin.
$R$ is the response matrix which encodes the mapping between true and reconstructed spaces (reported in Appendix~\ref{app:matrix}). $N-B$ is the background-subtracted number of events in data where the background is predicted using our base interaction model. $\phi$ is the integrated electron neutrino flux and $N_{\text{target}}$ is the number of target nucleons in the detector's fiducial volume. $\Delta x_i$ is the width of the true bin. The integrated electron neutrino flux is a constant reference value computed from the BNB flux predictions \cite{MiniBooNEFlux} following the nominal flux prescriptions from \cite{FluxPresc} and is scaled to POT. Here, the value is $\phi \simeq 5.21\times 10^9~\nu_e$~cm$^{-2}$. $N_{\text{target}}$ is computed to be $3.9107\times 10^{31}$ using pure argon with density $\rho = 1.3836 ~$g~cm$^{-3}$ and a fiducial volume defined by $x_{\text{min}}$= 10~cm, $x_{\text{max}}$= 246.4~cm, $y_{\text{min}}$= -101.5~cm, $y_{\text{max}}$= 101.5~cm, $z_{\text{min}}$= 10~cm, $z_{\text{max}}$= 986.8~cm. 
\\

We use the Wiener-SVD (WSVD) unfolding method \cite{WienerSVD} in order to report the final cross section.  The inverted response matrix $R_{ij}^{-1}$ is not well defined as written in Eq.~\ref{eq:XsecCalc}, so therefore requires regularization, given by the WSVD unfolding.  This regularization, along with the analysis binning and cross section extraction method, were determined using simulated data studies with an alternative interaction generator (NuWro).  These studies, described in more detail in Appendix~\ref{sec:fakedata}, demonstrated the robustness of the analysis against model dependence. \\    
The choice of regularization level in the WSVD method does not impact the overall $\chi^2$ results. However, in order to minimize the impact of the regularization on the final unfolded spectrum shape, we choose the WSVD first derivative regularization. This regularization is also encoded within a matrix that is applied to cross-section predictions to compare them to the unfolded data result.

\section{Results}

\subsection{Distributions of events}
\label{sec:Distribs}
The distribution of selected events with the MicroBooNE tuned GENIE model and the data are presented in Fig.~\ref{fig:DistribRes} for the four 1e$0\pi$Np analysis observables and in Fig.~\ref{fig:DistribResEp} for proton kinetic energy.  The reported uncertainty on these figures is the total event rate uncertainty including the full cross section uncertainties on the signal event prediction.  
The data consists of 232 1eNp0$\pi$ candidate events and 31 1e0p0$\pi$ candidate events that pass the selections.
Good overall agreement between the data and the base model is observed, with $p$-values between 0.12 and 0.90, and with similar features as those observed in the low energy excess (LEE) search analysis \cite{PELEEnew}.

The background model was validated with two sideband data samples.
The first one addresses the modeling of the $\pi^0$ background and requires at least two reconstructed showers. The other one is designed to check the modeling of any signal-adjacent background by inverting the BDT selection criteria while keeping the rest of the selection the same. Good agreement was found between the data and MC in both sidebands with $p$-values between 0.43 and 0.94. 
The sideband distributions are shown in Appendix~\ref{app:sidebands}.

\begin{figure*}
\subfigure[]{
    \includegraphics[width=8cm]{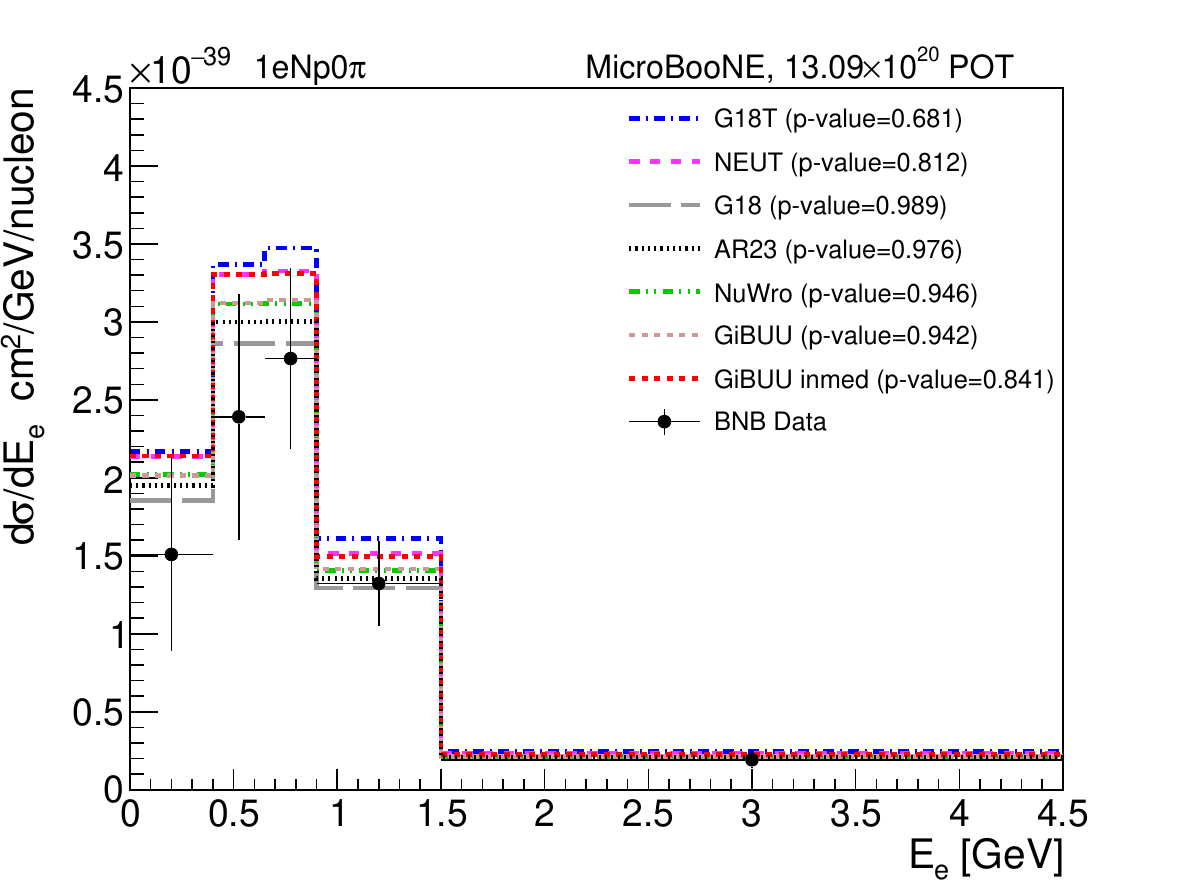}}
       \subfigure[]
       { \includegraphics[width=8cm]{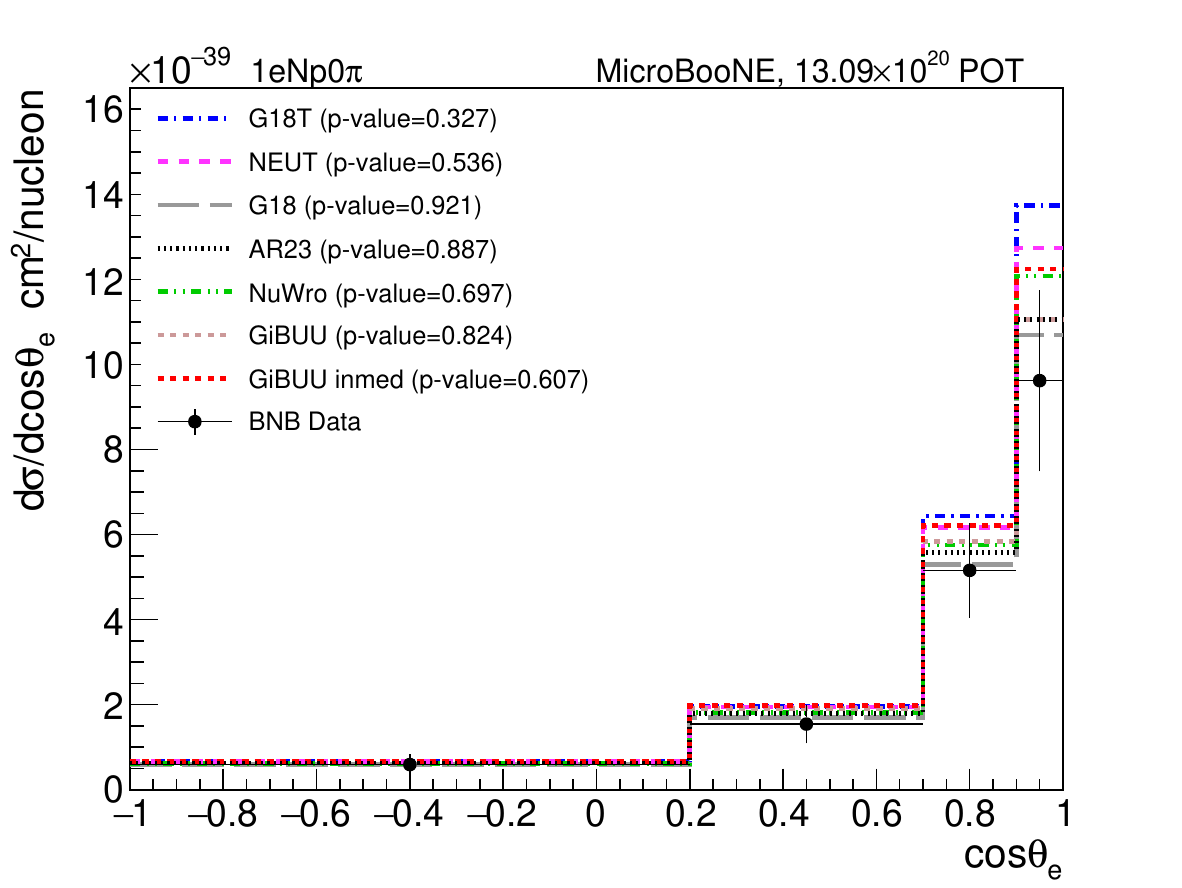}}
    \subfigure[]
    {\includegraphics[width=8cm]{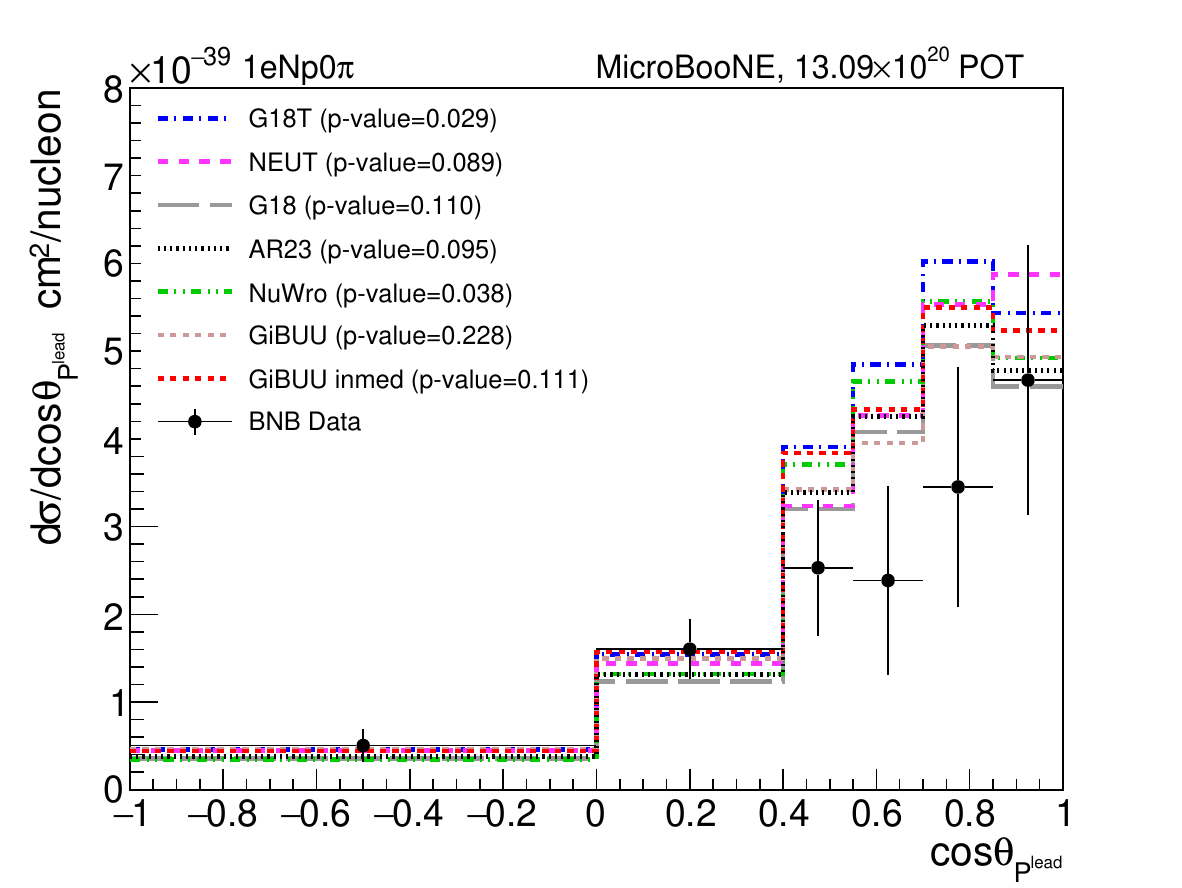}}
    \subfigure[]
    {\includegraphics[width=8cm]{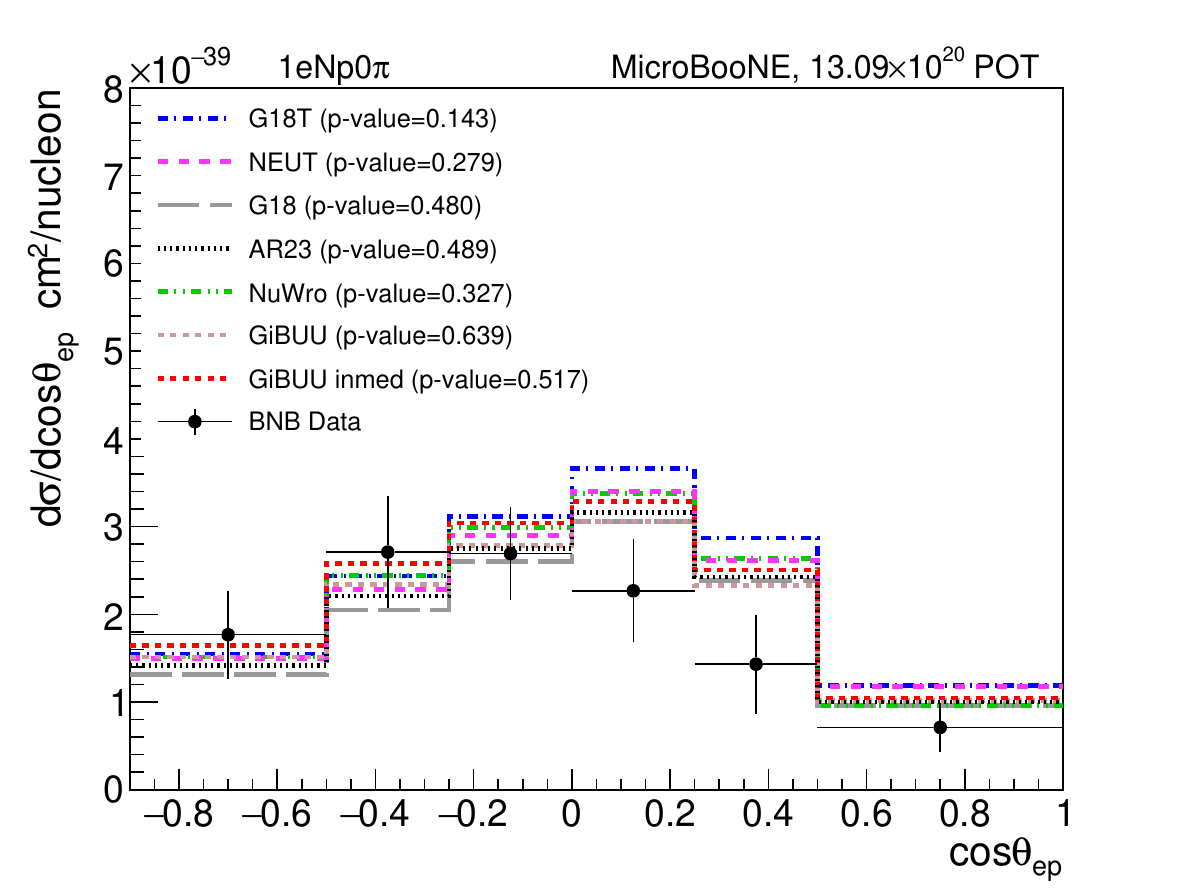}}

\caption{Unfolded data compared to the set of models described in section~\ref{sec:models}, for all analysis variables using the 1eNp0$\pi$ channel: (a) electron energy, (b) cosine of electron angle, (c) proton angle, (d) cosine of electron-proton opening angle.} \label{fig:UnfRes}
\end{figure*}

\begin{figure*}
   \subfigure[]
   { \includegraphics[width=8.5cm]{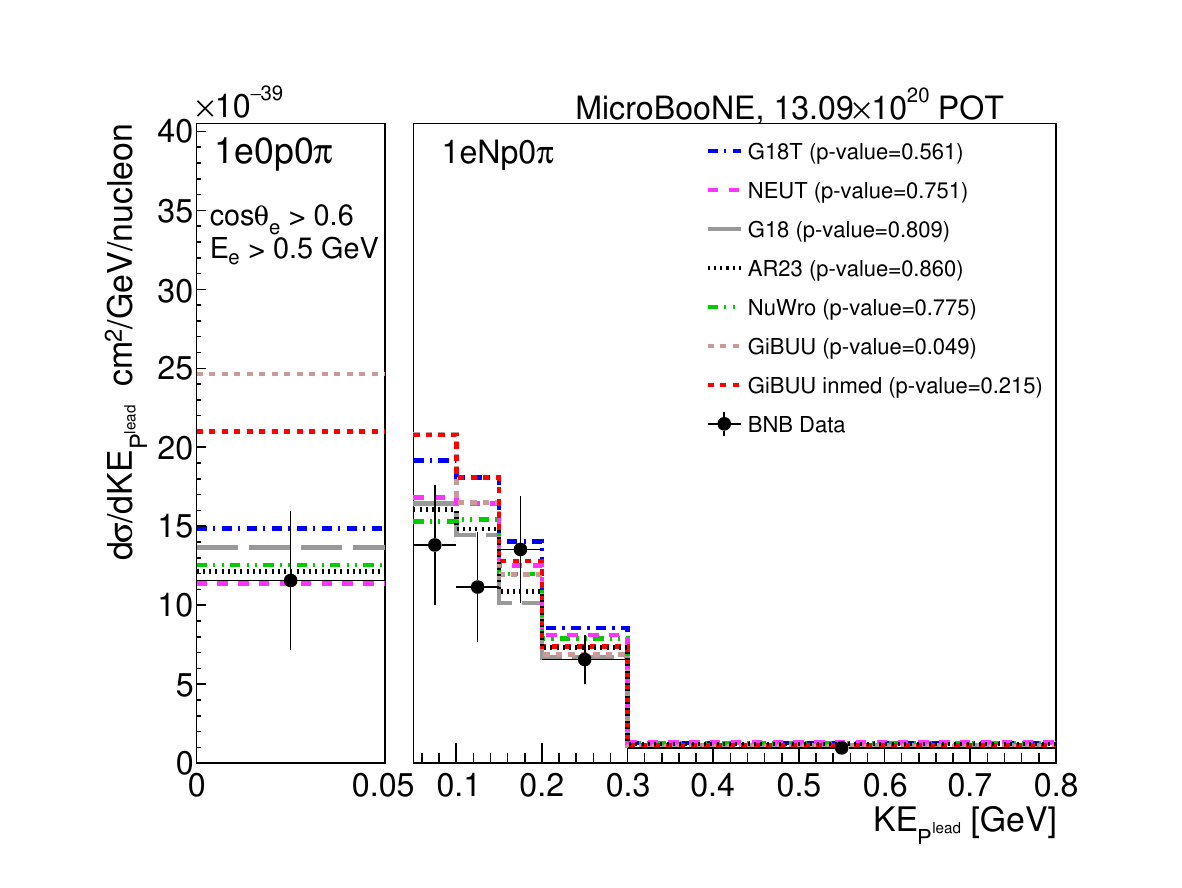}}
   \subfigure[]
   { \includegraphics[width=8cm]{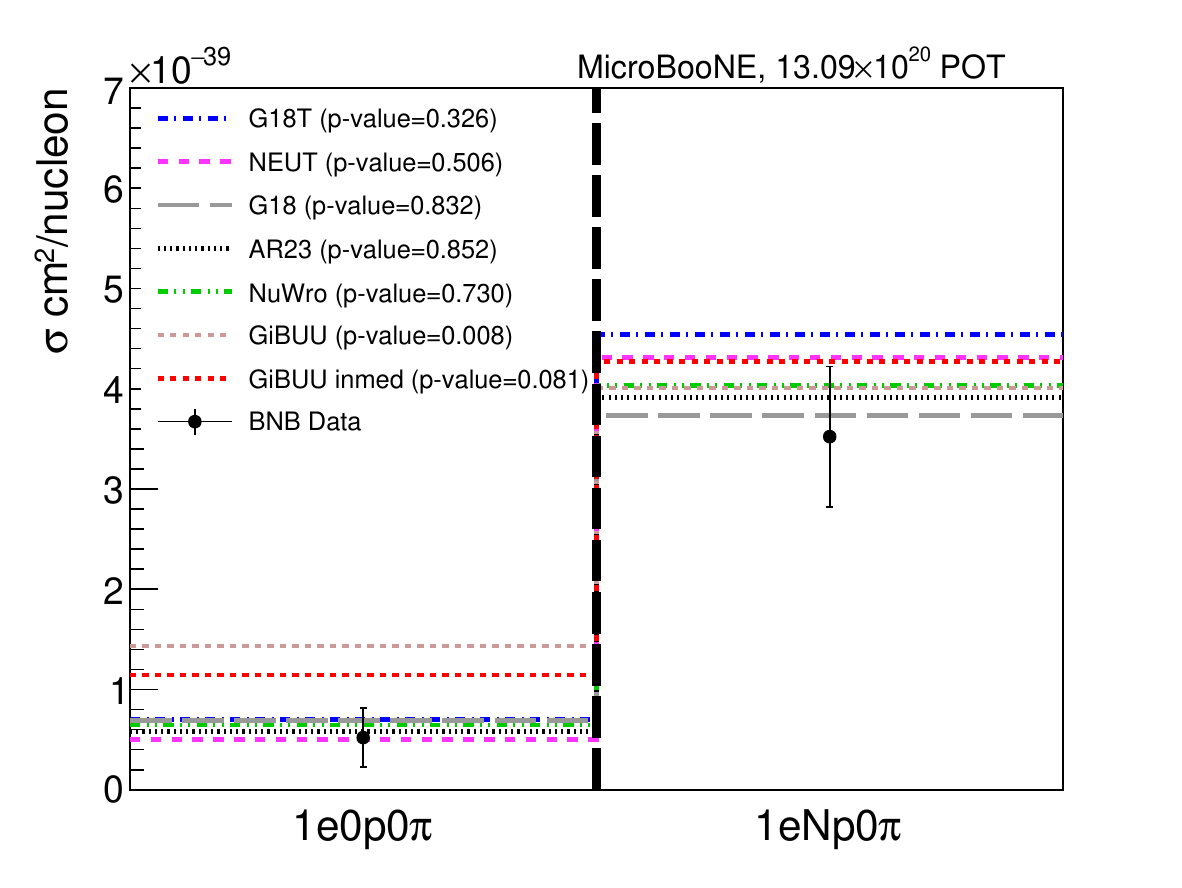} 
      \label{fig:sub:UnfResPetwobin}
   }
\caption{Unfolded data compared to the set of models described in section~\ref{sec:models}, for proton kinetic energy for the 1eXp0$\pi$ channel. (a) shows this distribution with multiple bins for 1eNp0$\pi$ events, and (b) the same distribution with one bin per channel such that all 1eNp0$\pi$ events are in a single bin from 50 to 800 MeV.} \label{fig:UnfResPe}
\end{figure*}

\subsection{Model comparisons}
\label{sec:models}
After unfolding the data following the procedure described in Section~\ref{sec:unfold}, the cross section results are compared to a variety of available generator predictions using different combinations of interaction models: 
\begin{itemize}
    \item The base model GENIE v3.0.6 G18\_10a\_02\_11a with the MicroBooNE tune (G18T) described in section \ref{sec:F&I}.
    \item GENIE v3.0.6 G18\_10a\_02\_11a without the tune (G18).
    \item GENIE v3.6.0 AR23 20i 00 000 (AR23): The AR23 model is similar to G18\_10a\_02\_11a but uses the LFG ground state modeling with a correlated high-momentum nucleon tail; the z-expansion for the CCQE axial form factors (Z-exp) \cite{ZEXP}; the SuSAv2 modeling for MEC interactions \cite{SuSAv2}; emission of de-excitation photons for argon nuclei; and the free nucleon tune \cite{TUNEfreeNuc}. It is currently the base model for other LArTPC experiments, such as DUNE~\cite{DUNEnuscale}, ICARUS~\cite{ICARUSexotics} and SBND~\cite{SBNDPRISM}.
    \item NuWro v21.09.2 (NuWro) \cite{NuWro} uses the LFG model, the Llewellyn-Smith (LS) model for quasi-elastic events \cite{LLEWELLYNSMITH1972}, the Nieves model for MEC interactions \cite{2p2hNieves} and the Adler-Rarita-Schwinger (ARS) formalism \cite{AdlerRES} for the $\Delta$ resonance. NuWro uses a standard intranuclear cascade (INC) for FSI \cite{NucTransp,FSIEffects}. The DIS prediction is taken from the BY model.
    \item NEUT v5.7.0 (NEUT) \cite{NEUT} uses the same ground state, CCQE and MEC models as described for this analysis' base model v3.0.6 G18\_10a\_02\_11a. It also uses the BS RES \cite{BergerSehgal2007,QuarkRES,Auchincloss:1990tu} and BS coherent \cite{BSCOH} scattering models. The FSI model is based on intra-nuclear cascades with in-medium pion corrections \cite{SALCEDO1988557}. The DIS model is custom~\cite{Gl_ck_1998} at low hadronic energy with BY corrections, and uses PYTHIA~\cite{PYTHIA} for higher hadronic system energies (above 2~GeV)~\cite{Bronner_2016}.
    \item GiBUU 2025 \cite{GiBuu} with (GiBUU inmed) or without (GiBUU) in-medium nucleon-nucleon corrections \cite{INMEDIUMGIBUU} uses the LFG model and its own custom MEC and resonant calculations. These are implemented in a consistent way by solving the Boltzmann-Uehling-Uhlenbeck transport equation~\cite{Mosel_2019}. More specifically, it uses a standard CCQE expression \cite{GiBUUCCQE}, an empirical MEC model and a custom spin-dependent resonance calculation through the MAID analysis~\cite{Mosel_2019}. The FSI prediction is obtained through the propagation of hadrons in a nuclear potential. The DIS prediction comes from PYTHIA.
\end{itemize}
A summary table of the different models used by these generators is available in Appendix~\ref{app:models} Table~\ref{tbl:models}. Predictions from these generators are obtained using the NUISANCE framework \cite{NUISANCE}.

\subsection{Data unfolded cross section results}
\label{sec:Results}
Figures \ref{fig:UnfRes} and \ref{fig:UnfResPe} present the final data cross section results compared to the alternative predictions described above. The regularization from the unfolding process, in the form of an output matrix, is applied to the prediction to allow the comparison. All regularization tables are shown in Appendix~\ref{app:matrix}. \cref{tbl:ResEe,tbl:ResEA,tbl:ResPA,tbl:ResPe,tbl:ResPe2,tbl:ResOpAng} in Appendix~\ref{app:xsecres} summarize the background subtracted data event counts and the unfolded cross section results for all six configurations of variables and binning. Note that the two-bin proton energy measurement is treated slightly differently as we do not apply the division by the bin width in Eq.~\ref{eq:XsecCalc}; this prevents the result from appearing counter-intuitive as the bins representing the two channels have very different widths. Moreover, for this two-bin measurement, we use direct inversion of the response matrix without regularization instead of the Wiener-SVD unfolding method so that the results are more directly comparable to a total cross section. All $p$-values discussed in this section are summarized in Table~\ref{tbl:pvalues}, figures showing the various GENIE comparisons are available in Appendix~\ref{app:GENIE}, and an additional NEUT comparison is shown in Appendix~\ref{app:NEUT}.\\

The data generally agrees with the models tested here, and the $p$-values are higher than those observed in MicroBooNE's previous result \cite{MicroBooNEXsec}. This is particularly the case in terms of lepton modeling where, in electron energy and angle, the data $p$-values are highest when compared to GENIE v3.0.6 G18\_10a\_02\_11a without the MicroBooNE tune and with GENIE AR23 v3.6.0 with $p$-values above 0.887. Both of these models predict a slightly lower count of events than the tuned GENIE G18T. It is worth noting that the application of the tune results in an enhancement of the quasi-elastic and meson exchange components.\\ 

There are more differences between models and disagreements with the data in terms of the hadronic system modeling. The data strongly disagrees with the GiBUU model without in-medium corrections \cite{INMEDIUMGIBUU} in proton energy with a $p$-value of 0.008 in the two-bin measurement, particularly in the 1e0p0$\pi$ bin. As demonstrated in \cite{INMEDIUMGIBUU}, modifying the GiBUU model by adding in-medium nucleon-nucleon interactions can decrease the 1e0p0$\pi$ prediction and lead to lower cross sections, closer to the other models. In that case we observe better compatibility with our data with $p$-values of, respectively, 0.215 and 0.081 for the full binning and the two-bin measurement. It can also be noted, as shown in Fig.~\ref{fig:sub:UnfResPetwobin} presenting the spread of model comparisons across the proton visibility threshold, that models with the best agreement in the visible proton region are not the same as the models with best agreement in the invisible proton region.
In proton angle and electron-proton opening angle, all models produce lower $p$-values. The models overpredict the cross section at larger cosines of these angles, closer to the incoming neutrino direction, compared to the data. For these observables GiBUU without in-medium NN corrections produces the highest $p$-values, respectively, 0.228 and 0.639. \\

Within GENIE, we investigate a single model for one interaction type at a time to evaluate the impact of that specific model on the agreement with all other aspects being equal. Those variations are produced without the MicroBooNE tune and are compared to the base model configuration without the tune. Note that these comparisons are done within GENIE v3.6.0, which is a more recent GENIE version than the one used for the MicroBooNE simulation. The base model $p$-values in this newer GENIE version are also reported in Table~\ref{tbl:pvalues} and are consistent with the previous version. 
Our base GENIE model uses the hA18 FSI model. We can also set the FSI model to the hN18. hN is a traditional hadron-nucleon INC model with similar features to NuWro and NEUT. hA is a custom model that has a more empirical approach and relies on data to consider a single interaction instead of the multiple hadron-nucleon interactions in cascade models \cite{CompHAHN}.  We also compare to the G4Bertini \cite{BertiniG4} FSI cascade model. Both alternative models, denoted GENIE v3.6.0 G18\_10b\_02\_11a (G18\_10b) using hN18 and GENIE v3.6.0 G18\_10d\_02\_11a (G18\_10d) using G4Bertini, show consistently lower $p$-values across all variables compared to the base model with hA, especially the G4Bertini model. \\

We also test various ground states, QE, and MEC models within GENIE using v3.6.0 and without the MicroBooNE tune applied. The SuSAv2 \cite{SuSAv2} model for QE events lowers the model $p$-values with respect to the data, such as, for example, the model set G21\_11a\_00\_000 (G21\_11a), but other variations of QE or MEC models do not significantly impact the results and provide similar predictions. Two variations without a particular impact include GENIE AR23 and GENIE G18\_10i\_02\_11b (G18\_10i). These both use the z-expansion QE form factors and AR23 uses SuSAv2 for MEC instead of Nieves. Another model set that provides a similar prediction to the baseline is the G18\_02a\_02\_11a (G18\_02a) model set that uses the relativistic Fermi gas (RFG) ground state and the Dytman MEC model \cite{DytmanMECModel}. \\

\begin{table*}[t]
\caption{$p$-values of comparisons to the various models tested for all six analysis variables and binnings.}
\label{tbl:pvalues}
{\begin{center} \begin{tabular}{|c|c|c|c|c|c|c|c|c|}
\hline
Generator & Model& Label on Figs. & $E_e$ &$KE_p$ & $\cos{\theta_{e}}$ & $\cos{\theta_{p}}$&$\cos{\theta_{ep}}$ & 2-bin $KE_p$ \\
\hline
\hline
\multirow{2}{*}{GENIE v3.0.6 } &G18\_10a\_02\_11a Tuned & G18T &0.682& 0.561& 0.327& 0.029&  0.143& 0.326\\
\cline{2-9}
& G18\_10a\_02\_11a & G18&0.989& 0.809& 0.921& 0.110& 0.480& 0.832\\
\hline
\hline
\multirow{6}{*}{GENIE v3.6.0 }& G18\_10a\_02\_11a& v3.6 G18& 0.987  & 0.861 & 0.915& 0.122 &0.514 & 0.914 \\
\cline{2-9}
&AR23\_20i\_00\_000&AR23 & 0.976 & 0.860 & 0.887& 0.095& 0.489& 0.852\\
\cline{2-9}
&G18\_10b\_02\_11a& G18\_10b&  0.934 & 0.799& 0.700 &  0.078& 0.318& 0.667\\
\cline{2-9}
&G18\_10d\_02\_11a&G18\_10d & 0.797& 0.558& 0.548& 0.014& 0.096& 0.452\\
\cline{2-9}
&G18\_10i\_02\_11b&G18\_10i &  0.993& 0.861& 0.920& 0.112& 0.494& 0.917\\
\cline{2-9}
&G18\_02a\_02\_11a& G18\_02a&  0.992& 0.867& 0.813& 0.321& 0.665& 0.907\\
\cline{2-9}
&G21\_11a\_00\_000 &G21\_11a & 0.782& 0.608& 0.383& 0.064& 0.245& 0.417\\
\hline
\hline
\multirow{2}{*}{GiBUU 2025} & w/o In-med NN corr& GiBUU&0.942 & 0.049 &0.824 &0.228 & 0.639 &0.008 \\
\cline{2-9}
& w In-med NN corr & 
GiBUU inmed  &0.841& 0.215& 0.607& 0.111& 0.517& 0.081\\
\hline
\hline
\multirow{2}{*}{NEUT } &v5.7.0&NEUT &  0.812& 0.751& 0.536& 0.089& 0.279& 0.506\\
\cline{2-9}
& v6.1.4 & NEUT v6.1.4 & 0.832&0.674 &0.547 &0.103 &0.294 &0.481 \\
\hline
\hline
NuWro & v21.09.2&NuWro &0.946& 0.775& 0.697& 0.038& 0.327& 0.730\\
\hline
\end{tabular} \end{center}}
\end{table*}

Within NEUT, we test the impact of using the energy dependent relativistic mean field (EDRMF)~\cite{NEUTEDRMF,EDRMF} ground state model instead of LFG. This model is the one implemented by default in NEUT version 6.1.4. We find a negligible impact on the agreement with data. \\

Finally, our results can be compared to a few relevant previous measurements. The first iteration of this measurement, using half of the MicroBooNE BNB dataset~\cite{MicroBooNEXsec}, found the same hierarchy of agreement between models, but with overall lower $p$-values and more differences between NEUT and the other models. An overprediction was observed at intermediate shower energy in the MicroBooNE search for an anomalous excess of electron-neutrino events \cite{PELEEnew}, but such an effect is not as obvious in the leptonic variable distributions in this measurement's coarser binning. A similar exclusive 1eNp0$\pi$ cross section measurement was performed with MicroBooNE NuMI data~\cite{nueCCNuMI} in electron energy, visible energy and the cosine of the opening angle.  In this measurement, a slight underprediction was observed leading to models with higher cross sections, such as NEUT, being favored.  This measurement uses a different beam and has different dominant systematic uncertainties. However, just like in this measurement, good overall agreement with most models tested was found. We also note that a shape discrepancy in lepton-proton opening angle is emerging as a common feature across MicroBooNE measurements including one in muon flavor~\cite{Stevenpaper} and in the electron neutrinos in NuMI~\cite{nueCCNuMI}. Finally, MicroBooNE has also made a muon neutrino measurement exploring the cross section across the proton visibility threshold in kinetic energy~\cite{BensAna}. It found a preference for a higher prediction of events without a proton above the visibility threshold, and therefore the GiBUU generator, unlike in the measurement presented here.  It is possible that this is driven by differences in the design of these measurements, including the particle kinetic energy thresholds, or the fact that the muon neutrino measurement is inclusive while this measurement does not include visible pions.  
\\

In the future, new models will be able to provide additional insight on neutrino interactions on argon. An updated Valencia MEC model~\cite{MECValenciaUpdate} that implements a more consistent treatment of the nucleon self-energy and a simplified estimation of the $\Delta$ resonance self energy is being included in the neutrino generators. 
Moreover, another FSI model will be available for comparisons within GENIE: the Li\`ege Intranuclear Cascade model (INCL$++$)~\cite{INCL,INCL++}. INCL$++$ has a more sophisticated nuclear model and in-medium corrections than GENIE hA and hN, NEUT and NuWro. 
Future publications will compare to these and other new models.
\section{Summary and conclusions}
In conclusion, this paper presents an electron neutrino cross section measurement on argon without pions in the final state in the MicroBooNE detector using $13.09\times10^{20}$ POT of BNB data. The measurement is provided in variables that describe the leptonic and hadronic systems: electron and leading proton kinetic energy and angle with respect to the beam direction as well as the opening angle between the leading proton and the electron directions. In proton kinetic energy, we separate events below and above the proton visibility threshold of 50~MeV. In general, agreement with the different models is good in lepton kinematics and shows some overprediction of the tuned base model (GENIE G18 with MicroBooNE tune) in the hadronic system. This is especially the case in proton angle where the agreement is lower, particularly in the forward region. This is also the case in electron-proton opening angle but to a lesser extent. This measurement did not show any strong sensitivity to different QE or MEC models except for a lower agreement when using the SUSAv2 QE model. Within GENIE, the hA18 FSI model is the preferred model.
This measurement provides insight for further tuning of generators and understanding of electron neutrino cross sections on argon in view of LArTPC experiments such as the SBN program \cite{SBN} and the future long baseline neutrino oscillation experiment DUNE \cite{DUNE_ND,DUNE_FD}.
\FloatBarrier
\section{Acknowledgments}
This document was prepared by the MicroBooNE Collaboration using the resources of the Fermi National Accelerator Laboratory (Fermilab), a U.S. Department of Energy, Office of Science, Office of High Energy Physics HEP User Facility. Fermilab is managed by Fermi Forward Discovery Group, LLC, acting under Contract No. 89243024CSC000002. MicroBooNE is supported by the
following: 
the U.S. Department of Energy, Office of Science, Offices of High Energy Physics and Nuclear Physics; 
the U.S. National Science Foundation; 
the Swiss National Science Foundation; 
the Science and Technology Facilities Council (STFC), part of United Kingdom Research and Innovation (UKRI);
the Royal Society (United Kingdom);
the UKRI Future Leaders Fellowship;
the NSF AI Institute for Artificial Intelligence and Fundamental Interactions;
and the European Union’s Horizon 2020 research and innovation programme under the Marie Sk\l{}odowska-Curie grant agreement No. 101003460 (PROBES).

Additional support for the laser calibration system and cosmic ray tagger was provided by the Albert Einstein Center for Fundamental Physics, Bern, Switzerland. We also acknowledge the contributions of technical and scientific staff to the design, construction, and operation of the MicroBooNE detector as well as the contributions of past collaborators to the development of MicroBooNE analyses, without whom this work would not have been possible. 

For the purpose of open access, the authors have applied a Creative Commons Attribution (CC BY) public copyright license to any Author Accepted Manuscript version arising from this submission.

\FloatBarrier
\onecolumngrid
\clearpage

\appendix{}
\onecolumngrid
\section{1e0p0$\pi$ phase space distributions}
\label{app:cuts}
\begin{figure}[H]
  \subfigure[]
  {
    \includegraphics[width=8cm]{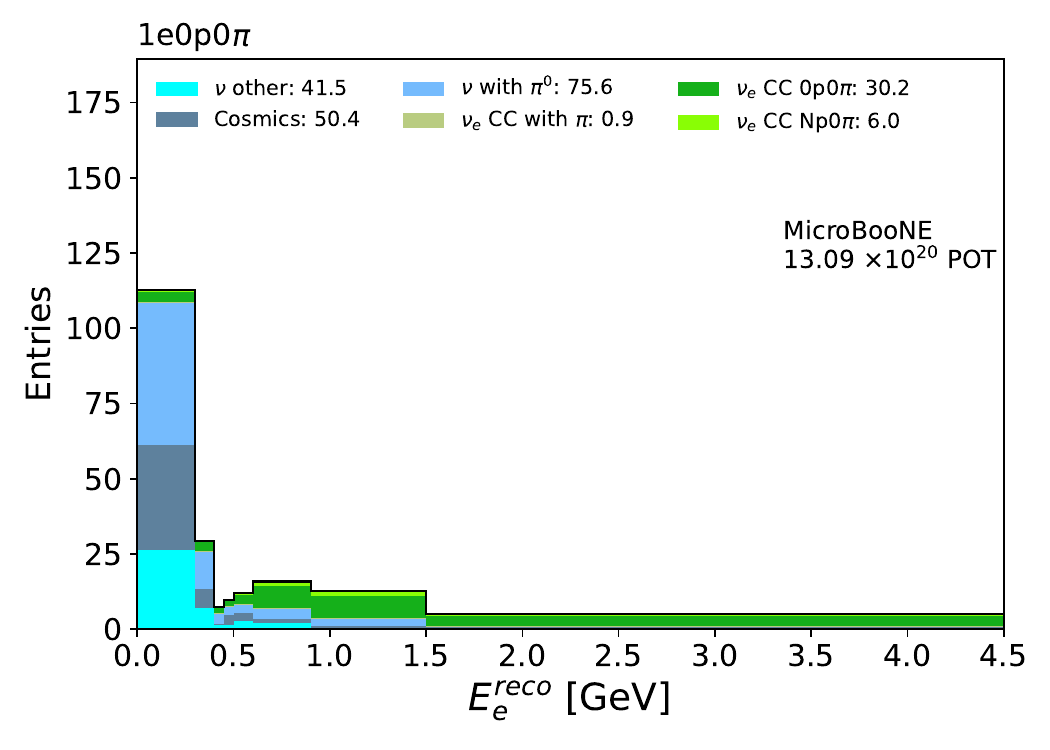}}
  \subfigure[]  
   { \includegraphics[width=8cm]{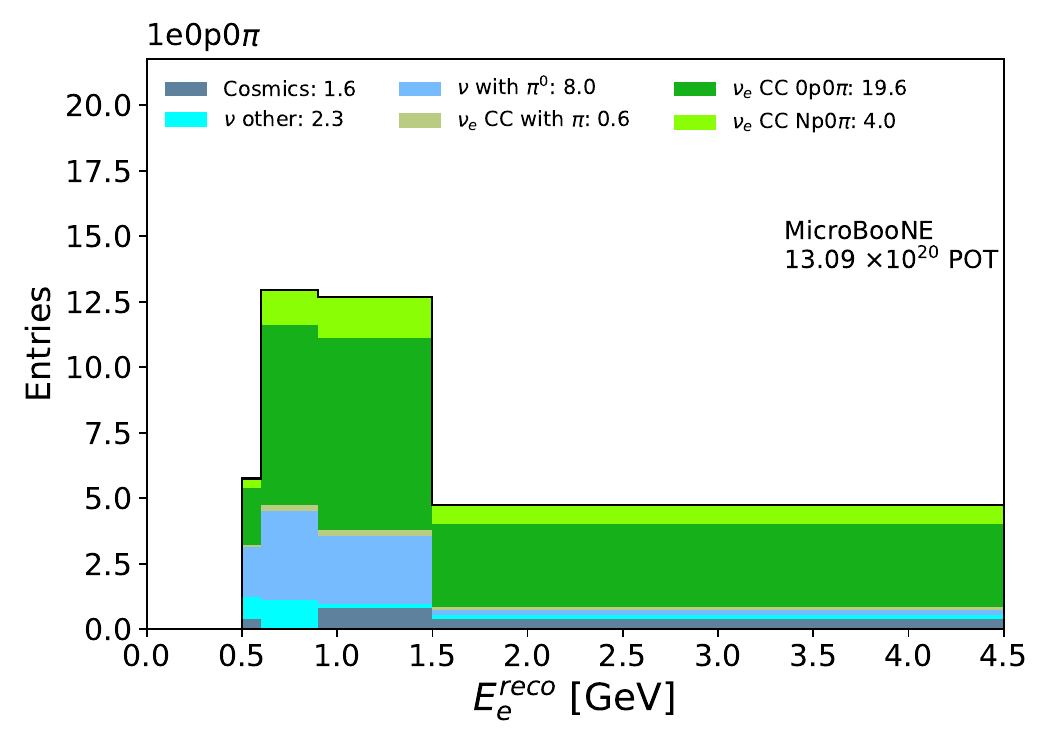}}
\caption{1e0p0$\pi$ selection predicted event rates (a) without and (b) with phase space cuts, as a function of electron energy.} 
\label{fig:app0pE}
\end{figure}

\begin{figure}[H]
  \subfigure[]
  {
    \includegraphics[width=8cm]{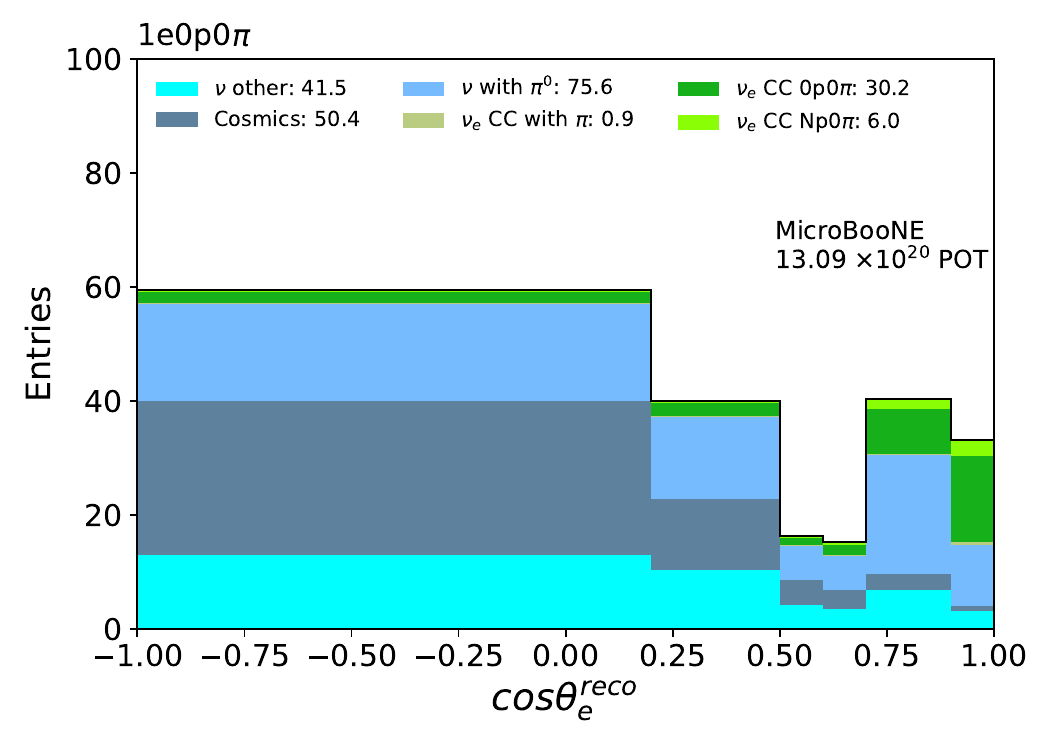}}
  \subfigure[]
   { \includegraphics[width=8cm]{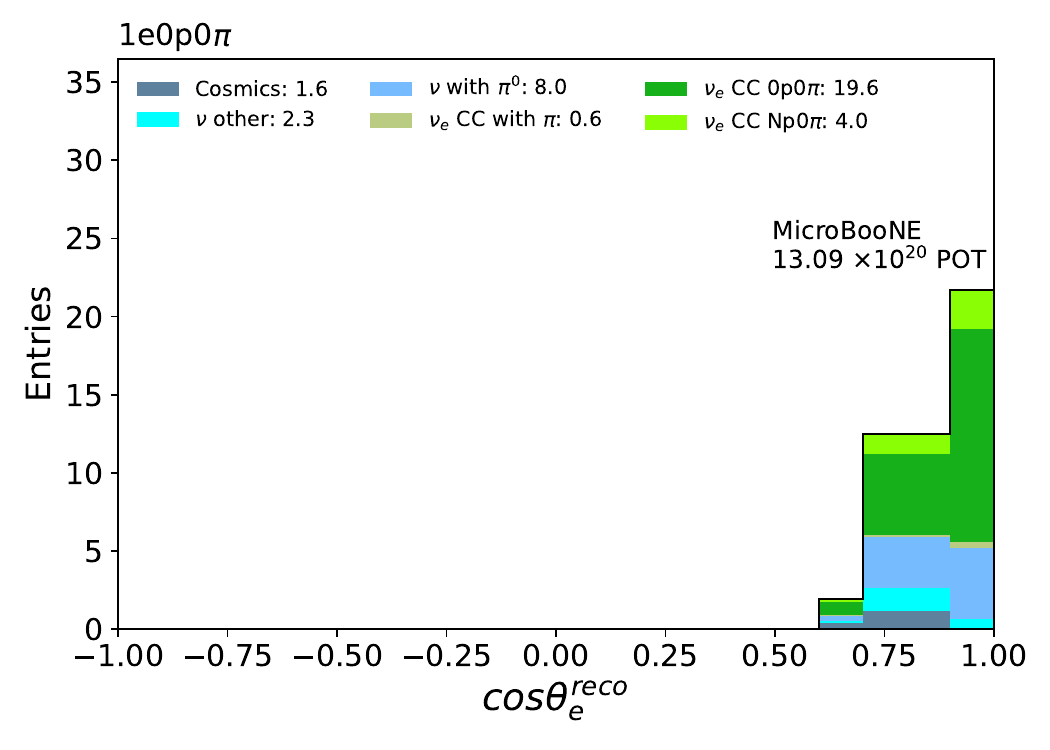}}
\caption{1e0p0$\pi$ selection predicted event rates (a) without and (b) with phase space cuts, as a function of electron angle.} 
\label{fig:app0pAng}
\end{figure}

\FloatBarrier
The two phase space cuts required for the 1e0p0$\pi$ events are: the electron energy is greater than 0.5~GeV and the cosine of the electron angle is greater than 0.6. Figure \ref{fig:app0pE} shows the predicted 1e0p0$\pi$ candidate  events before and after the phase space cut on electron energy. Similarly, Fig.~\ref{fig:app0pAng} shows the candidate events before and after the electron angle phase space cut. All other selection cuts are applied. This improves the selection purity from 12\% to 57\%.

\FloatBarrier
\section{Breakdown of the fractional systematic uncertainties}
\label{app:Systs}
\begin{figure}[H]
  \subfigure[]
  {\includegraphics[width=7cm]{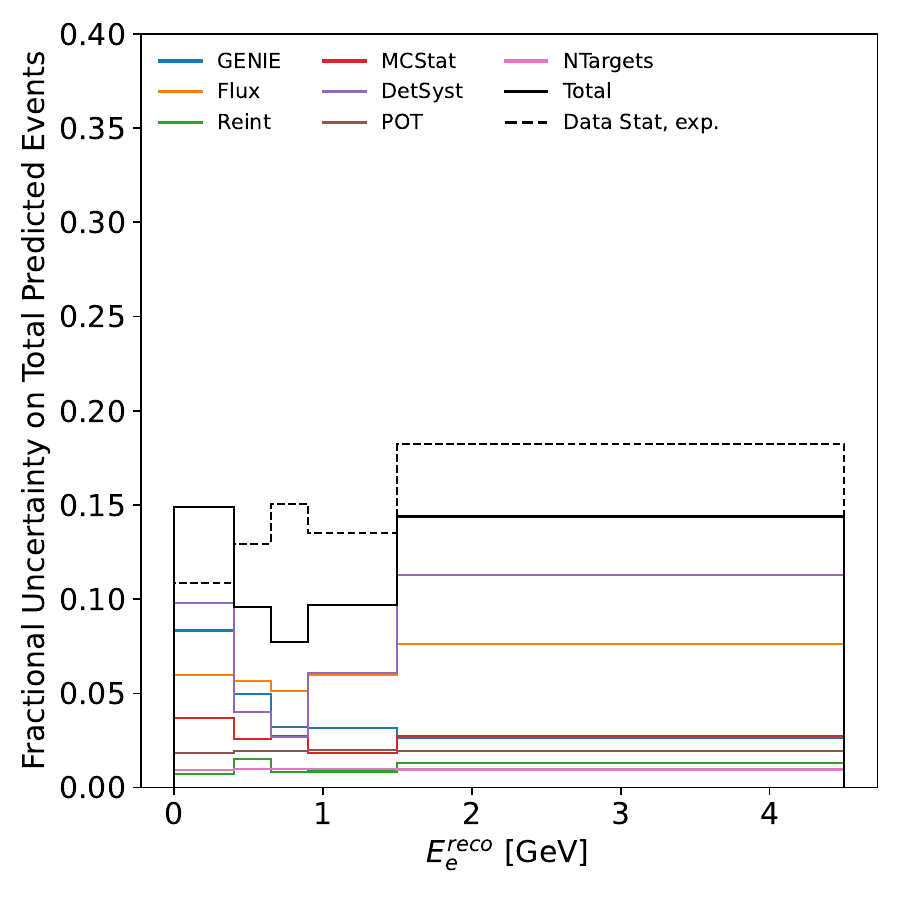}}
  \subfigure[]
   {\includegraphics[width=7cm]{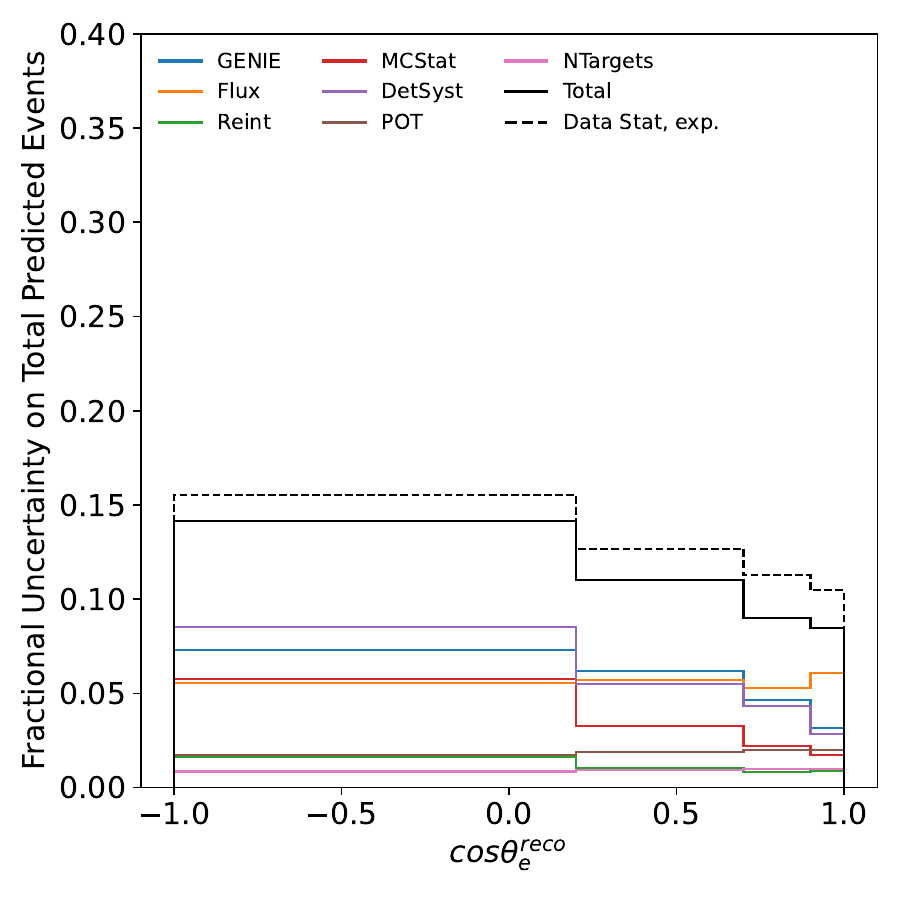}}
     \subfigure[]
   {\includegraphics[width=7cm]{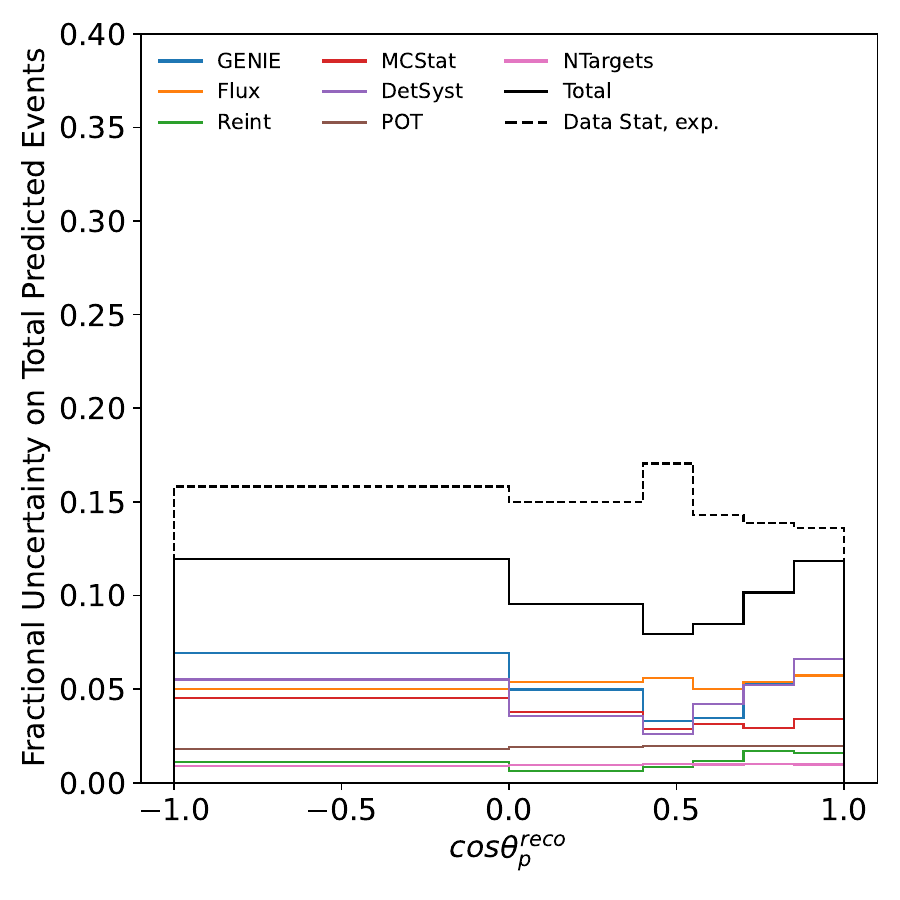}}
     \subfigure[]
   {\includegraphics[width=7cm]{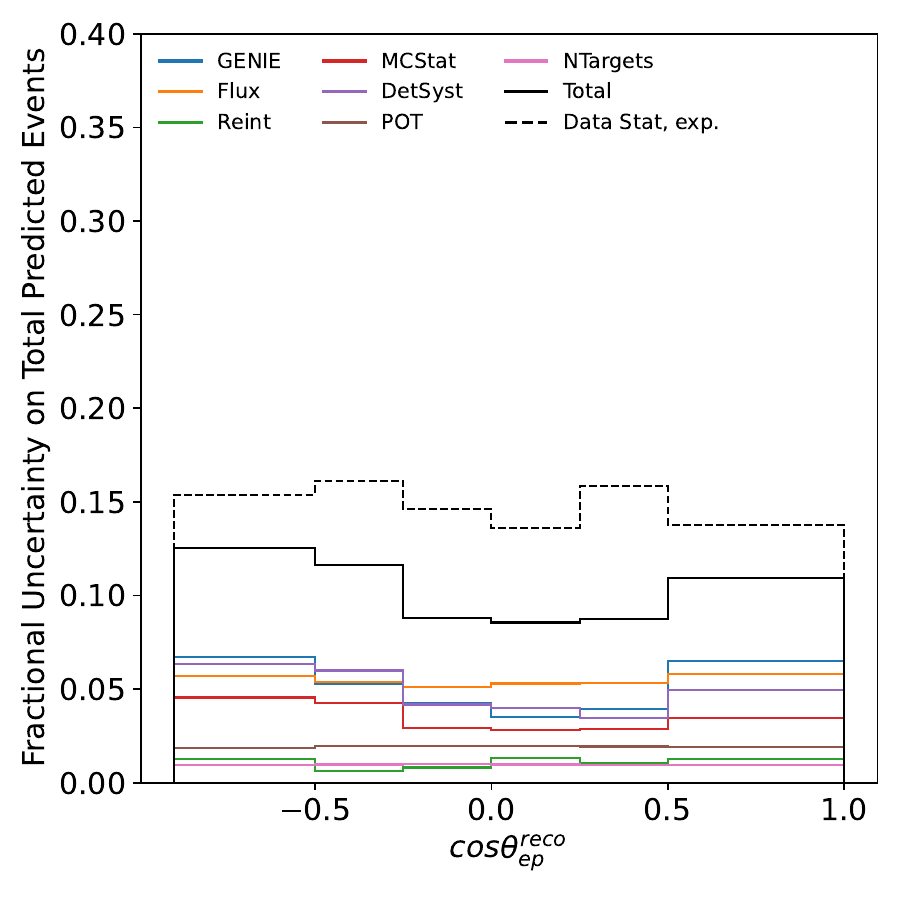}}
  \subfigure[]
  {\includegraphics[width=7cm]{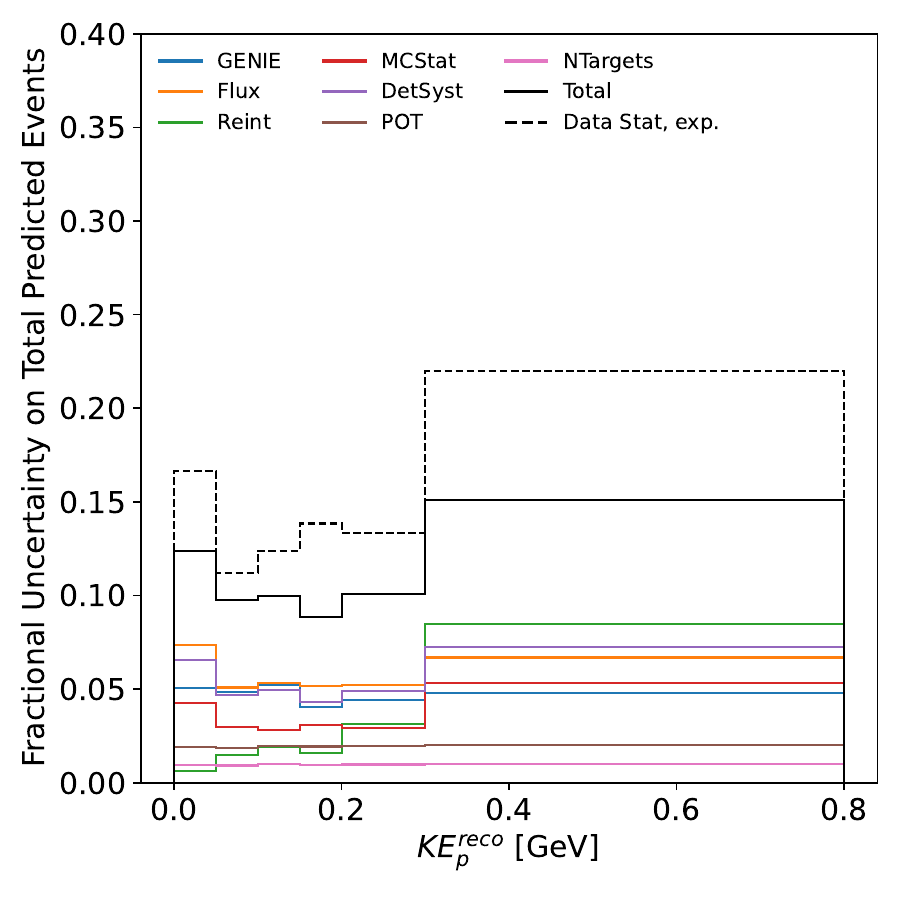}}
   \hspace*{\fill} 
  \subfigure[]
   {\includegraphics[width=7cm]{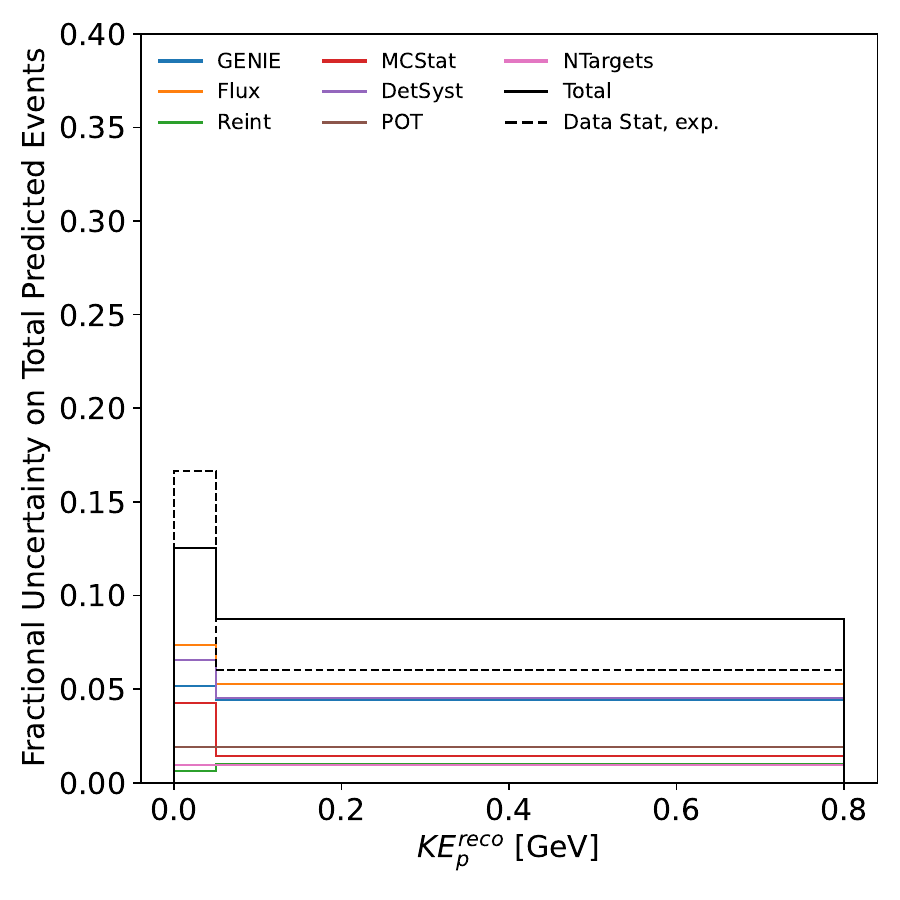}}
\caption{Breakdown of the fractional systematic uncertainties for: the 1eNp0$\pi$ channel in (a) electron energy, (b) electron angle, (c) proton angle and (d) opening angle; and for the 1eXp0$\pi$ channel in  proton kinetic energy (e) full binning and (f) measurement across the proton visibility threshold.} 
\label{fig:appsysts}
\end{figure}

Figure \ref{fig:appsysts} presents the distributions of fractional systematic uncertainties broken down by categories for each variable.

\FloatBarrier
\section{Sideband distributions}
\label{app:sidebands}
\begin{figure}[H]
  \subfigure[]{
    \includegraphics[width=8cm]{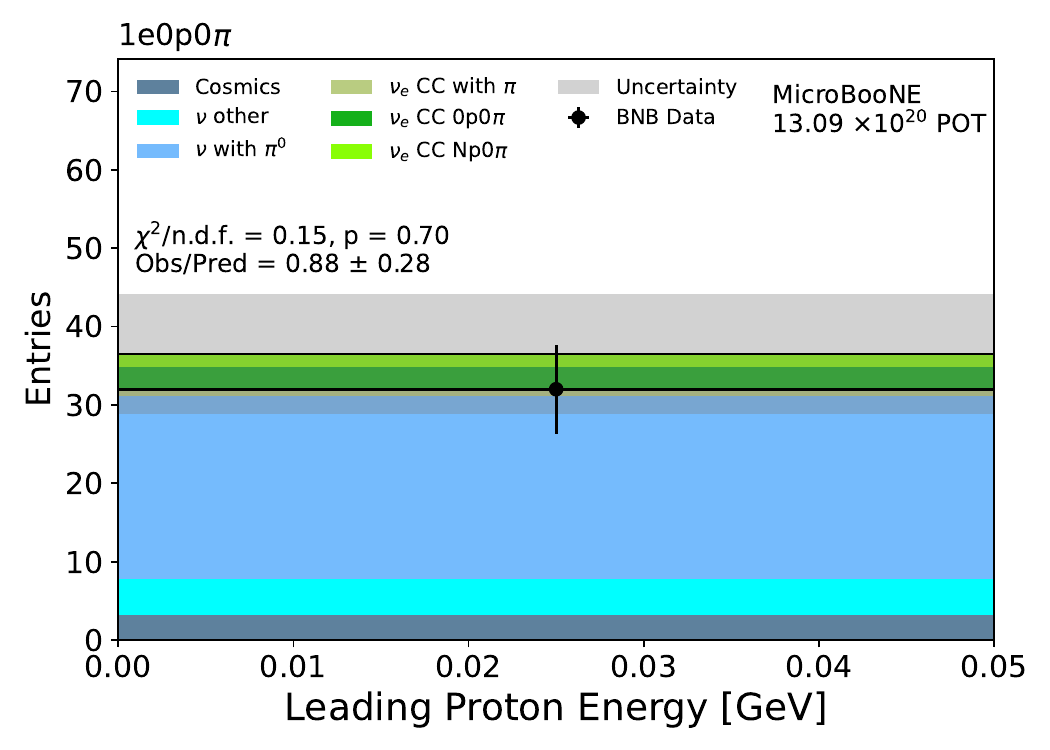} \label{subfig:0pBDTEp}}
  \subfigure[]
   { \includegraphics[width=8cm]{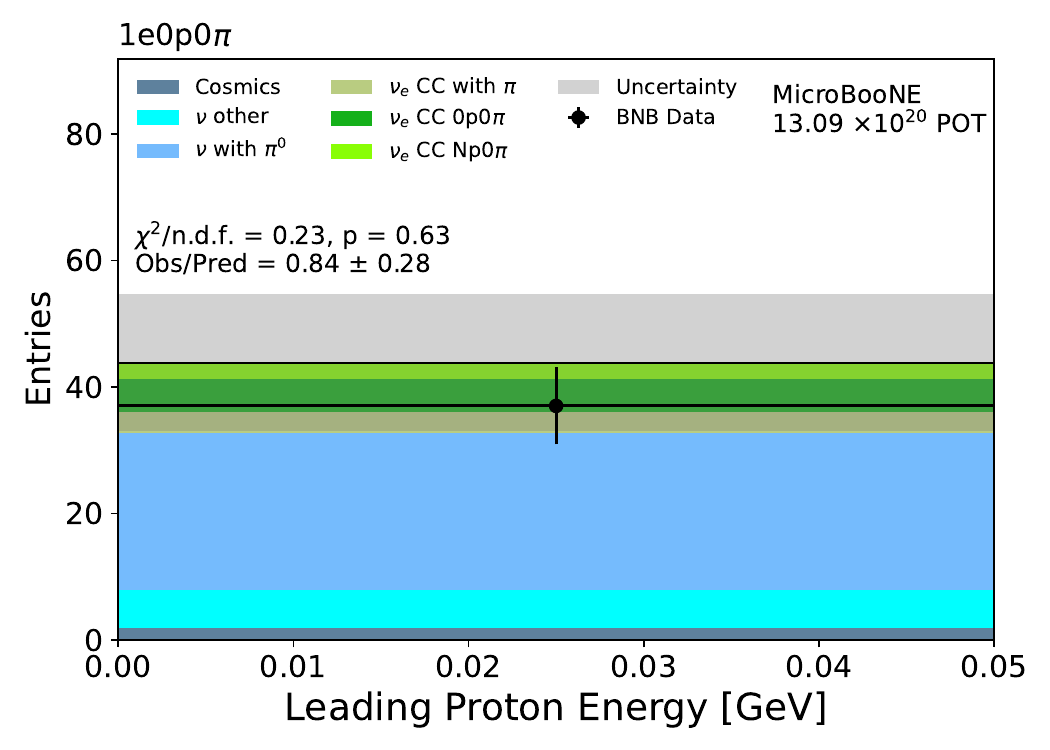}\label{subfig:0p2+Ep}}
\caption{Inverted BDT sideband (a) and sideband with at least two showers (b) for the 1e0p0$\pi$ bin.} 
\label{fig:Side0P}
\end{figure}

\begin{figure}[H]
  \subfigure[]{
    \includegraphics[width=8cm]{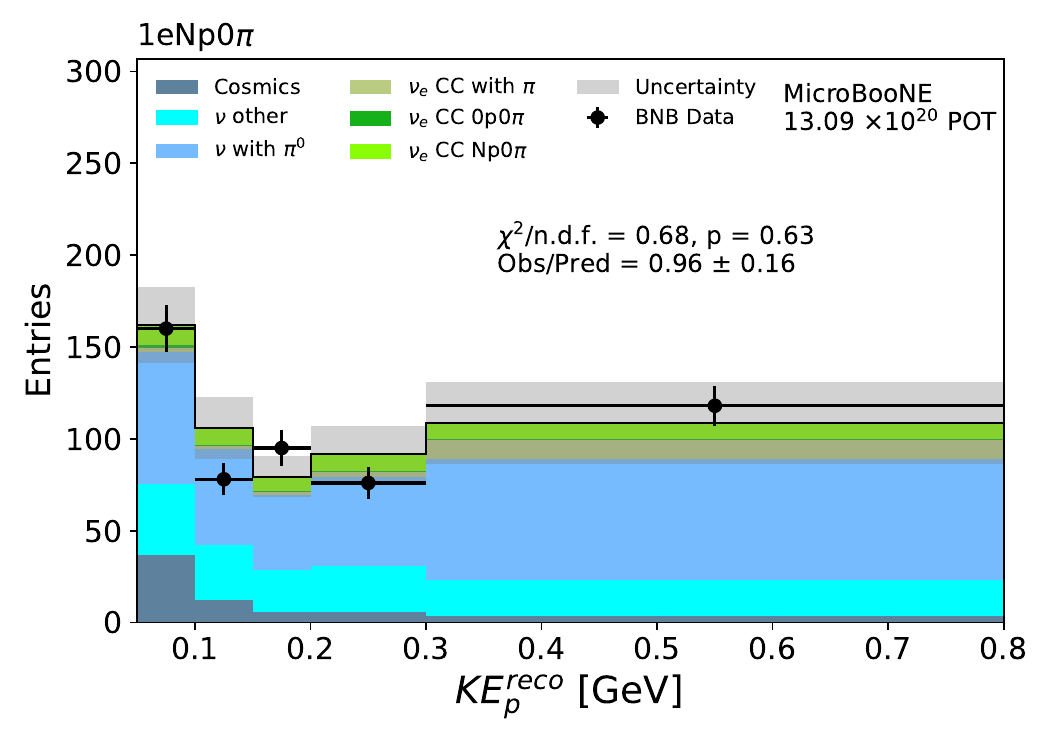}\label{subfig:NpBDTEp}}
  \subfigure[]
   { \includegraphics[width=8cm]{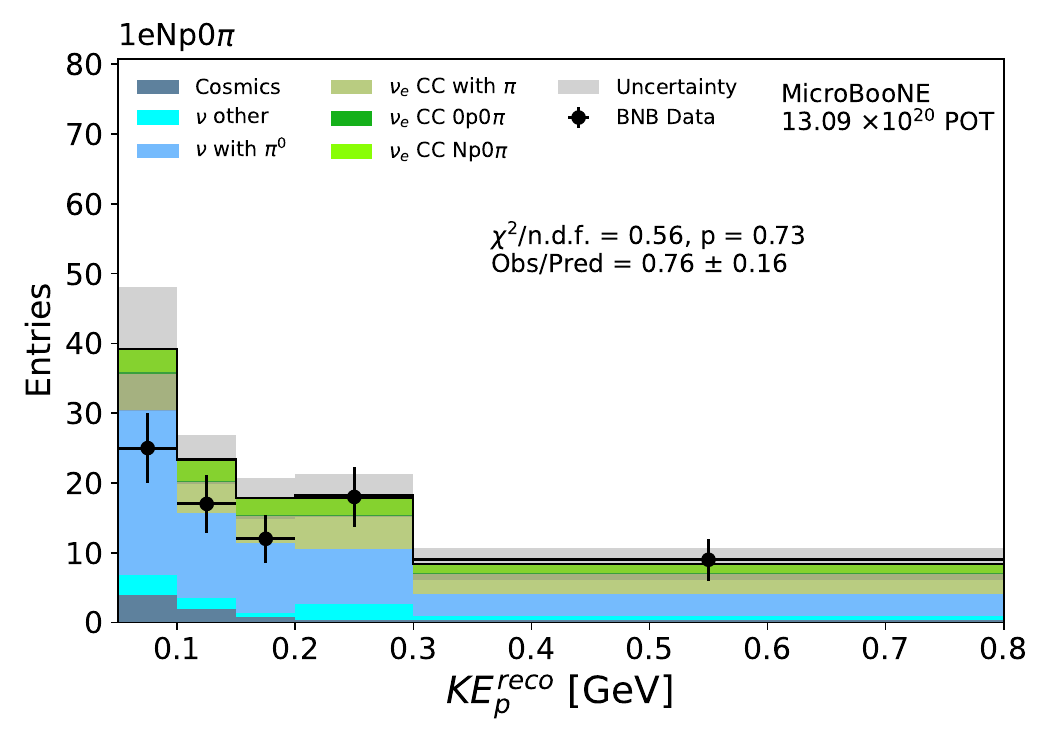}\label{subfig:Np2+Ep}}
\caption{Inverted BDT sideband (a) and sideband with at least two showers (b) for the 1eNp0$\pi$ channel in proton energy.} 
\label{fig:sideprotonE}
\end{figure}

\begin{figure}[H]
  \subfigure[]{
    \includegraphics[width=8cm]{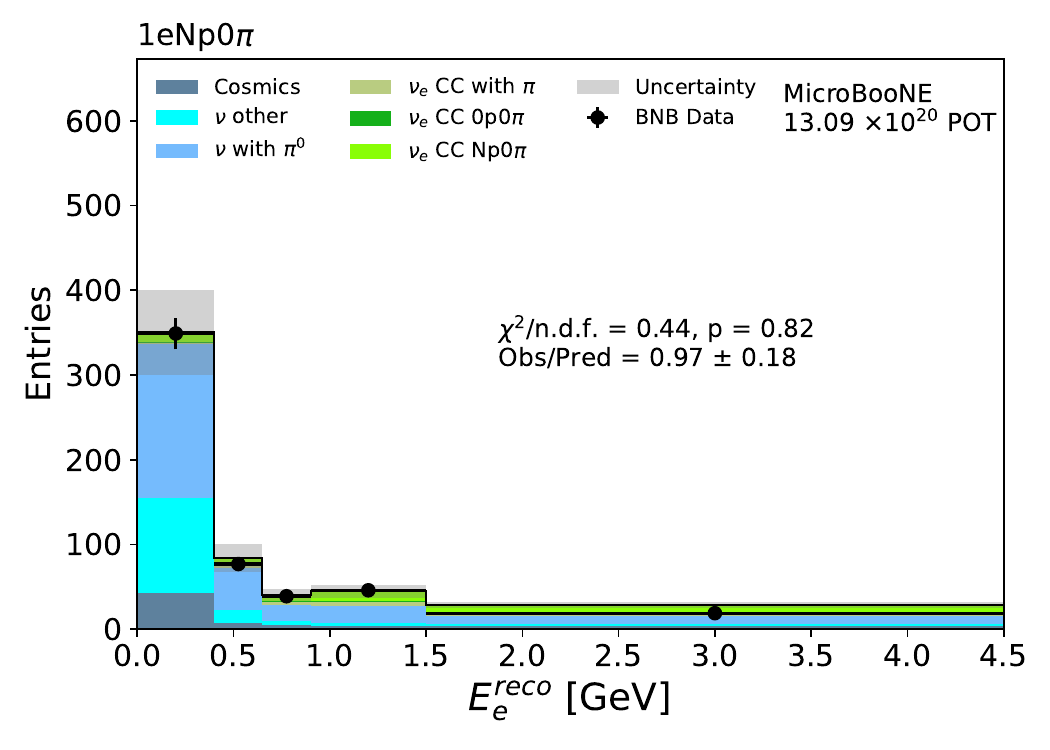}}
  \subfigure[]
   { \includegraphics[width=8cm]{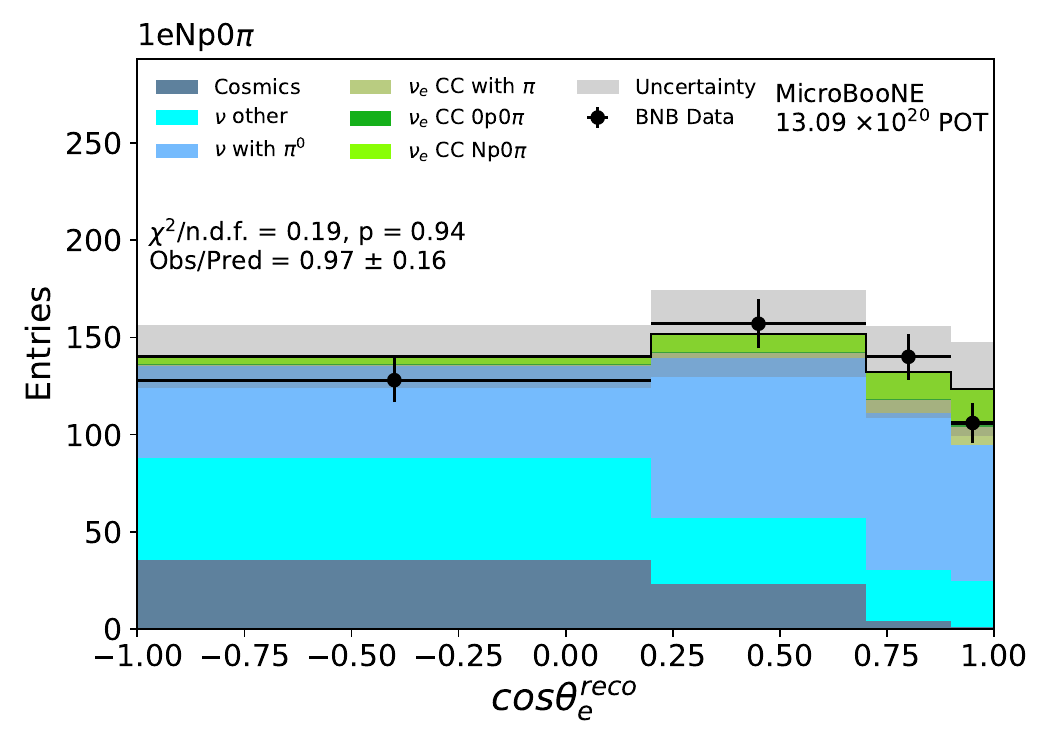}}
     \subfigure[]
   { \includegraphics[width=8cm]{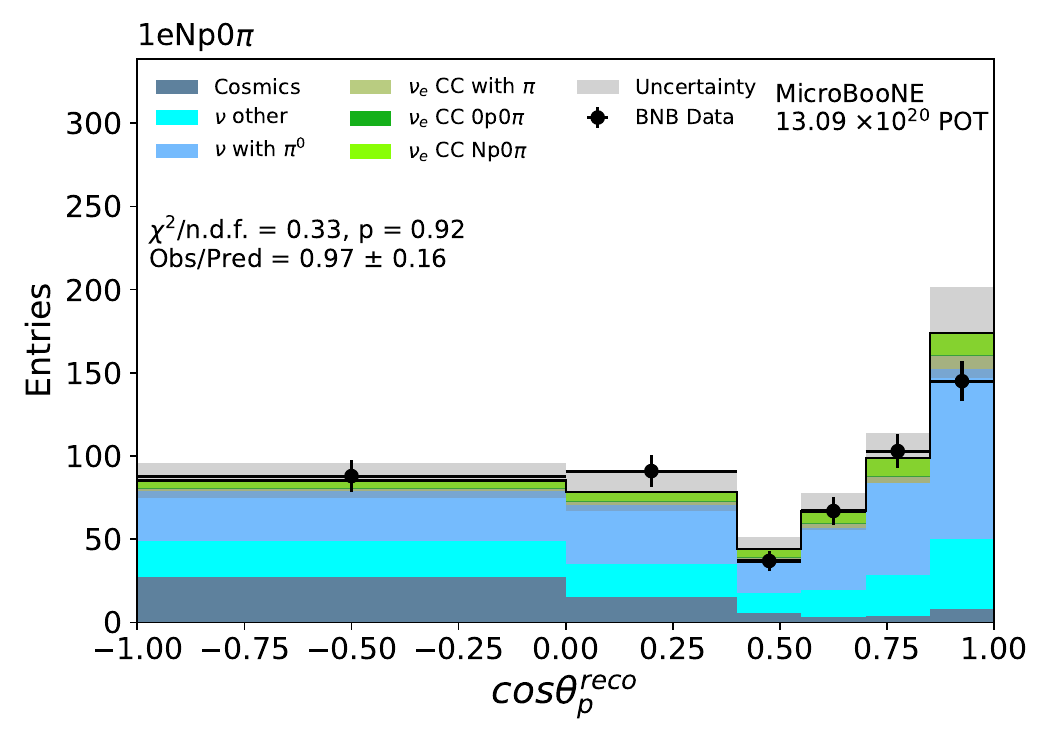}}
    \hspace*{\fill} 
     \subfigure[]
   { \includegraphics[width=8cm]{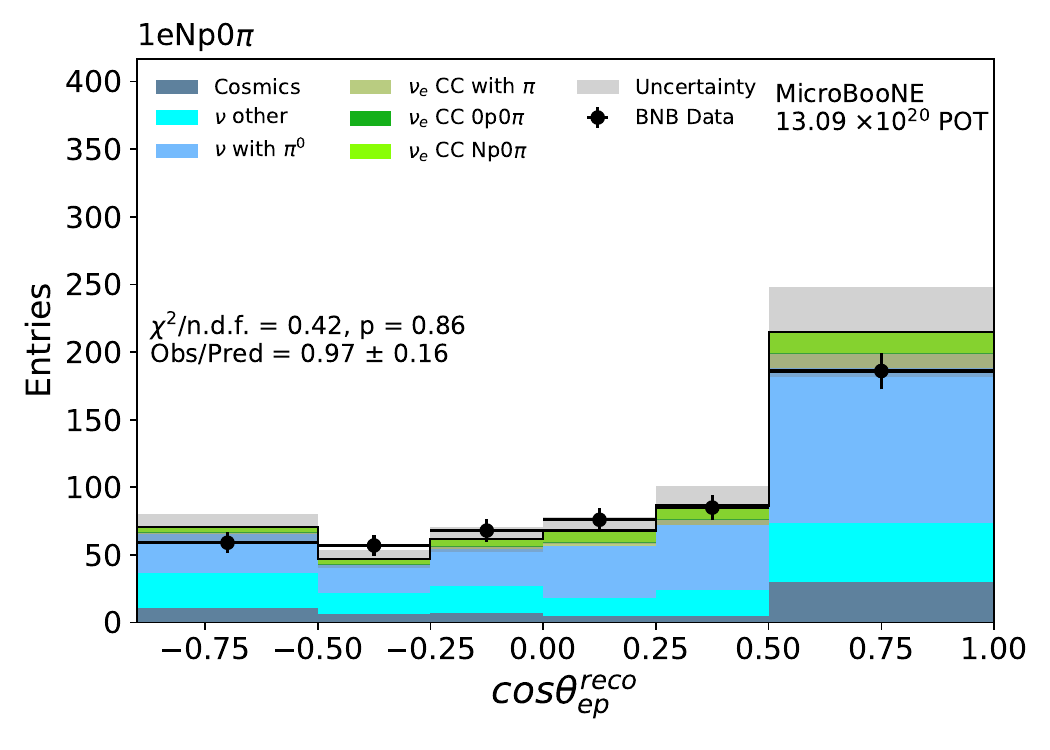}}
\caption{Inverted BDT sideband in the 1eNp0$\pi$ channel for (a) electron energy, (b) electron angle, (c) proton angle and (d) opening angle.} 
\label{fig:sideInvBDT}
\end{figure}

\begin{figure}[H]
  \subfigure[]{
    \includegraphics[width=8cm]{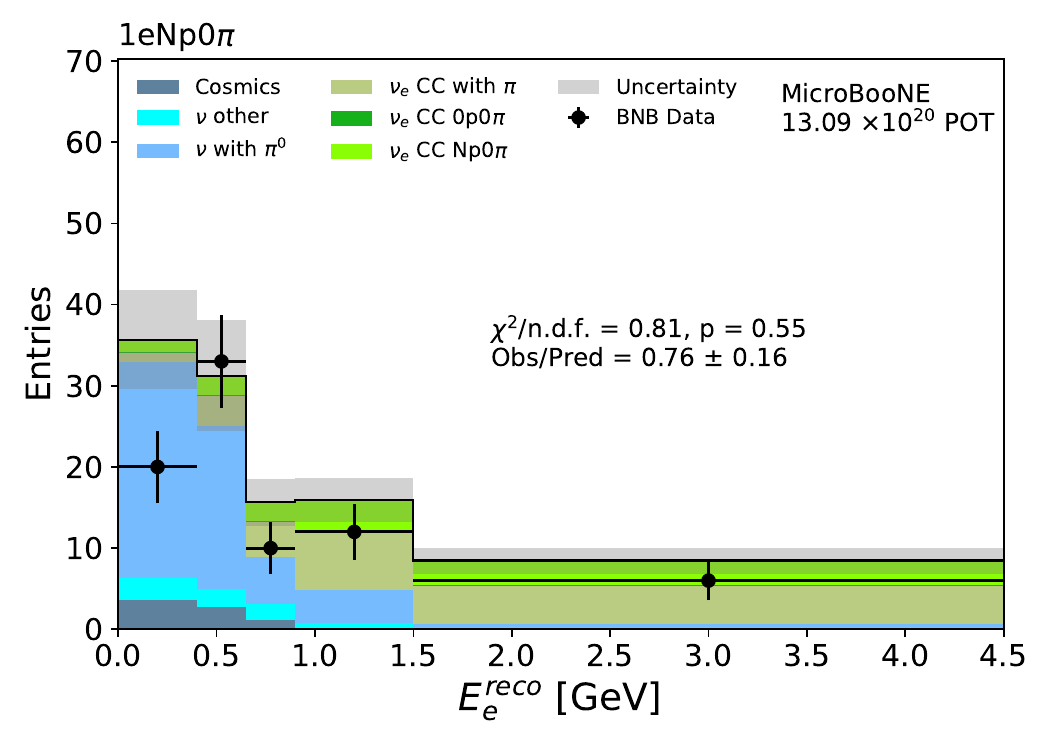}}
  \subfigure[]
   { \includegraphics[width=8cm]{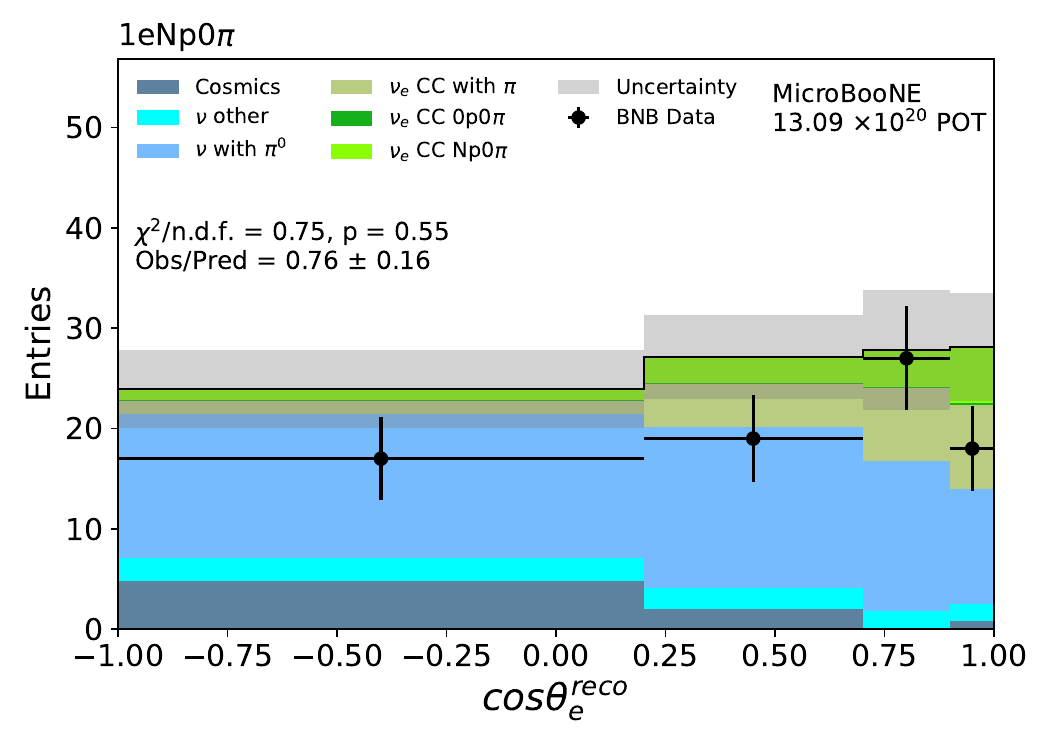}}
     \subfigure[]
   { \includegraphics[width=8cm]{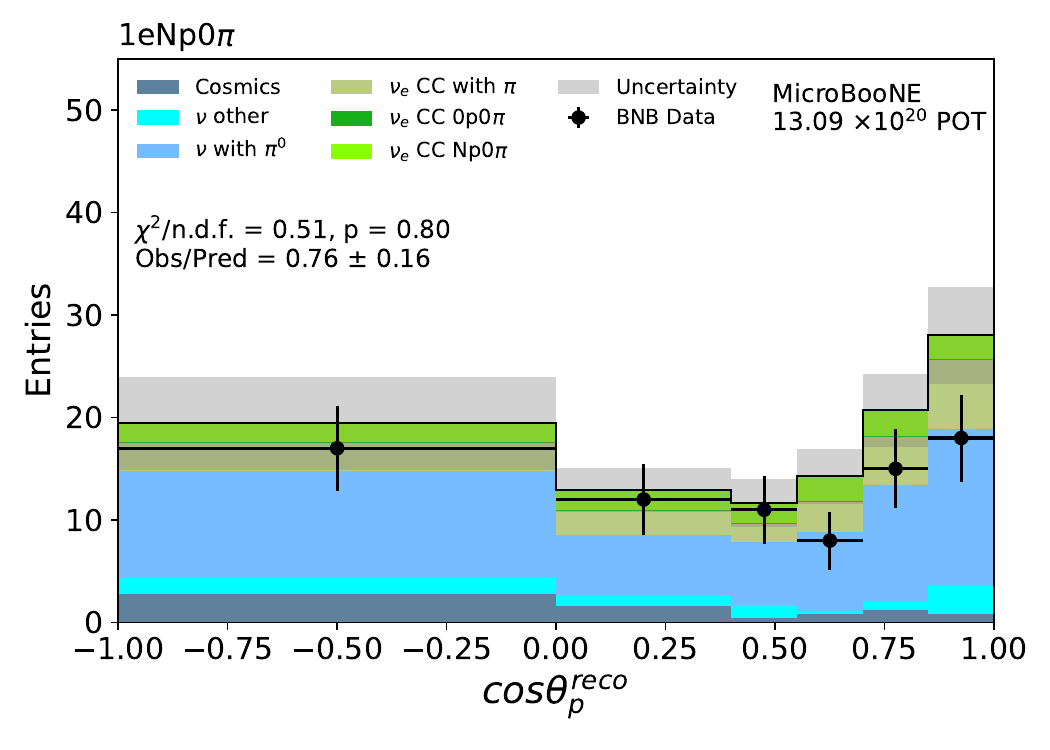}}
    \hspace*{\fill} 
     \subfigure[]
   { \includegraphics[width=8cm]{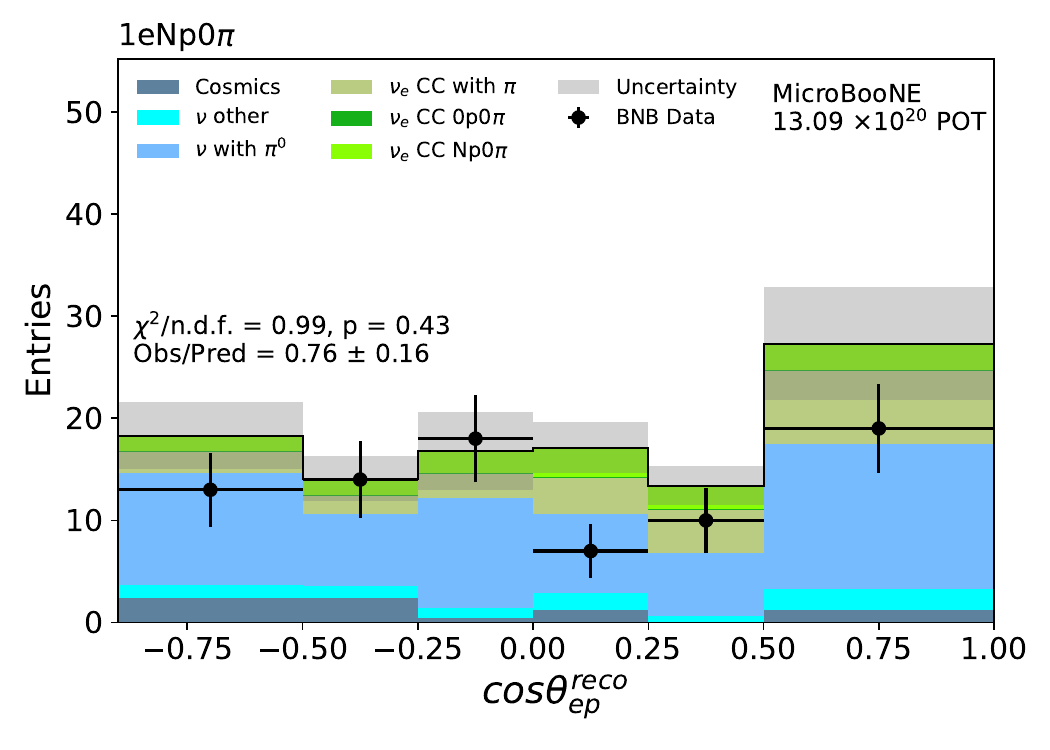}}
\caption{Sideband with at least two showers in the 1eNp0$\pi$ channel for (a) electron energy, (b) electron angle, (c) proton angle and (d) opening angle.} 
\label{fig:side2pp}
\end{figure}

As described in section \ref{sec:Distribs}, we use two sidebands to assess the quality of the background modeling for this measurement. 
In the first sideband we invert the BDT selection criteria to validate the modeling of the background adjacent to our selection; all of the other selection criteria remain the same.
This sideband is shown in Fig.~\ref{subfig:0pBDTEp} for the 1e0p0$\pi$ channel. For the 1eNp0$\pi$ channel, Fig.~\ref{subfig:NpBDTEp} shows the sideband for proton energy and Fig.~\ref{fig:sideInvBDT} for other variables. The data-simulation agreement in this sideband, taking into account the total event rate uncertainties, is very good, with $p$-values between 0.63 and 0.94. The other sideband uses the analysis selection, replacing the one contained shower requirement with a requirement that there be at least two showers, to address the main background: events with neutral pions. Figure~\ref{subfig:0p2+Ep} shows this sideband for the 1e0p0$\pi$ channel. Figure~\ref{subfig:Np2+Ep} and Fig.\ref{fig:side2pp} show the distribution of events in this sideband for the different 1eNp0$\pi$ channel variables. The data/MC agreement is also very good with $p$-values between 0.44 and 0.80.

\FloatBarrier
\section{Fake data studies}
\label{sec:fakedata}
Before unblinding the data, a number of choices had to be made regarding the analysis binning and the regularization method. Robustness tests of the analysis method were also performed. First, closure tests where we input the predicted signal from the nominal simulation into the unfolding procedure as the background-subtracted data were performed. Perfect ($p$-values = 1) post-unfolding agreement with the truth-level predicted cross section was found for all variables validating the overall procedure. Then, fake data produced using an alternative neutrino interaction generator, NuWro \cite{NuWro}, were analyzed as if they were real data in order to prove the robustness of the procedure against any base GENIE model-dependence. 
The analysis was found to be robust as better agreement with the NuWro predicted cross section than the base GENIE model was found for all variables when using the NuWro fake data.   
The NuWro fake data in proton energy was also reweighted in three different ways as additional tests of the analysis robustness: with the normalization decreased by 15\%, increased by 15\%, and normalized to the GiBUU prediction. The GiBUU normalization is obtained using the ratio to the NuWro prediction at truth level. This study was performed because of the large discrepancy in the 1e0p0$\pi$ bin and to ensure that the analysis would be robust if the data were to have a significant disagreement with the GENIE prediction.
In all cases the unfolded fake data showed better agreement with the input data rather than the base GENIE model. Finally, studies with this fake data were performed in order to optimize the binning in terms of model separation and predicted uncertainty per bin after unfolding as well as to choose a regularization method with the least visual impact on model separation.

\FloatBarrier

\section{Neutrino generator main differences}

This appendix presents a table of the model differences for the generators used to compare to the measurement presented in this paper.

\label{app:models}
\begin{center}

\begin{table}[H]
\caption{Summary of the main model differences for the generators used in this paper.}
\label{tbl:models}
\begin{center} 
\begin{tabular}{|c|c|c|c|c|c|c|c|c|}
\hline
Generator & Model& Label on Figs. & Ground state & QE & MEC & FSI & RES & DIS\\
\hline
\hline
GENIE   &G18\_10a\_02\_11a Tuned & G18T &\multirow{2}{*}{LFG }&\multirow{2}{*}{Valencia }&\multirow{2}{*}{Valencia }&\multirow{2}{*}{hA18}&\multirow{2}{*}{BS}& \multirow{2}{*}{BY}\\
\cline{2-3}
v3.0.6 & G18\_10a\_02\_11a & G18& & & & & & \\
\hline
\hline
\multirow{2}{*}& G18\_10a\_02\_11a& v3.6 G18&  LFG& Valencia& Valencia& hA18& BS& BY \\
\cline{2-9}
&AR23\_20i\_00\_000&AR23 & LFG + tail & Valencia $+$ Z$-$exp &SuSAv2 &hA18 & BS & BY\\
\cline{2-9}
GENIE &G18\_10b\_02\_11a& G18\_10b&  LFG& Valencia& Valencia& hN18& BS& BY  \\
\cline{2-9}
v3.6.0&G18\_10d\_02\_11a&G18\_10d & LFG& Valencia& Valencia& G4Bertini& BS& BY  \\
\cline{2-9}
\multirow{3}{*}&G18\_10i\_02\_11b&G18\_10i & LFG &Valencia$+$ Z$-$exp & Valencia& hN18& BS& BY \\
\cline{2-9}
&G18\_02a\_02\_11a& G18\_02a& RFG &LS &Dytman & hA18& BS&BY \\
\cline{2-9}
&G21\_11a\_00\_000 &G21\_11a & LFG& SuSAv2& SuSAv2& hA18& BS& BY  \\
\hline
\hline
GiBUU & w/o In-med NN corr& GiBUU& \multirow{2}{*}{LFG} &\multirow{2}{*}{custom} & \multirow{2}{*}{empirical} & \multirow{2}{*}{custom} &\multirow{2}{*}{custom/MAID} & \multirow{2}{*}{PYTHIA}\\
\cline{2-3}
2025 & w In-med NN corr &
GiBUU inmed  & & & & & & \\
\hline
\hline
NEUT   &v5.7.0 & NEUT &LFG&\multirow{2}{*}{Valencia }&\multirow{2}{*}{Valencia }&\multirow{2}{*}{INC}&\multirow{2}{*}{BS}& \multirow{2}{*}{custom}\\
\cline{2-4}
 & v6.1.4 & NEUT v6.1.4 &EDRMF & & & & & \\
\hline
\hline
NuWro & v21.09.2&NuWro &LFG&LS & Valencia& INC &ARS &BY \\
\hline
\end{tabular} \end{center}
\end{table}
\end{center}

\FloatBarrier
\section{Unfolded results compared to various GENIE tunes}
\label{app:GENIE}
\begin{figure}[H]
  \subfigure[]{
    \includegraphics[width=8cm]{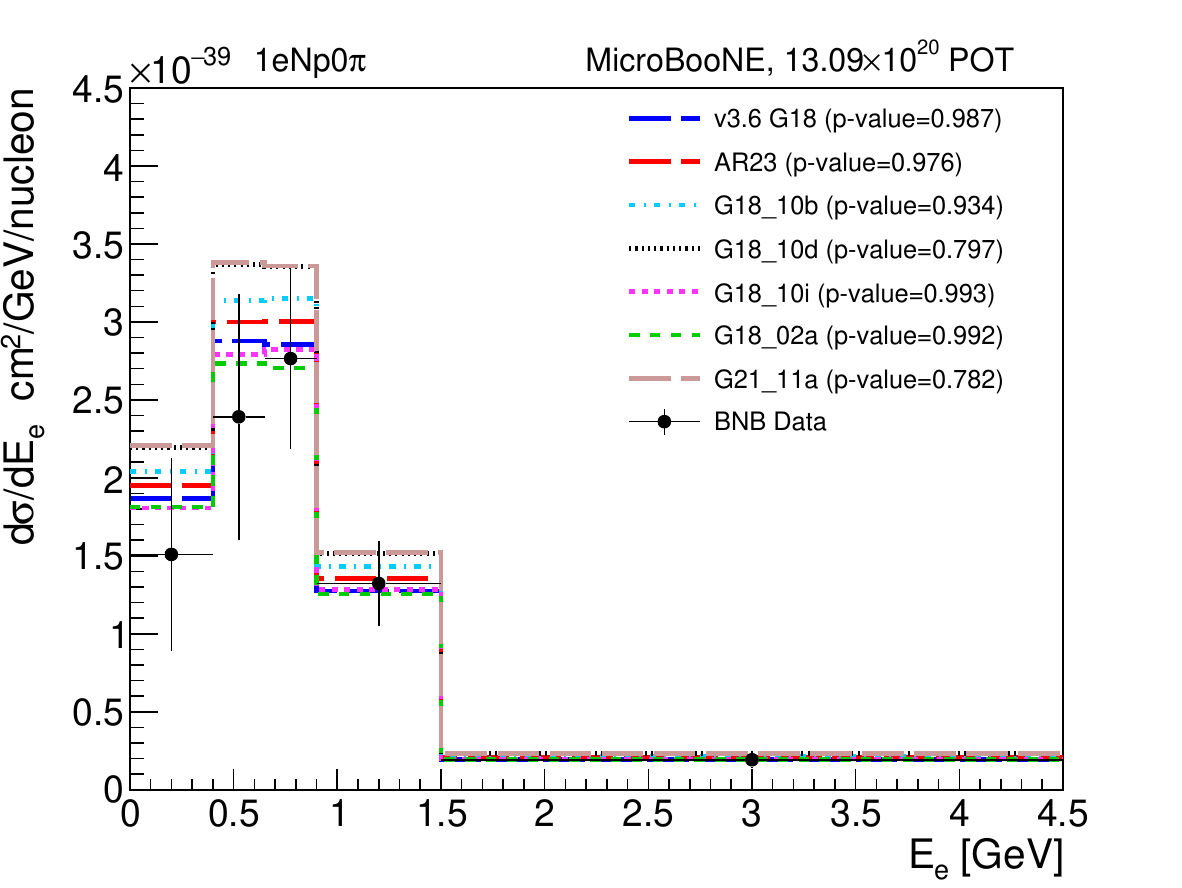}}
  \subfigure[]
   { \includegraphics[width=8cm]{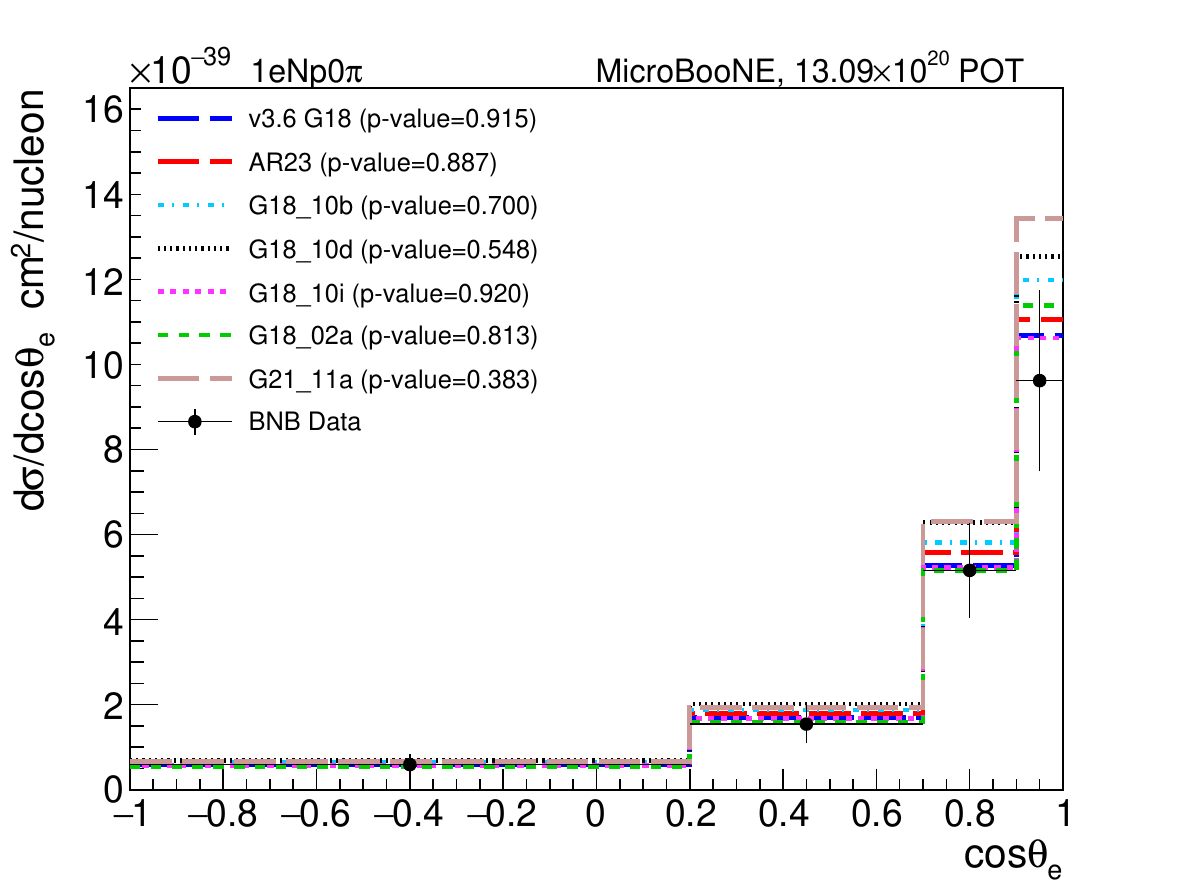}}
     \subfigure[]
   { \includegraphics[width=8cm]{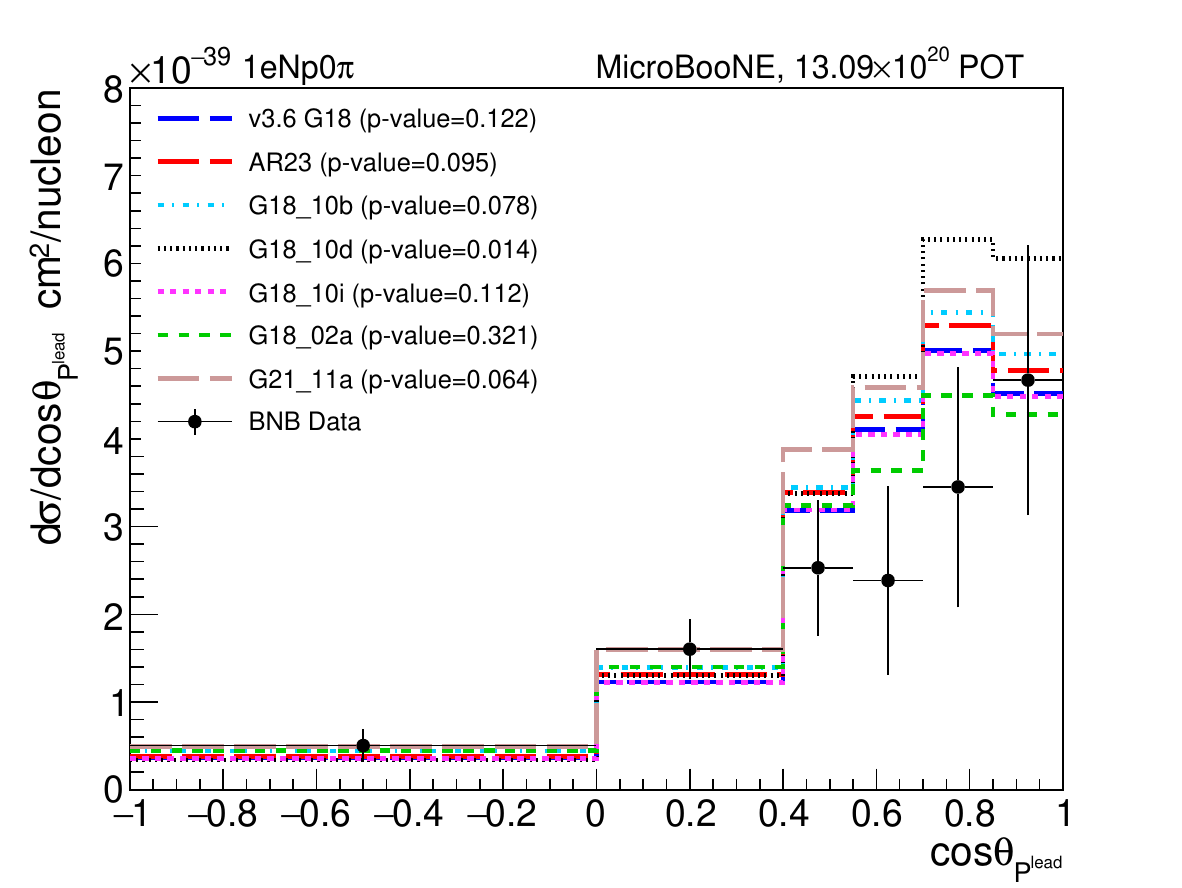}}
     \subfigure[]
   { \includegraphics[width=8cm]{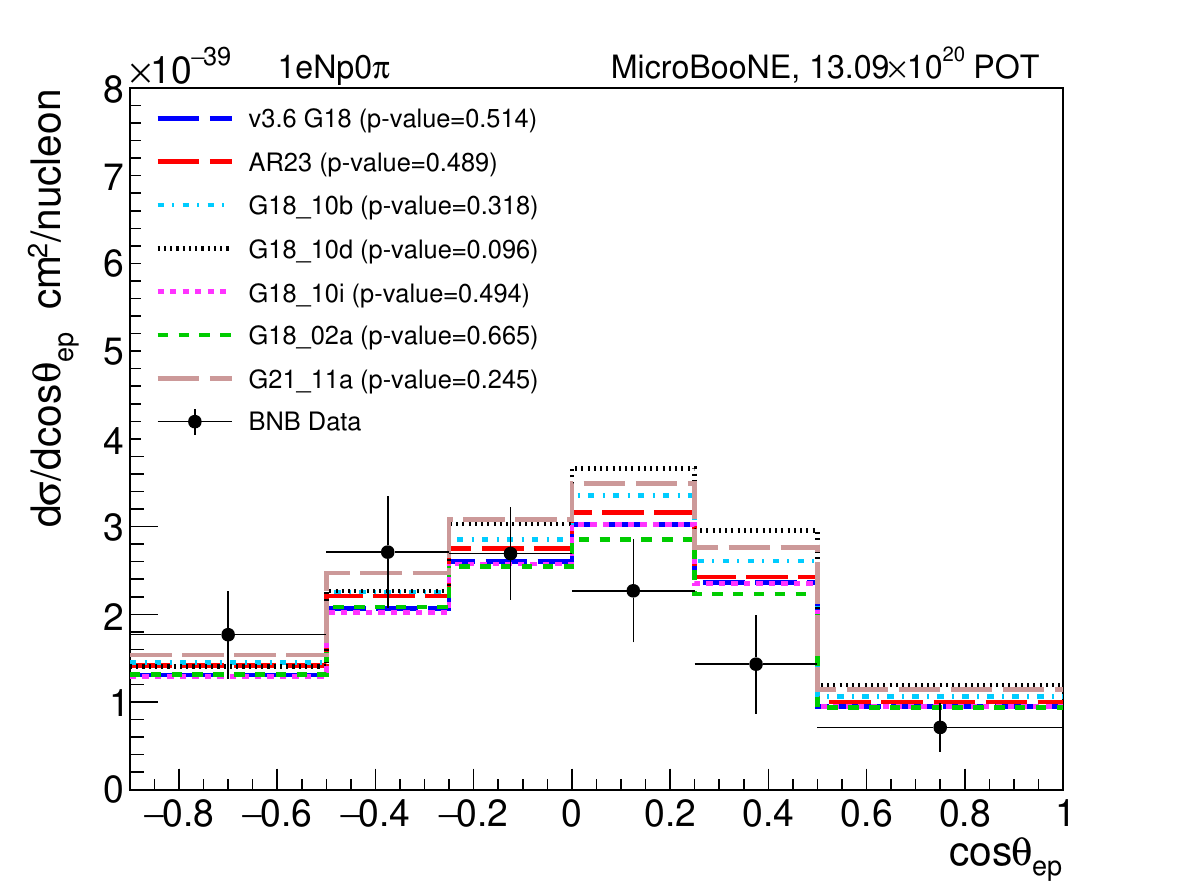}}
  \subfigure[]{
    \includegraphics[width=8cm]{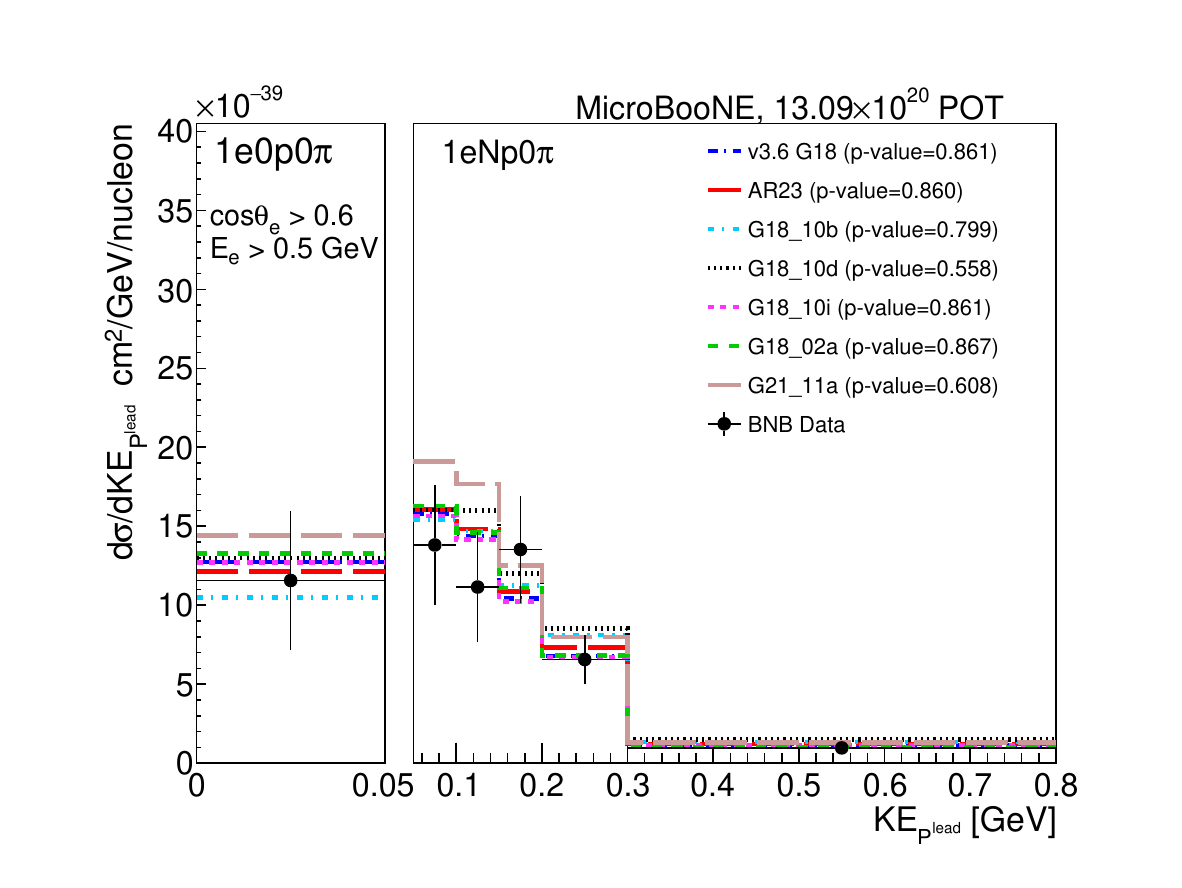}}
     \hspace*{\fill} 
  \subfigure[]
   { \includegraphics[width=8cm]{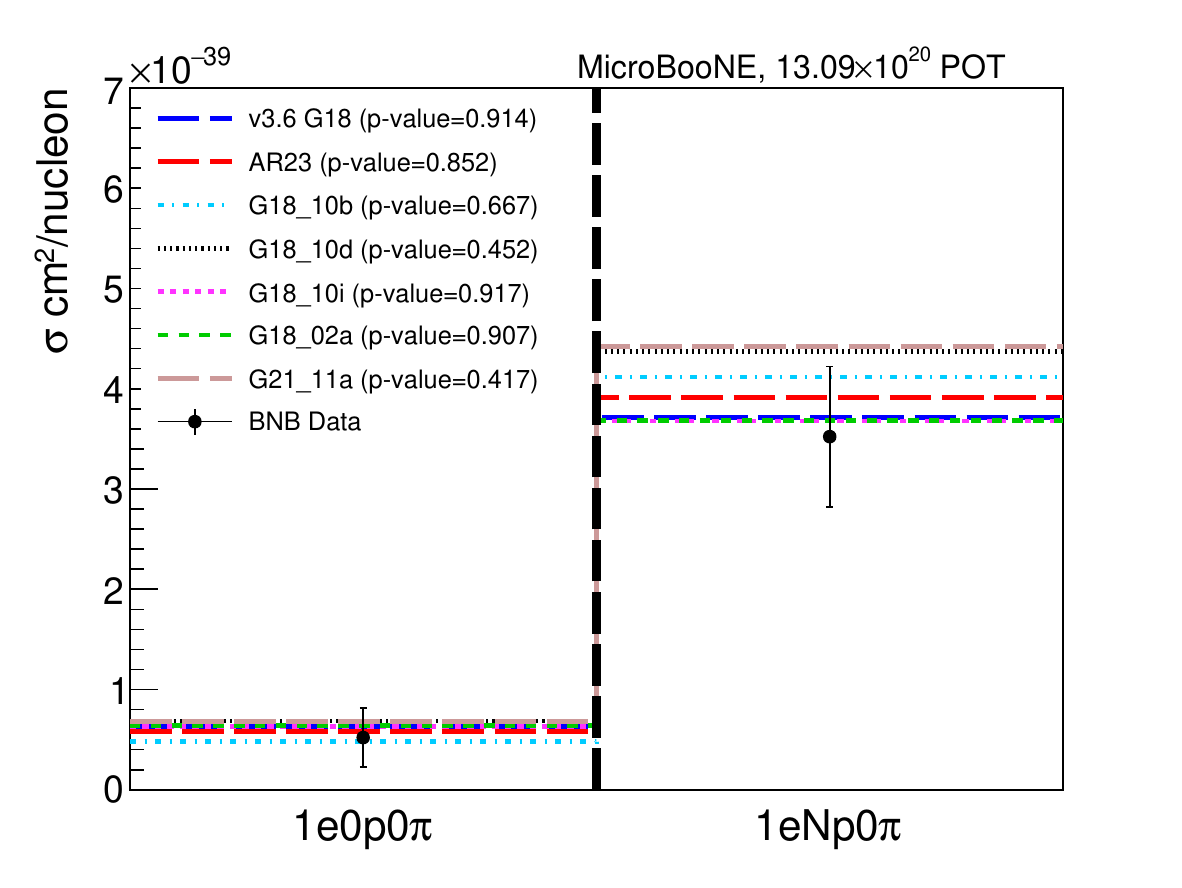}}
\caption{Unfolded cross section comparisons within GENIE v3.6.0 for all analysis variables using the 1eNp0$\pi$ channel: (a) electron energy, (b) electron angle, (c) proton angle, (d) opening angle. 
Also shown for proton kinetic energy and with 1eXp0$\pi$ events (e) proton energy, (f) measurement across the proton visibility threshold.}

\label{fig:appcompadd}
\end{figure}

This appendix presents additional comparisons between the unfolded cross sections and a variety of GENIE tunes, varying CCQE, MEC and FSI models, using GENIE version 3.6.0 as described in \ref{sec:Results}. These comparisons are shown for each variable in Fig.\ref{fig:appcompadd}.
\FloatBarrier

\FloatBarrier
\section{Unfolded results compared to two NEUT versions}
\label{app:NEUT}
\begin{figure}[H]
  \subfigure[]{
    \includegraphics[width=8cm]{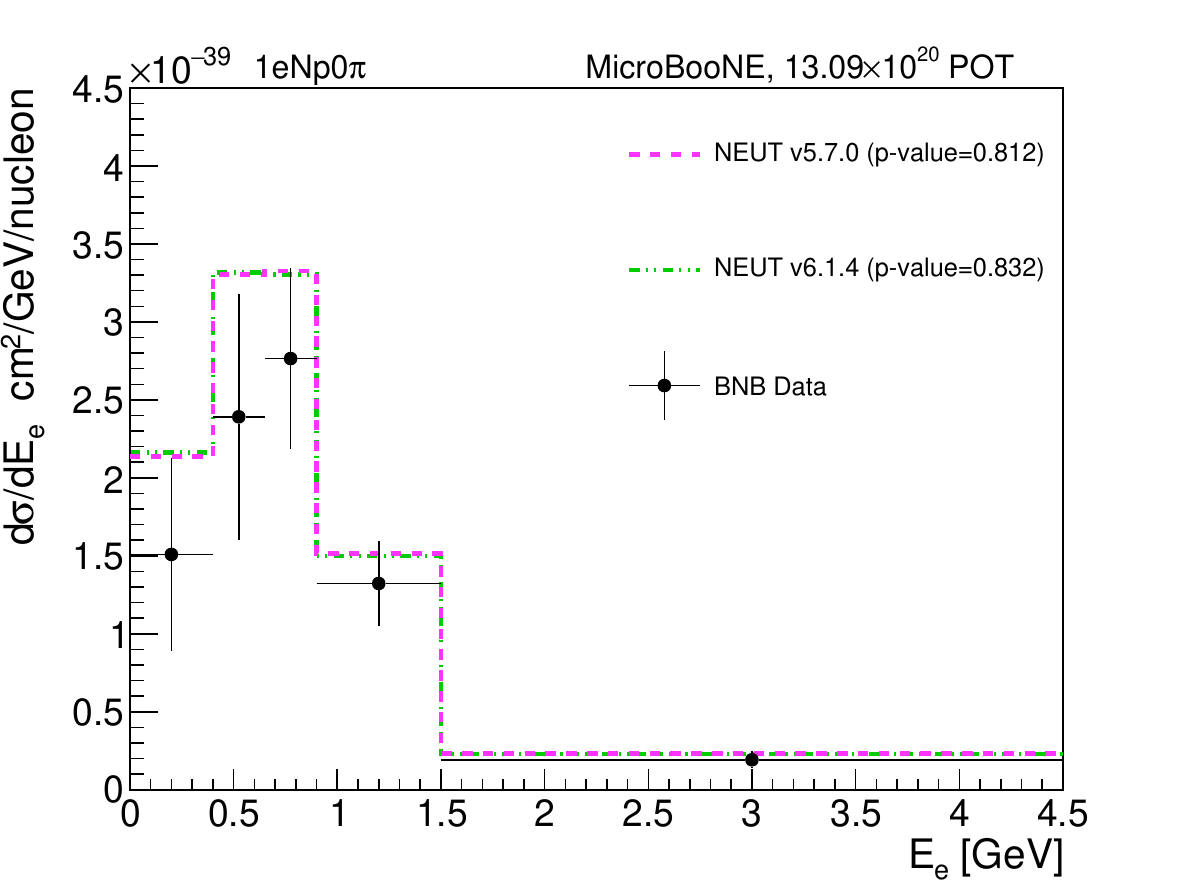}}
  \subfigure[]
   { \includegraphics[width=8cm]{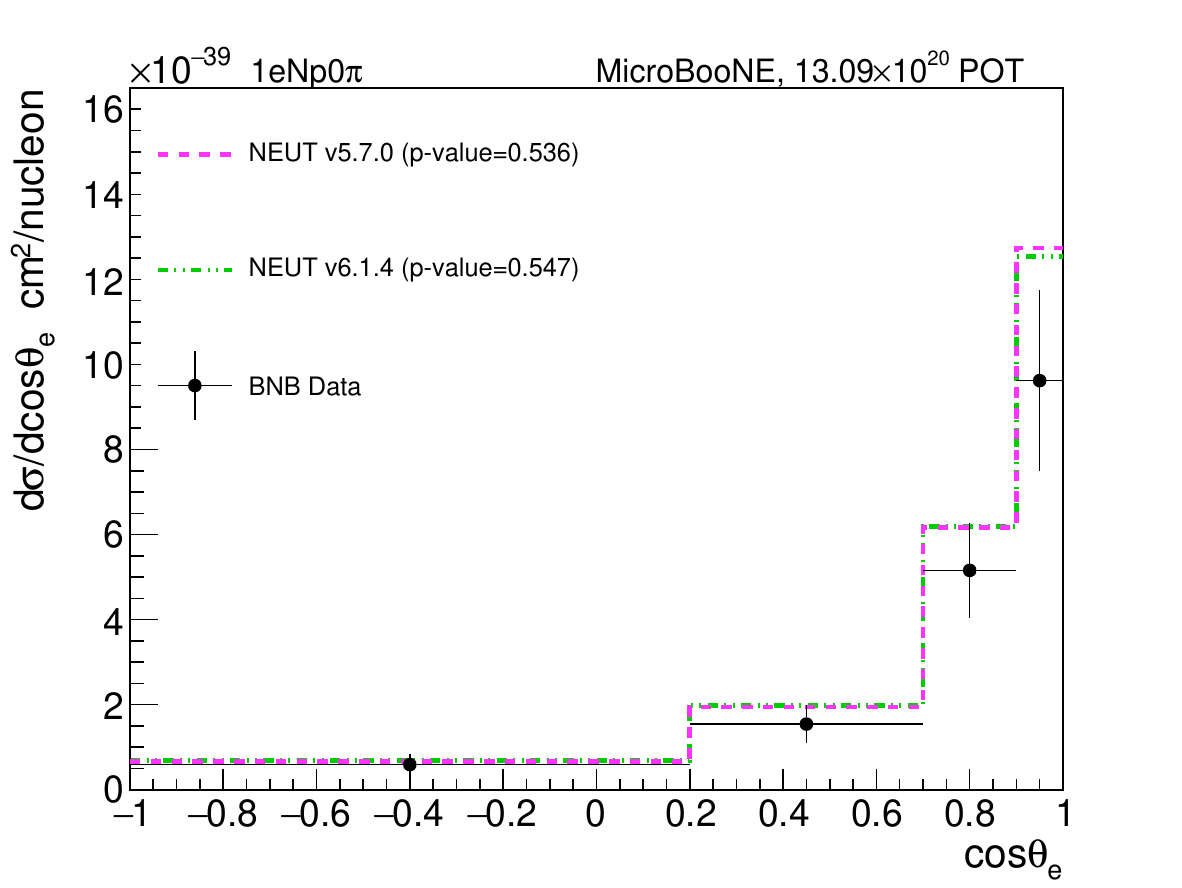}}
     \subfigure[]
   { \includegraphics[width=8cm]{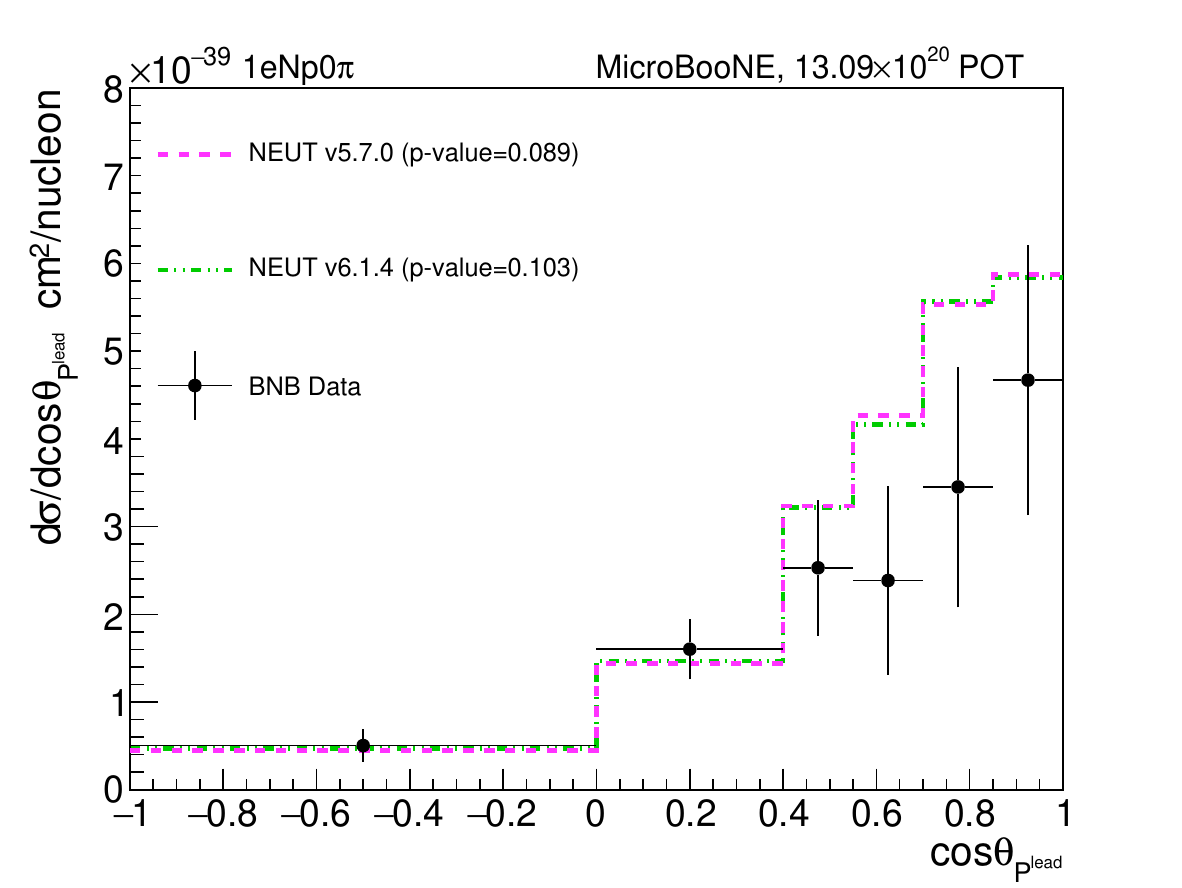}}
    \hspace*{\fill} 
     \subfigure[]
   { \includegraphics[width=8cm]{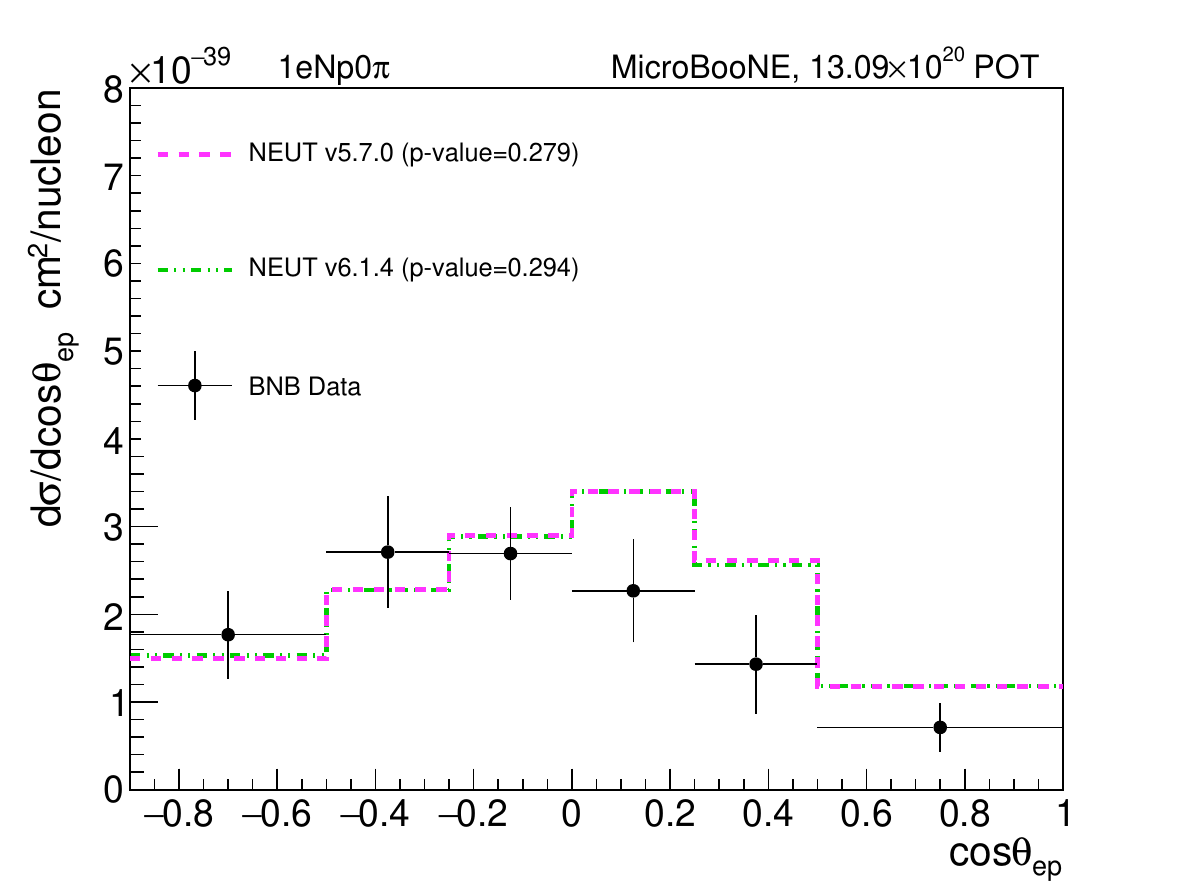}}

\caption{Unfolded cross section comparison between two NEUT versions for all analysis variables using the 1eNp0$\pi$ channel: (a) electron energy, (b) electron angle, (c) proton angle, (d) opening angle. } 
\label{fig:appcompaddNeut}
\end{figure}

\begin{figure}[H]
  \subfigure[]{
    \includegraphics[width=8cm]{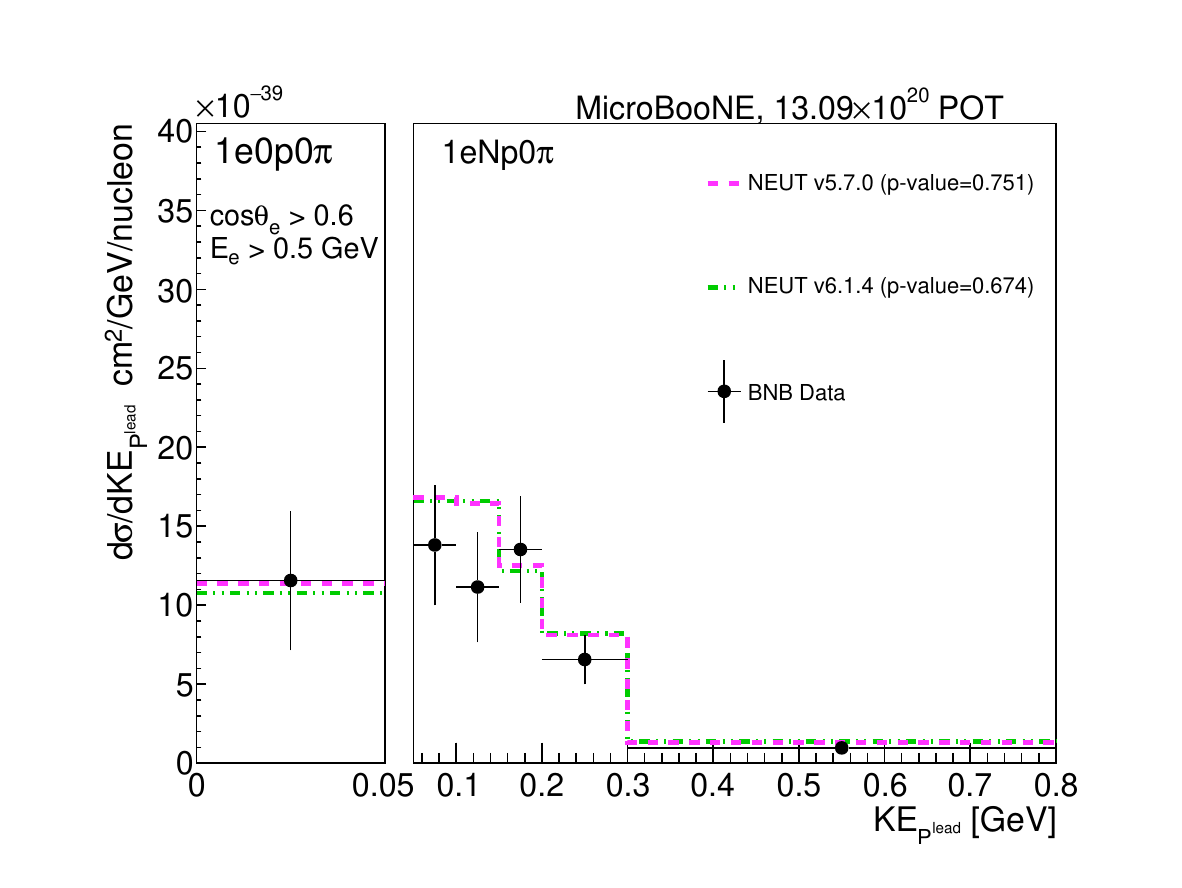}}
     \hspace*{\fill} 
  \subfigure[]
   { \includegraphics[width=8cm]{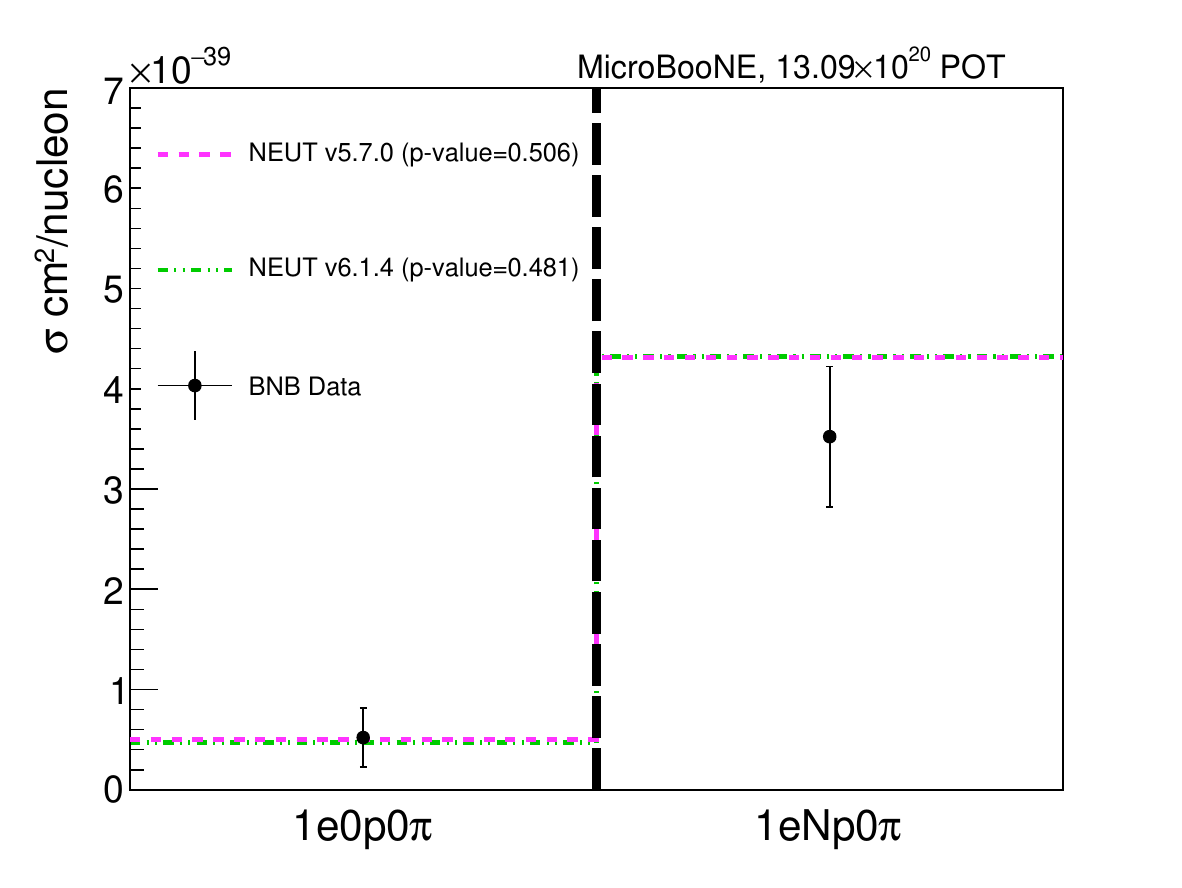}}
\caption{Unfolded cross section comparison between two NEUT versions for the leading proton kinetic energy variable using the 1eXp0$\pi$ and the (a) full proton energy binning, (b) measurement across the proton visibility threshold.} 
\label{fig:appcompaddNeut2}
\end{figure}

This appendix presents an additional comparison between the unfolded cross sections and two NEUT versions: v5.7.0 that uses an LFG ground state and v6.1.4. that uses an EDRMF~\cite{NEUTEDRMF,EDRMF} ground state model. This comparison is shown for each variable in Fig.\ref{fig:appcompaddNeut} and Fig.\ref{fig:appcompaddNeut2} .
\FloatBarrier

\FloatBarrier
\section{Tables of cross section results}
\label{app:xsecres}
\cref{tbl:ResEe,tbl:ResPe,tbl:ResPe2,tbl:ResEA,tbl:ResPA,tbl:ResOpAng} list the extracted cross section results, the corresponding uncertainty, and the background-subtracted number of candidate events per bin, for all six variables and configurations used in this analysis.
\FloatBarrier

\begin{table}[H]
\caption{Background-subtracted data events and unfolded cross sections for the electron energy measurement.}
\label{tbl:ResEe}
\begin{center}
\begin{tabular}{|c|c|c|c|c|c|}
\hline
Bin \# & $E_e$ low edge & $E_e$ high edge  & Data-background&  cross section  & Uncertainty   \\
Units& [GeV]&[GeV]&&[$10^{-39}\frac{\mathrm{cm^2}}{\mathrm{GeV nucleon}}$]&[$10^{-39}\frac{\mathrm{cm^2}}{\mathrm{GeV nucleon}}$]\\
&&&&&\\
0&0.&0.4& 27.267&1.51&0.62\\
1&0.4&0.65&29.887&2.39&0.79\\
2&0.65&0.9&28.75&2.76&0.58\\
3&0.9&1.5&36.568&1.32&0.27\\
4&1.5&4.5&18.5&0.192&0.054\\

\hline
\end{tabular} 
\end{center}
\end{table}\FloatBarrier

\begin{table}[H]
\caption{Background-subtracted data events and unfolded cross sections for the proton kinetic energy measurement.}
\label{tbl:ResPe}
\begin{center} 
\begin{tabular}{|c|c|c|c|c|c|}
\hline
Bin \# & $KE_{p}$ low edge & $KE_{p}$ high edge  & Data-background&  cross section  & Uncertainty   \\
Units& [GeV]&[GeV]&&[$10^{-39}\frac{\mathrm{cm^2}}{\mathrm{GeV nucleon}}$]&[$10^{-39}\frac{\mathrm{cm^2}}{\mathrm{GeV nucleon}}$]\\
&&&&&\\
0&0.&0.05& 15.468&11.6&4.43\\
1&0.05&0.1&38.016&13.8&3.82\\
2&0.1&0.15&25.998&11.2&3.50\\
3&0.15&0.2&36.843&13.5&3.38\\
4&0.2&0.3&29.55&6.56&1.57\\
5&0.3&0.8&13.374&0.965&0.299\\
\hline
\end{tabular} \end{center}
\end{table}\FloatBarrier

\begin{table}[H]
\caption{Background-subtracted data events and unfolded cross sections for the proton kinetic energy two-bin measurement.}
\label{tbl:ResPe2}
\begin{center} \begin{tabular}{|c|c|c|c|c|c|}
\hline
Bin \# & $KE_{p}$ low edge & $KE_{p}$ high edge  & Data-background&  cross section  & Uncertainty\\
Units& [GeV]&[GeV]&&[$10^{-39}\frac{\mathrm{cm^2}}{\mathrm{GeV nucleon}}$]&[$10^{-39}\frac{\mathrm{cm^2}}{\mathrm{GeV nucleon}}$]\\
&&&&&\\
0&0.&0.05& 15.468&0.521&0.294\\
1&0.05&0.8&143.781&3.52&0.70\\
\hline
\end{tabular} \end{center}
\end{table}

\begin{table}[H]
\caption{Background-subtracted data events and unfolded cross sections for the cosine of the electron angle measurement.}
\label{tbl:ResEA}
\begin{center} 
\begin{tabular}{|c|c|c|c|c|c|}
\hline
Bin \# & $\cos\theta_e$low edge & $\cos\theta_e$ high edge  & Data-background&  cross section  & Uncertainty   \\
Units& &&&[$10^{-39}\frac{\mathrm{cm^2}}{\mathrm{nucleon}}$]&[$10^{-39}\frac{\mathrm{cm^2}}{\mathrm{nucleon}}$]\\
&&&&&\\
0&-1&0.2& 17.237&0.593&0.254\\
1&0.2&0.7&28.138&1.54&0.46\\
2&0.7&0.9&46.665&5.16&1.11\\
3&0.9&1.&48.929&9.62&2.13\\
\hline
\end{tabular} \end{center}
\end{table}\FloatBarrier

\begin{table}[H]
\caption{Background-subtracted data events and unfolded cross sections for the cosine of the proton angle measurement.}
\label{tbl:ResPA}
\begin{center} 
\begin{tabular}{|c|c|c|c|c|c|}
\hline
Bin \# & $\cos\theta_p$ low edge & $\cos\theta_p$ high edge  & Data-background&  cross section  & Uncertainty   \\
Units& &&&[$10^{-39}\frac{\mathrm{cm^2}}{\mathrm{nucleon}}$]&[$10^{-39}\frac{\mathrm{cm^2}}{\mathrm{nucleon}}$]\\
&&&&&\\
0&-1&0.& 16.899&0.504&0.186\\
1&0&0.4&42.821&1.60&0.34\\
2&0.4&0.55&14.304&2.53&0.77\\
3&0.55&0.7&22.784&2.39&1.07\\
4&0.7&0.85&19.884&3.45&1.37\\
5&0.85&1.&24.277&4.67&1.54\\
\hline
\end{tabular} \end{center}
\end{table}\FloatBarrier
\begin{table}[H]
\caption{Background-subtracted data events and unfolded cross sections for the cosine of the opening angle measurement.}
\label{tbl:ResOpAng}
\begin{center} \begin{tabular}{|c|c|c|c|c|c|c|}
\hline

Bin \# & $\cos\theta_{ep}$ low edge & $\cos\theta_{ep}$ high edge  & Data-background&  cross section & Uncertainty  \\
Units& &&&[$10^{-39}\frac{\mathrm{cm^2}}{\mathrm{nucleon}}$]&[$10^{-39}\frac{\mathrm{cm^2}}{\mathrm{nucleon}}$]\\
&&&&&\\
0&-0.9&-0.5& 25.5&1.77&0.50\\
1&-0.5&-0.25&22.523&2.71&0.64\\
2&-0.25&0.&32.731&2.69&0.53\\
3&0.&0.25&23.465&2.27&0.59\\
4&0.25&0.5&16.32&1.43&0.57\\
5&0.5&1.&20.429&0.711&0.278\\
\hline
\end{tabular} \end{center}
\end{table}\FloatBarrier

\FloatBarrier
\section{Matrices for cross section calculation}
\label{app:matrix}
\cref{tbl:CovEe,tbl:CovPe,tbl:CovPe2,tbl:CovEA,tbl:CovPA,tbl:CovepA} provide the covariance matrices built as input to the unfolding procedure for each measurement. \cref{tbl:RepEe,tbl:RepPe,tbl:RepPe2,tbl:RepAe,tbl:RepAp,tbl:RepAep} are the response matrices used for the unfolding. \cref{tbl:ACEe,tbl:ACPe,tbl:ACPe2,tbl:ACAe,tbl:ACAp,tbl:ACAep} list the regularization obtained as part of the unfolding procedure output and applied to the generator predictions for comparison with the unfolded data cross sections.
\FloatBarrier
\begin{table}[H]
\caption{Covariance matrix for the electron energy cross section measurement.}
\label{tbl:CovEe}
\begin{center} \begin{tabular}{|c|c|c|c|c|c|}
\hline
Bins in GeV & [0.,0.4] & [0.4,0.65]& [0.65,0.9]& [0.9,1.5]& [1.5,4.5]\\\hline
{[0.,0.4]} &235.52 &40.29& 17.23 &9.844& 0.6409 	\\
{[0.4,0.65]}  &40.29& 79.81& 11.65& 10.85& 4.872  \\
{[0.65,0.9]}  &17.23& 11.65& 50.65& 12.56& 7.809   \\
{[0.9,1.5]}  &9.844& 10.85& 12.56& 74.28& 20.34   \\
{[1.5,4.5]}  & 0.6409& 4.872& 7.809& 20.34& 42.71 \\
\hline
\end{tabular} \end{center}
\end{table}
\FloatBarrier
\begin{table}[H]
\caption{Covariance matrix for the proton kinetic energy full-binning cross section measurement.} \label{tbl:CovPe}
\begin{center} \begin{tabular}{|c|c|c|c|c|c|c|}
\hline
Bins in GeV & [0.0,0.05] & [0.05,0.1]& [0.1,0.15]& [0.15,0.2]& [0.2,0.3]& [0.3,0.8]\\\hline
{[0.0,0.05]}  &51.05& 19.03& 15.51& 10.08& 10.07& 4.028 \\
{[0.05,0.1]}  &19.03& 128.3& 32.05& 20.30& 16.33& 5.248  \\
{[0.1,0.15]}  &15.51& 32.05& 88.89& 18.29& 16.56& 5.648  \\
{[0.15,0.2]}  &10.08& 20.30& 18.29& 73.40& 15.13& 6.169 \\
{[0.2,0.3]}  &10.07& 16.33& 16.56& 15.13& 78.16& 11.07  \\
{[0.3,0.8]}  &4.028& 5.248& 5.648& 6.169& 11.07& 28.70 \\
\hline
\end{tabular} \end{center}
\end{table}
\FloatBarrier
\begin{table}[H]
\caption{Covariance matrix for the two-bin proton kinetic energy cross section measurement.} \label{tbl:CovPe2}
\begin{center} \begin{tabular}{|c|c|c|}
\hline
Bins in GeV & [0.0,0.05] & [0.05,0.8]\\ \hline
{[0.0,0.05]} & 51.44& 60.04		\\
{[0.05,0.8]}  &60.04& 805.2   \\
\hline
\end{tabular} \end{center}
\end{table}
\FloatBarrier
\begin{table}[H]
\caption{Covariance matrix for the cosine of the electron angle cross section measurement.} \label{tbl:CovEA}
\begin{center} \begin{tabular}{|c|c|c|c|c|}
\hline
Bins & [ -1.,0.2] & [0.2,0.7]& [0.7,0.9]& [0.9,1.]\\\hline
{[ -1.,0.2]} &  76.36& 19.35& 17.87&13.77 		\\
{[0.2,0.7]}  &19.35& 98.13& 29.37 &23.90    \\
{[0.7,0.9]}  &17.87& 29.37& 120.4& 38.51    \\
{[0.9,1.]}  &13.771 &23.90& 38.51& 128.2  \\
\hline
\end{tabular} \end{center}
\end{table}
\FloatBarrier
\begin{table}[H]
\caption{Covariance matrix for the cosine of the proton angle cross section measurement.} \label{tbl:CovPA}
\begin{center} \begin{tabular}{|c|c|c|c|c|c|c|}
\hline
Bins & [-1,0] & [0,0.4]& [0.4,0.55]& [0.55,0.7]& [0.7,0.85]& [0.85,1]\\\hline
{[-1,0]} &  58.81& 10.75& 6.345& 8.869& 11.26& 13.86 	\\
{[0,0.4]}  &10.75& 74.10& 6.708& 9.347& 12.39& 14.52 \\
{[0.4,0.55]}  &6.345& 6.708& 28.51& 7.226& 8.176& 8.883 \\
{[0.55,0.7]}  &8.869& 9.347& 7.226& 53.33& 13.80& 15.02 \\
{[0.7,0.85]}  &11.26& 12.39& 8.176& 13.80& 63.83& 21.86 \\
{[0.85,1]}  &13.86& 14.52& 8.883& 15.02& 21.86& 87.79\\
\hline
\end{tabular} \end{center}
\end{table}
\FloatBarrier
\begin{table}[H]
\caption{Covariance matrix for the cosine of the opening angle cross section measurement.} \label{tbl:CovepA}
\begin{center} \begin{tabular}{|c|c|c|c|c|c|c|}
\hline
Bins & [-0.9,-0.5] & [-0.5,-0.25]& [-0.25,0.]& [0.,0.25]& [0.25,0.5]& [0.5,1.]\\\hline
{[-0.9,-0.5]} &  72.99& 11.88& 10.89& 11.67& 8.967& 15.95 \\
{[-0.5,-0.25]}  &11.88& 56.91& 9.714& 10.44& 6.254& 12.40  \\
{[-0.25,0.]}  &10.89& 9.714& 61.86& 12.94& 8.824& 13.94  \\
{[0.,0.25]}  &11.67& 10.44& 12.94& 58.35& 11.07& 16.21  \\
{[0.25,0.5]}  &8.967& 6.254& 8.824& 11.07& 38.15& 13.20  \\
{[0.5,1.]}  &15.95& 12.40& 13.94& 16.21& 13.20& 75.14 \\
\hline
\end{tabular} \end{center}
\end{table}
\FloatBarrier
\begin{table}[H]
\caption{Response matrix for the electron energy measurement.}
\label{tbl:RepEe}
\begin{center} \begin{tabular}{|c|c|c|c|c|c|}
\hline
Bins in GeV & [0.,0.4] & [0.4,0.65]& [0.65,0.9]& [0.9,1.5]& [1.5,4.5]\\\hline
{[0.,0.4]} &  0.1171& 0.05126& 0.01161& 0.003379& 0.0004370  	\\
{[0.4,0.65]}  &0.01134& 0.1343& 0.06994& 0.01692& 0.001655  \\
{[0.65,0.9]}  &3.383e-05& 0.01448& 0.1264& 0.05174& 0.004976   \\
{[0.9,1.5]}  &9.657e-06& 0.0001096& 0.01769& 0.1521& 0.05275   \\
{[1.5,4.5]}  & 9.563e-06& 1.716e-05& 0.0001218& 0.005967& 0.1336 \\
\hline
\end{tabular} \end{center}
\end{table}
\FloatBarrier
\begin{table}[H]
\caption{Response matrix for the proton kinetic energy full-binning measurement.} \label{tbl:RepPe}
\begin{center} \begin{tabular}{|c|c|c|c|c|c|c|}
\hline
Bins in GeV & [0.0,0.05] & [0.05,0.1]& [0.1,0.15]& [0.15,0.2]& [0.2,0.3]& [0.3,0.8]\\\hline
{[0.0,0.05]} &  0.1175& 0.01110& 0.004014& 0.001891& 0.001401& 0.0009824 \\
{[0.05,0.1]}  &0.01751& 0.1669& 0.03891& 0.01185& 0.007343& 0.002402  \\
{[0.1,0.15]}  &0.0003940& 0.006581& 0.1922& 0.03524& 0.01393& 0.002882  \\
{[0.15,0.2]}  &0.0002581& 0.0002626&0.008669& 0.1925& 0.02959& 0.004179  \\
{[0.2,0.3]}  &9.957e-05& 0.0002511& 9.242e-05 &0.007594& 0.1789& 0.02111  \\
{[0.3,0.8]}  &7.644e-05 &3.100e-05& 4.360e-05 &5.067e-05 &0.0009817& 0.08001 \\
\hline
\end{tabular} \end{center}
\end{table}
\FloatBarrier
\begin{table}[H]
\caption{Response matrix for the two-bin proton kinetic energy measurement.} \label{tbl:RepPe2}
\begin{center} \begin{tabular}{|c|c|c|}
\hline
Bins in GeV & [0.0,0.05] & [0.05,0.8]\\ \hline
{[0.0,0.05]} &  0.1175 &0.004198		\\
{[0.05,0.8]}  &0.01834 &0.1978  \\
\hline
\end{tabular} \end{center}
\end{table}
\FloatBarrier
\begin{table}[H]
\caption{Response matrix for the cosine of the electron angle measurement.} \label{tbl:RepAe}
\begin{center} \begin{tabular}{|c|c|c|c|c|}
\hline
Bins & [ -1.,0.2] & [0.2,0.7]& [0.7,0.9]& [0.9,1.]\\\hline
{[ -1.,0.2]} &  0.1006& 0.003130& 0.0 &0.0 		\\
{[0.2,0.7]}  &0.004279& 0.1680& 0.006750& 0.0    \\
{[0.7,0.9]}  &0.0 &0.009439& 0.2006& 0.009193    \\
{[0.9,1.]}  &0.0& 0.0 &0.01194& 0.2216 \\
\hline
\end{tabular} \end{center}
\end{table}
\FloatBarrier
\begin{table}[H]
\caption{Response matrix for the cosine of the proton angle measurement.} \label{tbl:RepAp}
\begin{center} \begin{tabular}{|c|c|c|c|c|c|c|}
\hline
Bins & [-1,0] & [0,0.4]& [0.4,0.55]& [0.55,0.7]& [0.7,0.85]& [0.85,1]\\\hline
{[-1,0]} &  0.1615& 0.01644& 0.007344& 0.005298& 0.005486& 0.005661 	\\
{[0,0.4]}  &0.01184& 0.1694& 0.03008& 0.005471& 0.003338& 0.002521 \\
{[0.4,0.55]}  &0.002412& 0.01802& 0.1743& 0.02315& 0.002453& 0.001199  \\
{[0.55,0.7]}  &0.003793& 0.003992& 0.02900& 0.1697& 0.01597& 0.002063 \\
{[0.7,0.85]}  &0.004264& 0.002986& 0.003496& 0.02259& 0.1402& 0.01059 \\
{[0.85,1]}  &0.005873& 0.004275& 0.002612& 0.002658& 0.01652& 0.1175\\
\hline
\end{tabular} \end{center}
\end{table}
\FloatBarrier
\begin{table}[H]
\caption{Response matrix for the cosine of the opening angle measurement.} \label{tbl:RepAep}
\begin{center} \begin{tabular}{|c|c|c|c|c|c|c|}
\hline
Bins & [-0.9,-0.5] & [-0.5,-0.25]& [-0.25,0.]& [0.,0.25]& [0.25,0.5]& [0.5,1.]\\\hline
{[-0.9,-0.5]} &  0.1216& 0.02181& 0.004377& 0.003119& 0.003529& 0.008934 \\
{[-0.5,-0.25]}  &0.01330& 0.1337&0.02174&0.003311& 0.003114& 0.004561  \\
{[-0.25,0.]}  &0.002263& 0.02362& 0.1528& 0.02425&0.003779& 0.003682  \\
{[0.,0.25]}  &0.001810& 0.002591& 0.02588& 0.1630&0.02468& 0.003286  \\
{[0.25,0.5]}  &0.001893& 0.002242& 0.002401& 0.02172&0.1504& 0.01370  \\
{[0.5,1.]}  &0.003891& 0.004196& 0.003855&0.003808& 0.01811& 0.1531 \\
\hline
\end{tabular} \end{center}
\end{table}
\FloatBarrier
\begin{table}[H]
\caption{Regularization matrix from the Wiener-SVD unfolding procedure for the electron energy measurement.} \label{tbl:ACEe}
\begin{center} \begin{tabular}{|c|c|c|c|c|c|}
\hline
Bins in GeV & [0.,0.4] & [0.4,0.65]& [0.65,0.9]& [0.9,1.5]& [1.5,4.5]\\\hline
{[0.,0.4]} &0.2892 &	0.5694&	0.2073&-0.05049	&-0.06408 	\\
{[0.4,0.65]}  &0.1455& 0.551& 0.3596& -0.02374&-0.07217 \\
{[0.65,0.9]}  &-0.01651& 0.2589& 0.5335&0.2568 &-0.03474  \\
{[0.9,1.5]}  &-0.02614& -0.01178 & 0.2506& 0.5938&0.205  \\
{[1.5,4.5]}  &-0.009027 & -0.02865&0.02412 & 0.2013&0.636  \\
\hline
\end{tabular} \end{center}
\end{table}
\FloatBarrier
\begin{table}[H]
\caption{Regularization matrix from the Wiener-SVD unfolding procedure for the proton kinetic energy, full binning measurement.} \label{tbl:ACPe}
\begin{center} \begin{tabular}{|c|c|c|c|c|c|c|}
\hline
Bins in GeV & [0.0,0.05] & [0.05,0.1]& [0.1,0.15]& [0.15,0.2]& [0.2,0.3]& [0.3,0.8]\\\hline
{[0.0,0.05]} & 0.7068&0.2011	&0.03124	&	0.09849&-0.01358 	& -0.05892 \\
{[0.05,0.1]}  &0.2049& 0.5071&0.2653 &  -0.00505& 0.01654& 0.02961\\
{[0.1,0.15]}  &0.009013&0.1499 & 0.688& 0.0964&0.04788 & 0.02533\\
{[0.15,0.2]}  &0.04722& -0.01873 &0.09477 &0.7666 &0.1181 & -0.07806  \\
{[0.2,0.3]}  &-0.01811 &0.01763 & 0.07031&0.1312 &0.6016 &0.1316 \\
{[0.3,0.8]}  & -0.05332 & 0.05522&0.08472 &-0.09095 & 0.282&0.3788 \\
\hline
\end{tabular} \end{center}
\end{table}
\FloatBarrier
\begin{table}[H]
\caption{Regularization matrix from the direct inversion for the proton kinetic energy two-bin measurement.} \label{tbl:ACPe2}
\begin{center} \begin{tabular}{|c|c|c|}
\hline
Bins in GeV & [0.0,0.05] & [0.05,0.8]\\ \hline
{[0.0,0.05]} & 1&	6.334e-11 	\\
{[0.05,0.8]}  &-2.328e-10 &  1 \\
\hline
\end{tabular} \end{center}
\end{table}
\FloatBarrier
\begin{table}[H]
\caption{Regularization matrix from the Wiener-SVD unfolding procedure for the cosine of the electron angle measurement.} \label{tbl:ACAe}
\begin{center} \begin{tabular}{|c|c|c|c|c|}
\hline
Bins & [ -1.,0.2] & [0.2,0.7]& [0.7,0.9]& [0.9,1.]\\\hline
{[ -1.,0.2]} & 0.6102&0.4765	&-0.09791	&-0.04398 \\
{[0.2,0.7]}  &0.1849& 0.6668&0.1735 & -0.06713   \\
{[0.7,0.9]}  &-0.04991&0.1523 &0.769& 0.138  \\
{[0.9,1.]}  &-0.01338& -0.03798& 0.1416& 0.829  \\
\hline
\end{tabular} \end{center}
\end{table}
\FloatBarrier
\begin{table}[H]
\caption{Regularization matrix from the Wiener-SVD unfolding procedure for the cosine of the proton angle measurement.} \label{tbl:ACAp}
\begin{center} \begin{tabular}{|c|c|c|c|c|c|c|}
\hline
Bins & [-1,0] & [0,0.4]& [0.4,0.55]& [0.55,0.7]& [0.7,0.85]& [0.85,1]\\\hline
{[-1,0]} & 0.774&0.2421	&-0.07663	&0.07954	& -0.01644 	&-0.04892 \\
{[0,0.4]}   &0.2588&0.4301 &0.386 &-0.1412 &0.02296 & 0.06136 \\
{[0.4,0.55]}  & -0.06841&0.2021 & 0.7549&0.1215 & -0.01433&-0.04056  \\
{[0.55,0.7]}  &0.06694& -0.1195& 0.1917& 0.7567& 0.0691&-0.003742  \\
{[0.7,0.85]}  &-0.0184& 0.04217& 0.001104&0.1082 & 0.738& 0.05083\\
{[0.85,1]}  &-0.06434& 0.1272& -0.06882&0.04524 &0.1363 &0.5708 \\

\hline
\end{tabular} \end{center}
\end{table}

\FloatBarrier
\begin{table}[H]
\caption{Regularization matrix from the Wiener-SVD unfolding procedure for the cosine of the opening angle measurement.} \label{tbl:ACAep}
\begin{center} \begin{tabular}{|c|c|c|c|c|c|c|}
\hline
Bins & [-0.9,-0.5] & [-0.5,-0.25]& [-0.25,0.]& [0.,0.25]& [0.25,0.5]& [0.5,1.]\\\hline
{[-0.9,-0.5]} &0.4948 &	0.3469&0.1595	&0.2123	&-0.2505	& -0.0845 \\
{[-0.5,-0.25]}  & 0.2016&0.4716	&0.3731	&-0.01751  	&0.01182	&   -0.1334 \\
{[-0.25,0.]} &0.03307&0.301	&0.413	&0.1634	&0.09962&0.01067\\
{[0.,0.25]}  & 0.07817&-0.04324	&0.1391	&0.5639	&0.2358	&0.08229\\
{[0.25,0.5]}  & -0.1065 &0.02321	&0.08465	&0.2342	&0.6674	&-0.01191 \\
{[0.5,1.]}  & -0.02879 &-0.0849	&0.07857	&0.2&	0.05643&0.4485\\
\hline
\end{tabular} \end{center}
\end{table}

\FloatBarrier
\twocolumngrid
%


\begin{thebibliography}{96}%
\makeatletter
\providecommand \@ifxundefined [1]{%
 \@ifx{#1\undefined}
}%
\providecommand \@ifnum [1]{%
 \ifnum #1\expandafter \@firstoftwo
 \else \expandafter \@secondoftwo
 \fi
}%
\providecommand \@ifx [1]{%
 \ifx #1\expandafter \@firstoftwo
 \else \expandafter \@secondoftwo
 \fi
}%
\providecommand \natexlab [1]{#1}%
\providecommand \enquote  [1]{``#1''}%
\providecommand \bibnamefont  [1]{#1}%
\providecommand \bibfnamefont [1]{#1}%
\providecommand \citenamefont [1]{#1}%
\providecommand \href@noop [0]{\@secondoftwo}%
\providecommand \href [0]{\begingroup \@sanitize@url \@href}%
\providecommand \@href[1]{\@@startlink{#1}\@@href}%
\providecommand \@@href[1]{\endgroup#1\@@endlink}%
\providecommand \@sanitize@url [0]{\catcode `\\12\catcode `\$12\catcode `\&12\catcode `\#12\catcode `\^12\catcode `\_12\catcode `\%12\relax}%
\providecommand \@@startlink[1]{}%
\providecommand \@@endlink[0]{}%
\providecommand \url  [0]{\begingroup\@sanitize@url \@url }%
\providecommand \@url [1]{\endgroup\@href {#1}{\urlprefix }}%
\providecommand \urlprefix  [0]{URL }%
\providecommand \Eprint [0]{\href }%
\providecommand \doibase [0]{https://doi.org/}%
\providecommand \selectlanguage [0]{\@gobble}%
\providecommand \bibinfo  [0]{\@secondoftwo}%
\providecommand \bibfield  [0]{\@secondoftwo}%
\providecommand \translation [1]{[#1]}%
\providecommand \BibitemOpen [0]{}%
\providecommand \bibitemStop [0]{}%
\providecommand \bibitemNoStop [0]{.\EOS\space}%
\providecommand \EOS [0]{\spacefactor3000\relax}%
\providecommand \BibitemShut  [1]{\csname bibitem#1\endcsname}%
\let\auto@bib@innerbib\@empty
\bibitem [{\citenamefont {Athar}\ \emph {et~al.}(2022)\citenamefont {Athar} \emph {et~al.}}]{ATHAR2022103947}%
  \BibitemOpen
  \bibfield  {author} {\bibinfo {author} {\bibfnamefont {M.~S.}\ \bibnamefont {Athar}} \emph {et~al.},\ }\bibfield  {title} {\bibinfo {title} {{Status and Perspectives of Neutrino Physics}},\ }\href {https://doi.org/https://doi.org/10.1016/j.ppnp.2022.103947} {\bibfield  {journal} {\bibinfo  {journal} {Prog. Part. Nucl. Phys.}\ }\textbf {\bibinfo {volume} {124}},\ \bibinfo {pages} {103947} (\bibinfo {year} {2022})}\BibitemShut {NoStop}%
\bibitem [{\citenamefont {Maki}\ \emph {et~al.}(1962)\citenamefont {Maki}, \citenamefont {Nakagawa},\ and\ \citenamefont {Sakata}}]{MAK62}%
  \BibitemOpen
  \bibfield  {author} {\bibinfo {author} {\bibfnamefont {Z.}~\bibnamefont {Maki}}, \bibinfo {author} {\bibfnamefont {M.}~\bibnamefont {Nakagawa}},\ and\ \bibinfo {author} {\bibfnamefont {S.}~\bibnamefont {Sakata}},\ }\bibfield  {title} {\bibinfo {title} {{Remarks on the Unified Model of Elementary Particles}},\ }\href {https://doi.org/10.1143/PTP.28.870} {\bibfield  {journal} {\bibinfo  {journal} {Prog. Theor. Phys.}\ }\textbf {\bibinfo {volume} {28}},\ \bibinfo {pages} {870} (\bibinfo {year} {1962})}\BibitemShut {NoStop}%
\bibitem [{\citenamefont {Gribov}\ and\ \citenamefont {Pontecorvo}(1969)}]{GRI69}%
  \BibitemOpen
  \bibfield  {author} {\bibinfo {author} {\bibfnamefont {V.}~\bibnamefont {Gribov}}\ and\ \bibinfo {author} {\bibfnamefont {B.}~\bibnamefont {Pontecorvo}},\ }\bibfield  {title} {\bibinfo {title} {{Neutrino Astronomy and Lepton Charge}},\ }\href {https://doi.org/https://doi.org/10.1016/0370-2693(69)90525-5} {\bibfield  {journal} {\bibinfo  {journal} {Phys.Lett. B}\ }\textbf {\bibinfo {volume} {28}},\ \bibinfo {pages} {493} (\bibinfo {year} {1969})}\BibitemShut {NoStop}%
\bibitem [{\citenamefont {Giganti}\ \emph {et~al.}(2018)\citenamefont {Giganti}, \citenamefont {Lavignac},\ and\ \citenamefont {Zito}}]{Giganti:2017fhf}%
  \BibitemOpen
  \bibfield  {author} {\bibinfo {author} {\bibfnamefont {C.}~\bibnamefont {Giganti}}, \bibinfo {author} {\bibfnamefont {S.}~\bibnamefont {Lavignac}},\ and\ \bibinfo {author} {\bibfnamefont {M.}~\bibnamefont {Zito}},\ }\bibfield  {title} {\bibinfo {title} {{Neutrino Oscillations: The Rise of the PMNS Paradigm}},\ }\href {https://doi.org/10.1016/j.ppnp.2017.10.001} {\bibfield  {journal} {\bibinfo  {journal} {Prog. Part. Nucl. Phys.}\ }\textbf {\bibinfo {volume} {98}},\ \bibinfo {pages} {1} (\bibinfo {year} {2018})}\BibitemShut {NoStop}%
\bibitem [{\citenamefont {Alvarez-Ruso}\ \emph {et~al.}(2018)\citenamefont {Alvarez-Ruso} \emph {et~al.}}]{NuSTEC:2017hzk}%
  \BibitemOpen
  \bibfield  {author} {\bibinfo {author} {\bibfnamefont {L.}~\bibnamefont {Alvarez-Ruso}} \emph {et~al.} (\bibinfo {collaboration} {NuSTEC}),\ }\bibfield  {title} {\bibinfo {title} {{NuSTEC White Paper: Status and challenges of neutrino{\textendash}nucleus scattering}},\ }\href {https://doi.org/10.1016/j.ppnp.2018.01.006} {\bibfield  {journal} {\bibinfo  {journal} {Prog. Part. Nucl. Phys.}\ }\textbf {\bibinfo {volume} {100}},\ \bibinfo {pages} {1} (\bibinfo {year} {2018})}\BibitemShut {NoStop}%
\bibitem [{\citenamefont {de~Salas}\ \emph {et~al.}(2018)\citenamefont {de~Salas}, \citenamefont {Gariazzo}, \citenamefont {Mena}, \citenamefont {Ternes},\ and\ \citenamefont {T√≥rtola}}]{massordering}%
  \BibitemOpen
  \bibfield  {author} {\bibinfo {author} {\bibfnamefont {P.~F.}\ \bibnamefont {de~Salas}}, \bibinfo {author} {\bibfnamefont {S.}~\bibnamefont {Gariazzo}}, \bibinfo {author} {\bibfnamefont {O.}~\bibnamefont {Mena}}, \bibinfo {author} {\bibfnamefont {C.~A.}\ \bibnamefont {Ternes}},\ and\ \bibinfo {author} {\bibfnamefont {M.}~\bibnamefont {T√≥rtola}},\ }\bibfield  {title} {\bibinfo {title} {{Neutrino Mass Ordering from Oscillations and Beyond: 2018 Status and Future Prospects}},\ }\href {http://dx.doi.org/10.3389/fspas.2018.00036} {\bibfield  {journal} {\bibinfo  {journal} {Front. Astron. Space Sci.}\ }\textbf {\bibinfo {volume} {5}},\ \bibinfo {pages} {36} (\bibinfo {year} {2018})}\BibitemShut {NoStop}%
\bibitem [{\citenamefont {de~Gouvea}(2016)}]{massmodelsreview}%
  \BibitemOpen
  \bibfield  {author} {\bibinfo {author} {\bibfnamefont {A.}~\bibnamefont {de~Gouvea}},\ }\bibfield  {title} {\bibinfo {title} {{Neutrino Mass Models}},\ }\href {https://doi.org/https://doi.org/10.1146/annurev-nucl-102115-044600} {\bibfield  {journal} {\bibinfo  {journal} {{Ann. Rev. Nucl. Part. Sci.}}\ }\textbf {\bibinfo {volume} {66}},\ \bibinfo {pages} {197} (\bibinfo {year} {2016})}\BibitemShut {NoStop}%
\bibitem [{\citenamefont {Kim}\ \emph {et~al.}(2022)\citenamefont {Kim}, \citenamefont {Murthy},\ and\ \citenamefont {Sahoo}}]{nunature}%
  \BibitemOpen
  \bibfield  {author} {\bibinfo {author} {\bibfnamefont {C.~S.}\ \bibnamefont {Kim}}, \bibinfo {author} {\bibfnamefont {M.~V.~N.}\ \bibnamefont {Murthy}},\ and\ \bibinfo {author} {\bibfnamefont {D.}~\bibnamefont {Sahoo}},\ }\bibfield  {title} {\bibinfo {title} {{Inferring the Nature of Active Neutrinos: Dirac or Majorana?}},\ }\href {https://doi.org/10.1103/PhysRevD.105.113006} {\bibfield  {journal} {\bibinfo  {journal} {Phys. Rev. D}\ }\textbf {\bibinfo {volume} {105}},\ \bibinfo {pages} {113006} (\bibinfo {year} {2022})}\BibitemShut {NoStop}%
\bibitem [{\citenamefont {Aguilar}\ \emph {et~al.}(2001)\citenamefont {Aguilar} \emph {et~al.}}]{LSND}%
  \BibitemOpen
  \bibfield  {author} {\bibinfo {author} {\bibfnamefont {A.}~\bibnamefont {Aguilar}} \emph {et~al.} (\bibinfo {collaboration} {LSND Collaboration}),\ }\bibfield  {title} {\bibinfo {title} {Evidence for neutrino oscillations from the observation of ${\overline{\ensuremath{\nu}}}_{e}$ appearance in a ${\overline{\ensuremath{\nu}}}_{\ensuremath{\mu}}$ beam},\ }\href {https://doi.org/10.1103/PhysRevD.64.112007} {\bibfield  {journal} {\bibinfo  {journal} {Phys. Rev. D}\ }\textbf {\bibinfo {volume} {64}},\ \bibinfo {pages} {112007} (\bibinfo {year} {2001})}\BibitemShut {NoStop}%
\bibitem [{\citenamefont {Aguilar-Arevalo}\ \emph {et~al.}(2021)\citenamefont {Aguilar-Arevalo} \emph {et~al.}}]{MiniBooNE}%
  \BibitemOpen
  \bibfield  {author} {\bibinfo {author} {\bibfnamefont {A.~A.}\ \bibnamefont {Aguilar-Arevalo}} \emph {et~al.} (\bibinfo {collaboration} {MiniBooNE Collaboration}),\ }\bibfield  {title} {\bibinfo {title} {Updated {MiniBooNE} neutrino oscillation results with increased data and new background studies},\ }\href {https://doi.org/10.1103/PhysRevD.103.052002} {\bibfield  {journal} {\bibinfo  {journal} {Phys. Rev. D}\ }\textbf {\bibinfo {volume} {103}},\ \bibinfo {pages} {052002} (\bibinfo {year} {2021})}\BibitemShut {NoStop}%
\bibitem [{\citenamefont {Abi}\ \emph {et~al.}(2020{\natexlab{a}})\citenamefont {Abi} \emph {et~al.}}]{DUNE_FD}%
  \BibitemOpen
  \bibfield  {author} {\bibinfo {author} {\bibfnamefont {B.}~\bibnamefont {Abi}} \emph {et~al.} (\bibinfo {collaboration} {DUNE Collaboration}),\ }\bibfield  {title} {\bibinfo {title} {{Deep Underground Neutrino Experiment (DUNE), Far Detector Technical Design Report, Volume I Introduction to DUNE}},\ }\href {https://doi.org/10.1088/1748-0221/15/08/T08008} {\bibfield  {journal} {\bibinfo  {journal} {{J. Instrum.}}\ }\textbf {\bibinfo {volume} {15}},\ \bibinfo {pages} {T08008} (\bibinfo {year} {2020}{\natexlab{a}})}\BibitemShut {NoStop}%
\bibitem [{\citenamefont {Abud}\ \emph {et~al.}(2022)\citenamefont {Abud} \emph {et~al.}}]{LowexpDUNEsensi}%
  \BibitemOpen
  \bibfield  {author} {\bibinfo {author} {\bibfnamefont {A.~A.}\ \bibnamefont {Abud}} \emph {et~al.} (\bibinfo {collaboration} {DUNE Collaboration}),\ }\bibfield  {title} {\bibinfo {title} {{Low Exposure Long-Baseline Neutrino Oscillation Sensitivity of the DUNE Experiment}},\ }\href {https://doi.org/10.1103/PhysRevD.105.072006} {\bibfield  {journal} {\bibinfo  {journal} {Phys. Rev. D}\ }\textbf {\bibinfo {volume} {105}},\ \bibinfo {pages} {072006} (\bibinfo {year} {2022})}\BibitemShut {NoStop}%
\bibitem [{\citenamefont {Abi}\ \emph {et~al.}(2020{\natexlab{b}})\citenamefont {Abi} \emph {et~al.}}]{LBLsensiDUNE}%
  \BibitemOpen
  \bibfield  {author} {\bibinfo {author} {\bibfnamefont {B.}~\bibnamefont {Abi}} \emph {et~al.} (\bibinfo {collaboration} {DUNE Collaboration}),\ }\bibfield  {title} {\bibinfo {title} {{Long-Baseline Neutrino Oscillation Physics Potential of the DUNE Experiment}},\ }\href {http://dx.doi.org/10.1140/epjc/s10052-020-08456-z} {\bibfield  {journal} {\bibinfo  {journal} {Eur. Phys. J. C}\ }\textbf {\bibinfo {volume} {80}} (\bibinfo {year} {2020}{\natexlab{b}})}\BibitemShut {NoStop}%
\bibitem [{\citenamefont {Abud}\ \emph {et~al.}(2021)\citenamefont {Abud} \emph {et~al.}}]{DUNE_ND}%
  \BibitemOpen
  \bibfield  {author} {\bibinfo {author} {\bibfnamefont {A.~A.}\ \bibnamefont {Abud}} \emph {et~al.} (\bibinfo {collaboration} {DUNE Collaboration}),\ }\bibfield  {title} {\bibinfo {title} {{Deep Underground Neutrino Experiment ({DUNE}) Near Detector Conceptual Design Report}},\ }\href {https://www.mdpi.com/2410-390X/5/4/31} {\bibfield  {journal} {\bibinfo  {journal} {Instruments}\ }\textbf {\bibinfo {volume} {5}},\ \bibinfo {pages} {31} (\bibinfo {year} {2021})}\BibitemShut {NoStop}%
\bibitem [{\citenamefont {Aguilar-Arevalo}\ \emph {et~al.}(2009)\citenamefont {Aguilar-Arevalo} \emph {et~al.}}]{MiniBooNEFlux}%
  \BibitemOpen
  \bibfield  {author} {\bibinfo {author} {\bibfnamefont {A.~A.}\ \bibnamefont {Aguilar-Arevalo}} \emph {et~al.} (\bibinfo {collaboration} {MiniBooNE Collaboration}),\ }\bibfield  {title} {\bibinfo {title} {{Neutrino flux prediction at MiniBooNE}},\ }\href {https://doi.org/10.1103/PhysRevD.79.072002} {\bibfield  {journal} {\bibinfo  {journal} {Phys. Rev. D}\ }\textbf {\bibinfo {volume} {79}},\ \bibinfo {pages} {072002} (\bibinfo {year} {2009})}\BibitemShut {NoStop}%
\bibitem [{\citenamefont {Adamson}\ \emph {et~al.}(2016)\citenamefont {Adamson} \emph {et~al.}}]{NuMIbeam}%
  \BibitemOpen
  \bibfield  {author} {\bibinfo {author} {\bibfnamefont {P.}~\bibnamefont {Adamson}} \emph {et~al.},\ }\bibfield  {title} {\bibinfo {title} {{The NuMI Neutrino Beam}},\ }\href {https://doi.org/https://doi.org/10.1016/j.nima.2015.08.063} {\bibfield  {journal} {\bibinfo  {journal} {Nuc. Instrum. Meth. A}\ }\textbf {\bibinfo {volume} {806}},\ \bibinfo {pages} {279} (\bibinfo {year} {2016})}\BibitemShut {NoStop}%
\bibitem [{\citenamefont {Lalnuntluanga}\ \emph {et~al.}(2024)\citenamefont {Lalnuntluanga}, \citenamefont {Pradhan},\ and\ \citenamefont {Giri}}]{DUNEuBfluxComp}%
  \BibitemOpen
  \bibfield  {author} {\bibinfo {author} {\bibfnamefont {R.}~\bibnamefont {Lalnuntluanga}}, \bibinfo {author} {\bibfnamefont {R.}~\bibnamefont {Pradhan}},\ and\ \bibinfo {author} {\bibfnamefont {A.}~\bibnamefont {Giri}},\ }\bibfield  {title} {\bibinfo {title} {{Probing neutrino-nucleus interaction in DUNE and MicroBooNE}},\ }\href {https://doi.org/https://doi.org/10.1016/j.nuclphysb.2024.116703} {\bibfield  {journal} {\bibinfo  {journal} {Nucl. Phys. B}\ }\textbf {\bibinfo {volume} {1008}},\ \bibinfo {pages} {116703} (\bibinfo {year} {2024})}\BibitemShut {NoStop}%
\bibitem [{\citenamefont {Abe}\ \emph {et~al.}(2014)\citenamefont {Abe} \emph {et~al.}}]{T2Knuexsec1}%
  \BibitemOpen
  \bibfield  {author} {\bibinfo {author} {\bibfnamefont {K.}~\bibnamefont {Abe}} \emph {et~al.} (\bibinfo {collaboration} {T2K Collaboration}),\ }\bibfield  {title} {\bibinfo {title} {{Measurement of the Inclusive Electron Neutrino Charged Current Cross Section on Carbon with the T2K Near Detector}},\ }\href {https://doi.org/10.1103/PhysRevLett.113.241803} {\bibfield  {journal} {\bibinfo  {journal} {Phys. Rev. Lett.}\ }\textbf {\bibinfo {volume} {113}},\ \bibinfo {pages} {241803} (\bibinfo {year} {2014})}\BibitemShut {NoStop}%
\bibitem [{\citenamefont {Abe}\ \emph {et~al.}(2015)\citenamefont {Abe} \emph {et~al.}}]{T2Knuexsec2}%
  \BibitemOpen
  \bibfield  {author} {\bibinfo {author} {\bibfnamefont {K.}~\bibnamefont {Abe}} \emph {et~al.} (\bibinfo {collaboration} {T2K Collaboration}),\ }\bibfield  {title} {\bibinfo {title} {{Measurement of the Electron Neutrino Charged-Current Interaction Rate on Water with the T2K ND280 ${\ensuremath{\pi}}^{0}$ Detector}},\ }\href {https://doi.org/10.1103/PhysRevD.91.112010} {\bibfield  {journal} {\bibinfo  {journal} {Phys. Rev. D}\ }\textbf {\bibinfo {volume} {91}},\ \bibinfo {pages} {112010} (\bibinfo {year} {2015})}\BibitemShut {NoStop}%
\bibitem [{\citenamefont {Abe}\ \emph {et~al.}(2020)\citenamefont {Abe} \emph {et~al.}}]{T2Knuexsec3}%
  \BibitemOpen
  \bibfield  {author} {\bibinfo {author} {\bibfnamefont {K.}~\bibnamefont {Abe}} \emph {et~al.} (\bibinfo {collaboration} {T2K Collaboration}),\ }\bibfield  {title} {\bibinfo {title} {{Measurement of the Charged-Current Electron (Anti-)Neutrino Inclusive Cross-Sections at the T2K Off-Axis Near Detector ND280}},\ }\href {http://dx.doi.org/10.1007/JHEP10(2020)114} {\bibfield  {journal} {\bibinfo  {journal} {{J. High Energy Phys.}}\ }\textbf {\bibinfo {volume} {2020}},\ \bibinfo {pages} {114} (\bibinfo {year} {2020})}\BibitemShut {NoStop}%
\bibitem [{\citenamefont {Acero}\ \emph {et~al.}(2023)\citenamefont {Acero} \emph {et~al.}}]{NOvanueXsec}%
  \BibitemOpen
  \bibfield  {author} {\bibinfo {author} {\bibfnamefont {M.~A.}\ \bibnamefont {Acero}} \emph {et~al.} (\bibinfo {collaboration} {NOvA Collaboration}),\ }\bibfield  {title} {\bibinfo {title} {{Measurement of the ${\ensuremath{\nu}}_{e}$-Nucleus Charged-Current Double-Differential Cross Section at $‚ü®{E}_{\ensuremath{\nu}}‚ü©=2.4\text{ }\text{ }\mathrm{GeV}$ Using NOvA}},\ }\href {https://doi.org/10.1103/PhysRevLett.130.051802} {\bibfield  {journal} {\bibinfo  {journal} {Phys. Rev. Lett.}\ }\textbf {\bibinfo {volume} {130}},\ \bibinfo {pages} {051802} (\bibinfo {year} {2023})}\BibitemShut {NoStop}%
\bibitem [{\citenamefont {An}\ \emph {et~al.}(2023)\citenamefont {An} \emph {et~al.}}]{COHERENTnuexsec}%
  \BibitemOpen
  \bibfield  {author} {\bibinfo {author} {\bibfnamefont {P.}~\bibnamefont {An}} \emph {et~al.},\ }\bibfield  {title} {\bibinfo {title} {{Measurement of Electron-Neutrino Charged-Current Cross Sections on $^{127}$I with the COHERENT Detector}},\ }\href {http://dx.doi.org/10.1103/PhysRevLett.131.221801} {\bibfield  {journal} {\bibinfo  {journal} {Phys. Rev. Lett.}\ }\textbf {\bibinfo {volume} {131}},\ \bibinfo {pages} {221801} (\bibinfo {year} {2023})}\BibitemShut {NoStop}%
\bibitem [{\citenamefont {Abe}\ \emph {et~al.}(2025)\citenamefont {Abe} \emph {et~al.}}]{recentxsecT2K}%
  \BibitemOpen
  \bibfield  {author} {\bibinfo {author} {\bibfnamefont {K.}~\bibnamefont {Abe}} \emph {et~al.} (\bibinfo {collaboration} {T2K Collaboration}),\ }\bibfield  {title} {\bibinfo {title} {{First Measurement of the Electron-Neutrino Charged-Current Pion Production Cross Section on Carbon with the T2K Near Detector}},\ }\href {https://doi.org/10.1103/klhv-7t6h} {\bibfield  {journal} {\bibinfo  {journal} {Phys. Rev. Lett.}\ }\textbf {\bibinfo {volume} {135}},\ \bibinfo {pages} {151802} (\bibinfo {year} {2025})}\BibitemShut {NoStop}%
\bibitem [{\citenamefont {Wolcott}\ \emph {et~al.}(2016)\citenamefont {Wolcott} \emph {et~al.}}]{MinervaExcnuexsec}%
  \BibitemOpen
  \bibfield  {author} {\bibinfo {author} {\bibfnamefont {J.}~\bibnamefont {Wolcott}} \emph {et~al.} (\bibinfo {collaboration} {MINERvA Collaboration}),\ }\bibfield  {title} {\bibinfo {title} {{Measurement of Electron Neutrino Quasielastic and Quasielasticlike Scattering on Hydrocarbon at $‚ü®{E}_{\ensuremath{\nu}}‚ü©=3.6\text{ }\text{ }\mathrm{GeV}$}},\ }\href {https://doi.org/10.1103/PhysRevLett.116.081802} {\bibfield  {journal} {\bibinfo  {journal} {Phys. Rev. Lett.}\ }\textbf {\bibinfo {volume} {116}},\ \bibinfo {pages} {081802} (\bibinfo {year} {2016})}\BibitemShut {NoStop}%
\bibitem [{\citenamefont {Acciarri}\ \emph {et~al.}(2020)\citenamefont {Acciarri} \emph {et~al.}}]{argoneutnuexsec}%
  \BibitemOpen
  \bibfield  {author} {\bibinfo {author} {\bibfnamefont {R.}~\bibnamefont {Acciarri}} \emph {et~al.} (\bibinfo {collaboration} {ArgoNeuT Collaboration}),\ }\bibfield  {title} {\bibinfo {title} {{First Measurement of Electron Neutrino Scattering Cross Section on Argon}},\ }\href {https://doi.org/10.1103/PhysRevD.102.011101} {\bibfield  {journal} {\bibinfo  {journal} {Phys. Rev. D}\ }\textbf {\bibinfo {volume} {102}},\ \bibinfo {pages} {011101} (\bibinfo {year} {2020})}\BibitemShut {NoStop}%
\bibitem [{\citenamefont {Abratenko}\ \emph {et~al.}(2021{\natexlab{a}})\citenamefont {Abratenko} \emph {et~al.}}]{UBnueInc21}%
  \BibitemOpen
  \bibfield  {author} {\bibinfo {author} {\bibfnamefont {P.}~\bibnamefont {Abratenko}} \emph {et~al.} (\bibinfo {collaboration} {MicroBooNE Collaboration}),\ }\bibfield  {title} {\bibinfo {title} {{Measurement of the Flux-Averaged Inclusive Charged-Current Electron Neutrino and Antineutrino Cross Section on Argon Using the NuMI Beam and the MicroBooNE Detector}},\ }\href {https://doi.org/10.1103/PhysRevD.104.052002} {\bibfield  {journal} {\bibinfo  {journal} {Phys. Rev. D}\ }\textbf {\bibinfo {volume} {104}},\ \bibinfo {pages} {052002} (\bibinfo {year} {2021}{\natexlab{a}})}\BibitemShut {NoStop}%
\bibitem [{\citenamefont {Abratenko}\ \emph {et~al.}(2022{\natexlab{a}})\citenamefont {Abratenko} \emph {et~al.}}]{UBnueInc22}%
  \BibitemOpen
  \bibfield  {author} {\bibinfo {author} {\bibfnamefont {P.}~\bibnamefont {Abratenko}} \emph {et~al.} (\bibinfo {collaboration} {MicroBooNE Collaboration}),\ }\bibfield  {title} {\bibinfo {title} {{First Measurement of Inclusive Electron-Neutrino and Antineutrino Charged Current Differential Cross Sections in Charged Lepton Energy on Argon in MicroBooNE}},\ }\href {https://doi.org/10.1103/PhysRevD.105.L051102} {\bibfield  {journal} {\bibinfo  {journal} {Phys. Rev. D}\ }\textbf {\bibinfo {volume} {105}},\ \bibinfo {pages} {L051102} (\bibinfo {year} {2022}{\natexlab{a}})}\BibitemShut {NoStop}%
\bibitem [{\citenamefont {Abratenko}\ \emph {et~al.}(2025{\natexlab{a}})\citenamefont {Abratenko} \emph {et~al.}}]{UBnueExcpip}%
  \BibitemOpen
  \bibfield  {author} {\bibinfo {author} {\bibfnamefont {P.}~\bibnamefont {Abratenko}} \emph {et~al.} (\bibinfo {collaboration} {MicroBooNE Collaboration}),\ }\bibfield  {title} {\bibinfo {title} {{First Measurement of $\nu_e$ and $\overline{\nu_e}$ Charged-Current Single Charged-Pion Production Differential Cross Sections on Argon Using the MicroBooNE Detector}},\ }\href {http://dx.doi.org/10.1103/sfrz-c8m9} {\bibfield  {journal} {\bibinfo  {journal} {Physical Review Letters}\ }\textbf {\bibinfo {volume} {135}},\ \bibinfo {pages} {061802} (\bibinfo {year} {2025}{\natexlab{a}})}\BibitemShut {NoStop}%
\bibitem [{\citenamefont {Abratenko}\ \emph {et~al.}(2025{\natexlab{b}})\citenamefont {Abratenko} \emph {et~al.}}]{nueCCNuMI}%
  \BibitemOpen
  \bibfield  {author} {\bibinfo {author} {\bibfnamefont {P.}~\bibnamefont {Abratenko}} \emph {et~al.} (\bibinfo {collaboration} {{MicroBooNE collaboration}}),\ }\bibfield  {title} {\bibinfo {title} {Measurements of differential charged-current cross sections on argon for electron neutrinos with final-state protons in {MicroBooNE}},\ }\href {https://arxiv.org/abs/2511.17342} {\bibfield  {journal} {\bibinfo  {journal} {{Phys. Rev. D}}\ } (\bibinfo {year} {2025}{\natexlab{b}})},\ \Eprint {https://arxiv.org/abs/2511.17342} {arXiv:2511.17342 [hep-ex]} \BibitemShut {NoStop}%
\bibitem [{\citenamefont {Abratenko}\ \emph {et~al.}(2022{\natexlab{b}})\citenamefont {Abratenko} \emph {et~al.}}]{MicroBooNEXsec}%
  \BibitemOpen
  \bibfield  {author} {\bibinfo {author} {\bibfnamefont {P.}~\bibnamefont {Abratenko}} \emph {et~al.} (\bibinfo {collaboration} {MicroBooNE Collaboration}),\ }\bibfield  {title} {\bibinfo {title} {{Differential Cross Section Measurement of Charged Current ${\ensuremath{\nu}}_{e}$ Interactions without Final-State Pions in MicroBooNE}},\ }\href {https://doi.org/10.1103/PhysRevD.106.L051102} {\bibfield  {journal} {\bibinfo  {journal} {Phys. Rev. D}\ }\textbf {\bibinfo {volume} {106}},\ \bibinfo {pages} {L051102} (\bibinfo {year} {2022}{\natexlab{b}})}\BibitemShut {NoStop}%
\bibitem [{\citenamefont {Abratenko}\ \emph {et~al.}(2025{\natexlab{c}})\citenamefont {Abratenko} \emph {et~al.}}]{MicroBooNE:2025tcm}%
  \BibitemOpen
  \bibfield  {author} {\bibinfo {author} {\bibfnamefont {P.}~\bibnamefont {Abratenko}} \emph {et~al.} (\bibinfo {collaboration} {MicroBooNE Collaboration}),\ }\href@noop {} {\bibinfo {title} {{Measurement of Charged-Current Muon Neutrino-Argon Interactions without Pions in the Final State using the MicroBooNE Detector}}} (\bibinfo {year} {2025}{\natexlab{c}}),\ \Eprint {https://arxiv.org/abs/2507.00921} {arXiv:2507.00921 [hep-ex]} \BibitemShut {NoStop}%
\bibitem [{\citenamefont {Abratenko}\ \emph {et~al.}(2025{\natexlab{d}})\citenamefont {Abratenko} \emph {et~al.}}]{Stevenpaper}%
  \BibitemOpen
  \bibfield  {author} {\bibinfo {author} {\bibfnamefont {P.}~\bibnamefont {Abratenko}} \emph {et~al.} (\bibinfo {collaboration} {MicroBooNE Collaboration}),\ }\bibfield  {title} {\bibinfo {title} {{Measurement of single- and double-differential cross sections for mesonless charged-current muon neutrino interactions on argon with final-state protons using the MicroBooNE detector}},\ }\href {https://doi.org/10.1103/8v2y-l89l} {\bibfield  {journal} {\bibinfo  {journal} {Phys. Rev. D}\ }\textbf {\bibinfo {volume} {112}},\ \bibinfo {pages} {112004} (\bibinfo {year} {2025}{\natexlab{d}})}\BibitemShut {NoStop}%
\bibitem [{\citenamefont {Abratenko}(2024)}]{BensAna}%
  \BibitemOpen
  \bibfield  {author} {\bibinfo {author} {\bibfnamefont {P.~‚.}\ \bibnamefont {Abratenko}} (\bibinfo {collaboration} {MicroBooNE Collaboration}),\ }\bibfield  {title} {\bibinfo {title} {{Inclusive Cross Section Measurements in Final States With and Without Protons for Charged-Current ${\ensuremath{\nu}}_{\ensuremath{\mu}}$-Ar Scattering in MicroBooNE}},\ }\href {https://doi.org/10.1103/PhysRevD.110.013006} {\bibfield  {journal} {\bibinfo  {journal} {Phys. Rev. D}\ }\textbf {\bibinfo {volume} {110}},\ \bibinfo {pages} {013006} (\bibinfo {year} {2024})}\BibitemShut {NoStop}%
\bibitem [{\citenamefont {Abratenko}\ \emph {et~al.}(2024)\citenamefont {Abratenko} \emph {et~al.}}]{BensAnaPRL}%
  \BibitemOpen
  \bibfield  {author} {\bibinfo {author} {\bibfnamefont {P.}~\bibnamefont {Abratenko}} \emph {et~al.} (\bibinfo {collaboration} {MicroBooNE Collaboration}),\ }\bibfield  {title} {\bibinfo {title} {{First Simultaneous Measurement of Differential Muon-Neutrino Charged-Current Cross Sections on Argon for Final States with and without Protons Using MicroBooNE Data}},\ }\href {https://doi.org/10.1103/PhysRevLett.133.041801} {\bibfield  {journal} {\bibinfo  {journal} {Phys. Rev. Lett.}\ }\textbf {\bibinfo {volume} {133}},\ \bibinfo {pages} {041801} (\bibinfo {year} {2024})}\BibitemShut {NoStop}%
\bibitem [{\citenamefont {Day}\ and\ \citenamefont {McFarland}(2012)}]{Day:2012gb}%
  \BibitemOpen
  \bibfield  {author} {\bibinfo {author} {\bibfnamefont {M.}~\bibnamefont {Day}}\ and\ \bibinfo {author} {\bibfnamefont {K.~S.}\ \bibnamefont {McFarland}},\ }\bibfield  {title} {\bibinfo {title} {Differences in quasielastic cross sections of muon and electron neutrinos},\ }\href {https://doi.org/10.1103/PhysRevD.86.053003} {\bibfield  {journal} {\bibinfo  {journal} {Phys. Rev. D}\ }\textbf {\bibinfo {volume} {86}},\ \bibinfo {pages} {053003} (\bibinfo {year} {2012})}\BibitemShut {NoStop}%
\bibitem [{\citenamefont {Nikolakopoulos}\ \emph {et~al.}(2019)\citenamefont {Nikolakopoulos}, \citenamefont {Jachowicz}, \citenamefont {Van~Dessel}, \citenamefont {Niewczas}, \citenamefont {Gonz\'alez-Jim\'enez}, \citenamefont {Ud\'{\i}as},\ and\ \citenamefont {Pandey}}]{elecvsMuon2}%
  \BibitemOpen
  \bibfield  {author} {\bibinfo {author} {\bibfnamefont {A.}~\bibnamefont {Nikolakopoulos}}, \bibinfo {author} {\bibfnamefont {N.}~\bibnamefont {Jachowicz}}, \bibinfo {author} {\bibfnamefont {N.}~\bibnamefont {Van~Dessel}}, \bibinfo {author} {\bibfnamefont {K.}~\bibnamefont {Niewczas}}, \bibinfo {author} {\bibfnamefont {R.}~\bibnamefont {Gonz\'alez-Jim\'enez}}, \bibinfo {author} {\bibfnamefont {J.~M.}\ \bibnamefont {Ud\'{\i}as}},\ and\ \bibinfo {author} {\bibfnamefont {V.}~\bibnamefont {Pandey}},\ }\bibfield  {title} {\bibinfo {title} {Electron versus muon neutrino induced cross sections in charged current quasielastic processes},\ }\href {https://doi.org/10.1103/PhysRevLett.123.052501} {\bibfield  {journal} {\bibinfo  {journal} {Phys. Rev. Lett.}\ }\textbf {\bibinfo {volume} {123}},\ \bibinfo {pages} {052501} (\bibinfo {year} {2019})}\BibitemShut {NoStop}%
\bibitem [{\citenamefont {Martini}\ \emph {et~al.}(2016)\citenamefont {Martini}, \citenamefont {Jachowicz}, \citenamefont {Ericson}, \citenamefont {Pandey}, \citenamefont {Van~Cuyck},\ and\ \citenamefont {Van~Dessel}}]{EvsMuMartini}%
  \BibitemOpen
  \bibfield  {author} {\bibinfo {author} {\bibfnamefont {M.}~\bibnamefont {Martini}}, \bibinfo {author} {\bibfnamefont {N.}~\bibnamefont {Jachowicz}}, \bibinfo {author} {\bibfnamefont {M.}~\bibnamefont {Ericson}}, \bibinfo {author} {\bibfnamefont {V.}~\bibnamefont {Pandey}}, \bibinfo {author} {\bibfnamefont {T.}~\bibnamefont {Van~Cuyck}},\ and\ \bibinfo {author} {\bibfnamefont {N.}~\bibnamefont {Van~Dessel}},\ }\bibfield  {title} {\bibinfo {title} {Electron-neutrino scattering off nuclei from two different theoretical perspectives},\ }\href {https://doi.org/10.1103/PhysRevC.94.015501} {\bibfield  {journal} {\bibinfo  {journal} {Phys. Rev. C}\ }\textbf {\bibinfo {volume} {94}},\ \bibinfo {pages} {015501} (\bibinfo {year} {2016})}\BibitemShut {NoStop}%
\bibitem [{\citenamefont {Ankowski}(2017)}]{Ankowski:2017yvm}%
  \BibitemOpen
  \bibfield  {author} {\bibinfo {author} {\bibfnamefont {A.~M.}\ \bibnamefont {Ankowski}},\ }\bibfield  {title} {\bibinfo {title} {{Effect of the charged-lepton's mass on the quasielastic neutrino cross sections}},\ }\href {https://doi.org/10.1103/PhysRevC.96.035501} {\bibfield  {journal} {\bibinfo  {journal} {Phys. Rev. C}\ }\textbf {\bibinfo {volume} {96}},\ \bibinfo {pages} {035501} (\bibinfo {year} {2017})},\ \Eprint {https://arxiv.org/abs/1707.01014} {arXiv:1707.01014 [nucl-th]} \BibitemShut {NoStop}%
\bibitem [{\citenamefont {Acciarri}\ \emph {et~al.}(2017)\citenamefont {Acciarri} \emph {et~al.}}]{Microboonedetector}%
  \BibitemOpen
  \bibfield  {author} {\bibinfo {author} {\bibfnamefont {R.}~\bibnamefont {Acciarri}} \emph {et~al.} (\bibinfo {collaboration} {MicroBooNE Collaboration}),\ }\bibfield  {title} {\bibinfo {title} {{Design and Construction of the {MicroBooNE} Detector}},\ }\href {https://doi.org/10.1088/1748-0221/12/02/p02017} {\bibfield  {journal} {\bibinfo  {journal} {{J. Instrum}}\ }\textbf {\bibinfo {volume} {12}},\ \bibinfo {pages} {P02017} (\bibinfo {year} {2017})}\BibitemShut {NoStop}%
\bibitem [{\citenamefont {Abratenko}\ \emph {et~al.}(2021{\natexlab{b}})\citenamefont {Abratenko} \emph {et~al.}}]{Drifttime2}%
  \BibitemOpen
  \bibfield  {author} {\bibinfo {author} {\bibfnamefont {P.}~\bibnamefont {Abratenko}} \emph {et~al.} (\bibinfo {collaboration} {MicroBooNE}),\ }\bibfield  {title} {\bibinfo {title} {{Measurement of the longitudinal diffusion of ionization electrons in the MicroBooNE detector}},\ }\href {https://doi.org/10.1088/1748-0221/16/09/P09025} {\bibfield  {journal} {\bibinfo  {journal} {{J.Instrum.}}\ }\textbf {\bibinfo {volume} {16}},\ \bibinfo {pages} {P09025} (\bibinfo {year} {2021}{\natexlab{b}})}\BibitemShut {NoStop}%
\bibitem [{\citenamefont {Adams}\ \emph {et~al.}(2019)\citenamefont {Adams} \emph {et~al.}}]{CRT_UB}%
  \BibitemOpen
  \bibfield  {author} {\bibinfo {author} {\bibfnamefont {C.}~\bibnamefont {Adams}} \emph {et~al.} (\bibinfo {collaboration} {MicroBooNE Collaboration}),\ }\bibfield  {title} {\bibinfo {title} {{Design and Construction of the MicroBooNE Cosmic Ray Tagger System}},\ }\href {https://doi.org/10.1088/1748-0221/14/04/P04004} {\bibfield  {journal} {\bibinfo  {journal} {{J. Instrum.}}\ }\textbf {\bibinfo {volume} {14}},\ \bibinfo {pages} {P04004} (\bibinfo {year} {2019})}\BibitemShut {NoStop}%
\bibitem [{\citenamefont {Abratenko}\ \emph {et~al.}(2018)\citenamefont {Abratenko} \emph {et~al.}}]{UBBNBFlux}%
  \BibitemOpen
  \bibfield  {author} {\bibinfo {author} {\bibfnamefont {P.}~\bibnamefont {Abratenko}} \emph {et~al.} (\bibinfo {collaboration} {MicroBooNE Collaboration}),\ }\href {https://microboone.fnal.gov/wp-content/uploads/MICROBOONE-NOTE-1031-PUB.pdf} {\bibinfo {title} {Booster neutrino flux prediction at microboone}} (\bibinfo {year} {2018})\BibitemShut {NoStop}%
\bibitem [{\citenamefont {Abratenko}\ \emph {et~al.}(2022{\natexlab{c}})\citenamefont {Abratenko} \emph {et~al.}}]{TuneUB}%
  \BibitemOpen
  \bibfield  {author} {\bibinfo {author} {\bibfnamefont {P.}~\bibnamefont {Abratenko}} \emph {et~al.} (\bibinfo {collaboration} {MicroBooNE Collaboration}),\ }\bibfield  {title} {\bibinfo {title} {{New $\mathrm{CC}0\ensuremath{\pi}$ {GENIE} Model Tune for MicroBooNE}},\ }\href {https://doi.org/10.1103/PhysRevD.105.072001} {\bibfield  {journal} {\bibinfo  {journal} {Phys. Rev. D}\ }\textbf {\bibinfo {volume} {105}},\ \bibinfo {pages} {072001} (\bibinfo {year} {2022}{\natexlab{c}})}\BibitemShut {NoStop}%
\bibitem [{\citenamefont {Andreopoulos}\ \emph {et~al.}(2010)\citenamefont {Andreopoulos} \emph {et~al.}}]{GENIE}%
  \BibitemOpen
  \bibfield  {author} {\bibinfo {author} {\bibfnamefont {C.}~\bibnamefont {Andreopoulos}} \emph {et~al.},\ }\bibfield  {title} {\bibinfo {title} {{The GENIE Neutrino Monte Carlo Generator}},\ }\href {https://doi.org/https://doi.org/10.1016/j.nima.2009.12.009} {\bibfield  {journal} {\bibinfo  {journal} {Nuc. Instrum. Meth. A}\ }\textbf {\bibinfo {volume} {614}},\ \bibinfo {pages} {87} (\bibinfo {year} {2010})}\BibitemShut {NoStop}%
\bibitem [{\citenamefont {Alvarez-Ruso}\ \emph {et~al.}(2021)\citenamefont {Alvarez-Ruso} \emph {et~al.}}]{Geniev3}%
  \BibitemOpen
  \bibfield  {author} {\bibinfo {author} {\bibfnamefont {L.}~\bibnamefont {Alvarez-Ruso}} \emph {et~al.},\ }\bibfield  {title} {\bibinfo {title} {{Recent Highlights from {GENIE} v3}},\ }\href {https://doi.org/10.1140/epjs/s11734-021-00295-7} {\bibfield  {journal} {\bibinfo  {journal} {Eur. Phys. J Spec. Top.}\ }\textbf {\bibinfo {volume} {230}},\ \bibinfo {pages} {4449} (\bibinfo {year} {2021})}\BibitemShut {NoStop}%
\bibitem [{\citenamefont {Carrasco}\ and\ \citenamefont {Oset}(1992)}]{LFG}%
  \BibitemOpen
  \bibfield  {author} {\bibinfo {author} {\bibfnamefont {R.}~\bibnamefont {Carrasco}}\ and\ \bibinfo {author} {\bibfnamefont {E.}~\bibnamefont {Oset}},\ }\bibfield  {title} {\bibinfo {title} {{Interaction of Real Photons with Nuclei from 100 to 500 MeV}},\ }\href {https://doi.org/https://doi.org/10.1016/0375-9474(92)90109-W} {\bibfield  {journal} {\bibinfo  {journal} {Nucl. Phys. A}\ }\textbf {\bibinfo {volume} {536}},\ \bibinfo {pages} {445} (\bibinfo {year} {1992})}\BibitemShut {NoStop}%
\bibitem [{\citenamefont {Nieves}\ \emph {et~al.}(2004)\citenamefont {Nieves}, \citenamefont {Amaro},\ and\ \citenamefont {Valverde}}]{CCNieves}%
  \BibitemOpen
  \bibfield  {author} {\bibinfo {author} {\bibfnamefont {J.}~\bibnamefont {Nieves}}, \bibinfo {author} {\bibfnamefont {J.~E.}\ \bibnamefont {Amaro}},\ and\ \bibinfo {author} {\bibfnamefont {M.}~\bibnamefont {Valverde}},\ }\bibfield  {title} {\bibinfo {title} {{Inclusive Quasielastic Charged-Current Neutrino-Nucleus Reactions}},\ }\href {https://doi.org/10.1103/PhysRevC.70.055503} {\bibfield  {journal} {\bibinfo  {journal} {Phys. Rev. C}\ }\textbf {\bibinfo {volume} {70}},\ \bibinfo {pages} {055503} (\bibinfo {year} {2004})}\BibitemShut {NoStop}%
\bibitem [{\citenamefont {Gran}\ \emph {et~al.}(2013)\citenamefont {Gran}, \citenamefont {Nieves}, \citenamefont {Sanchez},\ and\ \citenamefont {Vacas}}]{2p2hNieves}%
  \BibitemOpen
  \bibfield  {author} {\bibinfo {author} {\bibfnamefont {R.}~\bibnamefont {Gran}}, \bibinfo {author} {\bibfnamefont {J.}~\bibnamefont {Nieves}}, \bibinfo {author} {\bibfnamefont {F.}~\bibnamefont {Sanchez}},\ and\ \bibinfo {author} {\bibfnamefont {M.~J.~V.}\ \bibnamefont {Vacas}},\ }\bibfield  {title} {\bibinfo {title} {{Neutrino-Nucleus Quasi-Elastic and 2p2h Interactions up to 10 GeV}},\ }\href {https://doi.org/10.1103/PhysRevD.88.113007} {\bibfield  {journal} {\bibinfo  {journal} {Phys. Rev. D}\ }\textbf {\bibinfo {volume} {88}},\ \bibinfo {pages} {113007} (\bibinfo {year} {2013})}\BibitemShut {NoStop}%
\bibitem [{\citenamefont {Nowak}\ \emph {et~al.}(2009)\citenamefont {Nowak} \emph {et~al.}}]{4MomMiniBooNE}%
  \BibitemOpen
  \bibfield  {author} {\bibinfo {author} {\bibfnamefont {J.~A.}\ \bibnamefont {Nowak}} \emph {et~al.},\ }\bibfield  {title} {\bibinfo {title} {Four momentum transfer discrepancy in the charged current $\pi^+$ production in the {MiniBooNE}: Data versus theory},\ }\href {https://www.osti.gov/biblio/965015} {\bibfield  {journal} {\bibinfo  {journal} {AIP Conf.Proc.\bf{1189},243}\ } (\bibinfo {year} {2009})}\BibitemShut {NoStop}%
\bibitem [{\citenamefont {Berger}\ and\ \citenamefont {Sehgal}(2007)}]{BergerSehgal2007}%
  \BibitemOpen
  \bibfield  {author} {\bibinfo {author} {\bibfnamefont {C.}~\bibnamefont {Berger}}\ and\ \bibinfo {author} {\bibfnamefont {L.~M.}\ \bibnamefont {Sehgal}},\ }\bibfield  {title} {\bibinfo {title} {{Lepton Mass Effects in Single Pion Production by Neutrinos}},\ }\href {https://doi.org/10.1103/PhysRevD.76.113004} {\bibfield  {journal} {\bibinfo  {journal} {Phys. Rev. D}\ }\textbf {\bibinfo {volume} {76}},\ \bibinfo {pages} {113004} (\bibinfo {year} {2007})}\BibitemShut {NoStop}%
\bibitem [{\citenamefont {Kuzmin}\ \emph {et~al.}(2004)\citenamefont {Kuzmin}, \citenamefont {Lyubushkin},\ and\ \citenamefont {Naumov}}]{LepPOL}%
  \BibitemOpen
  \bibfield  {author} {\bibinfo {author} {\bibfnamefont {K.~S.}\ \bibnamefont {Kuzmin}}, \bibinfo {author} {\bibfnamefont {V.~V.}\ \bibnamefont {Lyubushkin}},\ and\ \bibinfo {author} {\bibfnamefont {V.~A.}\ \bibnamefont {Naumov}},\ }\bibfield  {title} {\bibinfo {title} {{Lepton Polarization in Neutrino-Nucleon Interactions}},\ }\href {https://doi.org/10.1142/S0217732304016172} {\bibfield  {journal} {\bibinfo  {journal} {Mod. Phys. Lett. A}\ }\textbf {\bibinfo {volume} {19}},\ \bibinfo {pages} {2815} (\bibinfo {year} {2004})}\BibitemShut {NoStop}%
\bibitem [{\citenamefont {Graczyk}\ and\ \citenamefont {Sobczyk}(2008)}]{QuarkRES}%
  \BibitemOpen
  \bibfield  {author} {\bibinfo {author} {\bibfnamefont {K.~M.}\ \bibnamefont {Graczyk}}\ and\ \bibinfo {author} {\bibfnamefont {J.~T.}\ \bibnamefont {Sobczyk}},\ }\bibfield  {title} {\bibinfo {title} {{Form Factors in the Quark Resonance Model}},\ }\href {https://doi.org/10.1103/PhysRevD.77.053001} {\bibfield  {journal} {\bibinfo  {journal} {Phys. Rev. D}\ }\textbf {\bibinfo {volume} {77}},\ \bibinfo {pages} {053001} (\bibinfo {year} {2008})}\BibitemShut {NoStop}%
\bibitem [{\citenamefont {Berger}\ and\ \citenamefont {Sehgal}(2009)}]{BSCOH}%
  \BibitemOpen
  \bibfield  {author} {\bibinfo {author} {\bibfnamefont {C.}~\bibnamefont {Berger}}\ and\ \bibinfo {author} {\bibfnamefont {L.~M.}\ \bibnamefont {Sehgal}},\ }\bibfield  {title} {\bibinfo {title} {{Partially Conserved Axial Vector Current and Coherent Pion Production by Low Energy Neutrinos}},\ }\href {https://doi.org/10.1103/PhysRevD.79.053003} {\bibfield  {journal} {\bibinfo  {journal} {Phys. Rev. D}\ }\textbf {\bibinfo {volume} {79}},\ \bibinfo {pages} {053003} (\bibinfo {year} {2009})}\BibitemShut {NoStop}%
\bibitem [{\citenamefont {Yang}\ and\ \citenamefont {Bodek}(1999)}]{BYDIS}%
  \BibitemOpen
  \bibfield  {author} {\bibinfo {author} {\bibfnamefont {U.~K.}\ \bibnamefont {Yang}}\ and\ \bibinfo {author} {\bibfnamefont {A.}~\bibnamefont {Bodek}},\ }\bibfield  {title} {\bibinfo {title} {{Parton Distributions, $\mathit{d}/\mathit{u}$, and Higher Twist Effects at High $\mathit{x}$}},\ }\href {https://doi.org/10.1103/PhysRevLett.82.2467} {\bibfield  {journal} {\bibinfo  {journal} {Phys. Rev. Lett.}\ }\textbf {\bibinfo {volume} {82}},\ \bibinfo {pages} {2467} (\bibinfo {year} {1999})}\BibitemShut {NoStop}%
\bibitem [{\citenamefont {Sjostrand}\ \emph {et~al.}(2006)\citenamefont {Sjostrand}, \citenamefont {Mrenna},\ and\ \citenamefont {Skands}}]{PYTHIA}%
  \BibitemOpen
  \bibfield  {author} {\bibinfo {author} {\bibfnamefont {T.}~\bibnamefont {Sjostrand}}, \bibinfo {author} {\bibfnamefont {S.}~\bibnamefont {Mrenna}},\ and\ \bibinfo {author} {\bibfnamefont {P.}~\bibnamefont {Skands}},\ }\bibfield  {title} {\bibinfo {title} {{PYTHIA 6.4 Physics and Manual}},\ }\href {https://doi.org/10.1088/1126-6708/2006/05/026} {\bibfield  {journal} {\bibinfo  {journal} {{J. High Energy Phys.}}\ }\textbf {\bibinfo {volume} {2006}},\ \bibinfo {pages} {026} (\bibinfo {year} {2006})}\BibitemShut {NoStop}%
\bibitem [{\citenamefont {Agostinelli}\ \emph {et~al.}(2003)\citenamefont {Agostinelli} \emph {et~al.}}]{GEANT4}%
  \BibitemOpen
  \bibfield  {author} {\bibinfo {author} {\bibfnamefont {S.}~\bibnamefont {Agostinelli}} \emph {et~al.},\ }\bibfield  {title} {\bibinfo {title} {{Geant4 ‚Äî A Simulation Toolkit}},\ }\href {https://doi.org/https://doi.org/10.1016/S0168-9002(03)01368-8} {\bibfield  {journal} {\bibinfo  {journal} {Nucl. Instrum. Meth. A}\ }\textbf {\bibinfo {volume} {506}},\ \bibinfo {pages} {250} (\bibinfo {year} {2003})}\BibitemShut {NoStop}%
\bibitem [{\citenamefont {Snider}\ and\ \citenamefont {Petrillo}(2017)}]{LArSoft}%
  \BibitemOpen
  \bibfield  {author} {\bibinfo {author} {\bibfnamefont {E.}~\bibnamefont {Snider}}\ and\ \bibinfo {author} {\bibfnamefont {G.}~\bibnamefont {Petrillo}},\ }\bibfield  {title} {\bibinfo {title} {{LArSoft: Toolkit for Simulation, Reconstruction and Analysis of Liquid Argon TPC Neutrino Detectors}},\ }\href {https://doi.org/10.1088/1742-6596/898/4/042057} {\bibfield  {journal} {\bibinfo  {journal} {J. Phys.: Conf. Ser.}\ }\textbf {\bibinfo {volume} {898}},\ \bibinfo {pages} {042057} (\bibinfo {year} {2017})}\BibitemShut {NoStop}%
\bibitem [{\citenamefont {Acciarri}\ \emph {et~al.}(2018)\citenamefont {Acciarri} \emph {et~al.}}]{Pandora}%
  \BibitemOpen
  \bibfield  {author} {\bibinfo {author} {\bibfnamefont {R.}~\bibnamefont {Acciarri}} \emph {et~al.} (\bibinfo {collaboration} {{MicroBooNE Collaboration}}),\ }\bibfield  {title} {\bibinfo {title} {{The Pandora Multi-Algorithm Approach to Automated Pattern Recognition of Cosmic-Ray Muon and Neutrino Events in the MicroBooNE Detector}},\ }\href {https://doi.org/10.1140/epjc/s10052-017-5481-6} {\bibfield  {journal} {\bibinfo  {journal} {Eur. Phys. J. C}\ }\textbf {\bibinfo {volume} {78}},\ \bibinfo {pages} {82} (\bibinfo {year} {2018})}\BibitemShut {NoStop}%
\bibitem [{\citenamefont {Abratenko}\ \emph {et~al.}(2022{\natexlab{d}})\citenamefont {Abratenko} \emph {et~al.}}]{oldPELEE}%
  \BibitemOpen
  \bibfield  {author} {\bibinfo {author} {\bibfnamefont {P.}~\bibnamefont {Abratenko}} \emph {et~al.} (\bibinfo {collaboration} {MicroBooNE Collaboration}),\ }\bibfield  {title} {\bibinfo {title} {{Search for an Anomalous Excess of Charged-Current $\nu_{e}$ Interactions without Pions in the Final State with the MicroBooNE Experiment}},\ }\href {https://doi.org/10.1103/PhysRevD.105.112004} {\bibfield  {journal} {\bibinfo  {journal} {Phys. Rev. D}\ }\textbf {\bibinfo {volume} {105}},\ \bibinfo {pages} {112004} (\bibinfo {year} {2022}{\natexlab{d}})}\BibitemShut {NoStop}%
\bibitem [{\citenamefont {Abratenko}\ \emph {et~al.}(2025{\natexlab{e}})\citenamefont {Abratenko} \emph {et~al.}}]{PELEEnew}%
  \BibitemOpen
  \bibfield  {author} {\bibinfo {author} {\bibfnamefont {P.}~\bibnamefont {Abratenko}} \emph {et~al.} (\bibinfo {collaboration} {MicroBooNE Collaboration}),\ }\bibfield  {title} {\bibinfo {title} {{Search for an Anomalous Production of Charged-Current {\ensuremath{\nu_e}} Interactions without Visible Pions across Multiple Kinematic Observables in MicroBooNE}},\ }\href {https://doi.org/10.1103/x259-z6mf} {\bibfield  {journal} {\bibinfo  {journal} {Phys. Rev. Lett.}\ }\textbf {\bibinfo {volume} {135}},\ \bibinfo {pages} {081802} (\bibinfo {year} {2025}{\natexlab{e}})}\BibitemShut {NoStop}%
\bibitem [{\citenamefont {Abratenko}\ \emph {et~al.}(2022{\natexlab{e}})\citenamefont {Abratenko} \emph {et~al.}}]{UbNCDelta}%
  \BibitemOpen
  \bibfield  {author} {\bibinfo {author} {\bibfnamefont {P.}~\bibnamefont {Abratenko}} \emph {et~al.} (\bibinfo {collaboration} {MicroBooNE Collaboration}),\ }\bibfield  {title} {\bibinfo {title} {{Search for Neutrino-Induced Neutral-Current $\mathrm{\ensuremath{\Delta}}$ Radiative Decay in MicroBooNE and a First Test of the MiniBooNE Low Energy Excess under a Single-Photon Hypothesis}},\ }\href {https://doi.org/10.1103/PhysRevLett.128.111801} {\bibfield  {journal} {\bibinfo  {journal} {Phys. Rev. Lett.}\ }\textbf {\bibinfo {volume} {128}},\ \bibinfo {pages} {111801} (\bibinfo {year} {2022}{\natexlab{e}})}\BibitemShut {NoStop}%
\bibitem [{\citenamefont {Abratenko}\ \emph {et~al.}(2019)\citenamefont {Abratenko} \emph {et~al.}}]{Ubincnumu}%
  \BibitemOpen
  \bibfield  {author} {\bibinfo {author} {\bibfnamefont {P.}~\bibnamefont {Abratenko}} \emph {et~al.} (\bibinfo {collaboration} {MicroBooNE Collaboration}),\ }\bibfield  {title} {\bibinfo {title} {{First Measurement of Inclusive Muon Neutrino Charged Current Differential Cross Sections on Argon at ${E}_{\ensuremath{\nu}}\ensuremath{\sim}0.8\text{ }\text{ }\mathrm{GeV}$ with the MicroBooNE Detector}},\ }\href {https://doi.org/10.1103/PhysRevLett.123.131801} {\bibfield  {journal} {\bibinfo  {journal} {Phys. Rev. Lett.}\ }\textbf {\bibinfo {volume} {123}},\ \bibinfo {pages} {131801} (\bibinfo {year} {2019})}\BibitemShut {NoStop}%
\bibitem [{\citenamefont {Abratenko}\ \emph {et~al.}(2022{\natexlab{f}})\citenamefont {Abratenko} \emph {et~al.}}]{detsysts}%
  \BibitemOpen
  \bibfield  {author} {\bibinfo {author} {\bibfnamefont {P.}~\bibnamefont {Abratenko}} \emph {et~al.},\ }\bibfield  {title} {\bibinfo {title} {{Novel Approach for Evaluating Detector-Related Uncertainties in a {LArTPC} using MicroBooNE Data}},\ }\href {http://dx.doi.org/10.1140/epjc/s10052-022-10270-8} {\bibfield  {journal} {\bibinfo  {journal} {Eur. Phys. J. C}\ }\textbf {\bibinfo {volume} {82}} (\bibinfo {year} {2022}{\natexlab{f}})}\BibitemShut {NoStop}%
\bibitem [{\citenamefont {Calcutt}\ \emph {et~al.}(2021)\citenamefont {Calcutt}, \citenamefont {Thorpe}, \citenamefont {Mahn},\ and\ \citenamefont {Fields}}]{ReINT}%
  \BibitemOpen
  \bibfield  {author} {\bibinfo {author} {\bibfnamefont {J.}~\bibnamefont {Calcutt}}, \bibinfo {author} {\bibfnamefont {C.}~\bibnamefont {Thorpe}}, \bibinfo {author} {\bibfnamefont {K.}~\bibnamefont {Mahn}},\ and\ \bibinfo {author} {\bibfnamefont {L.}~\bibnamefont {Fields}},\ }\bibfield  {title} {\bibinfo {title} {{Geant4Reweight: a Framework for Evaluating and Propagating Hadronic Interaction Uncertainties in Geant4}},\ }\href {https://doi.org/10.1088/1748-0221/16/08/P08042} {\bibfield  {journal} {\bibinfo  {journal} {{J. Instrum.}}\ }\textbf {\bibinfo {volume} {16}},\ \bibinfo {pages} {P08042} (\bibinfo {year} {2021})}\BibitemShut {NoStop}%
\bibitem [{\citenamefont {Koch}\ and\ \citenamefont {Dolan}(2020)}]{FluxPresc}%
  \BibitemOpen
  \bibfield  {author} {\bibinfo {author} {\bibfnamefont {L.}~\bibnamefont {Koch}}\ and\ \bibinfo {author} {\bibfnamefont {S.}~\bibnamefont {Dolan}},\ }\bibfield  {title} {\bibinfo {title} {{Treatment of Flux Shape Uncertainties in Unfolded, Flux-Averaged Neutrino Cross-Section Measurements}},\ }\href {https://doi.org/10.1103/PhysRevD.102.113012} {\bibfield  {journal} {\bibinfo  {journal} {Phys. Rev. D}\ }\textbf {\bibinfo {volume} {102}},\ \bibinfo {pages} {113012} (\bibinfo {year} {2020})}\BibitemShut {NoStop}%
\bibitem [{\citenamefont {Tang}\ \emph {et~al.}(2017)\citenamefont {Tang}, \citenamefont {Li}, \citenamefont {Qian}, \citenamefont {Wei},\ and\ \citenamefont {Zhang}}]{WienerSVD}%
  \BibitemOpen
  \bibfield  {author} {\bibinfo {author} {\bibfnamefont {W.}~\bibnamefont {Tang}}, \bibinfo {author} {\bibfnamefont {X.}~\bibnamefont {Li}}, \bibinfo {author} {\bibfnamefont {X.}~\bibnamefont {Qian}}, \bibinfo {author} {\bibfnamefont {H.}~\bibnamefont {Wei}},\ and\ \bibinfo {author} {\bibfnamefont {C.}~\bibnamefont {Zhang}},\ }\bibfield  {title} {\bibinfo {title} {{Data Unfolding with {Wiener-SVD} Method}},\ }\href {https://doi.org/10.1088/1748-0221/12/10/P10002} {\bibfield  {journal} {\bibinfo  {journal} {{J. Instrum.}}\ }\textbf {\bibinfo {volume} {12}},\ \bibinfo {pages} {P10002} (\bibinfo {year} {2017})}\BibitemShut {NoStop}%
\bibitem [{\citenamefont {Meyer}\ \emph {et~al.}(2016)\citenamefont {Meyer}, \citenamefont {Betancourt}, \citenamefont {Gran},\ and\ \citenamefont {Hill}}]{ZEXP}%
  \BibitemOpen
  \bibfield  {author} {\bibinfo {author} {\bibfnamefont {A.~S.}\ \bibnamefont {Meyer}}, \bibinfo {author} {\bibfnamefont {M.}~\bibnamefont {Betancourt}}, \bibinfo {author} {\bibfnamefont {R.}~\bibnamefont {Gran}},\ and\ \bibinfo {author} {\bibfnamefont {R.~J.}\ \bibnamefont {Hill}},\ }\bibfield  {title} {\bibinfo {title} {{Deuterium Target Data for Precision Neutrino-Nucleus Cross Sections}},\ }\href {https://doi.org/10.1103/PhysRevD.93.113015} {\bibfield  {journal} {\bibinfo  {journal} {Phys. Rev. D}\ }\textbf {\bibinfo {volume} {93}},\ \bibinfo {pages} {113015} (\bibinfo {year} {2016})}\BibitemShut {NoStop}%
\bibitem [{\citenamefont {Dolan}\ \emph {et~al.}(2020)\citenamefont {Dolan}, \citenamefont {Megias},\ and\ \citenamefont {Bolognesi}}]{SuSAv2}%
  \BibitemOpen
  \bibfield  {author} {\bibinfo {author} {\bibfnamefont {S.}~\bibnamefont {Dolan}}, \bibinfo {author} {\bibfnamefont {G.~D.}\ \bibnamefont {Megias}},\ and\ \bibinfo {author} {\bibfnamefont {S.}~\bibnamefont {Bolognesi}},\ }\bibfield  {title} {\bibinfo {title} {{Implementation of the SuSAv2-Meson Exchange Current 1p1h and 2p2h Models in GENIE and Analysis of Nuclear Effects in T2K Measurements}},\ }\href {https://doi.org/10.1103/PhysRevD.101.033003} {\bibfield  {journal} {\bibinfo  {journal} {Phys. Rev. D}\ }\textbf {\bibinfo {volume} {101}},\ \bibinfo {pages} {033003} (\bibinfo {year} {2020})}\BibitemShut {NoStop}%
\bibitem [{\citenamefont {Tena-Vidal}\ \emph {et~al.}(2021)\citenamefont {Tena-Vidal} \emph {et~al.}}]{TUNEfreeNuc}%
  \BibitemOpen
  \bibfield  {author} {\bibinfo {author} {\bibfnamefont {J.}~\bibnamefont {Tena-Vidal}} \emph {et~al.} (\bibinfo {collaboration} {GENIE Collaboration}),\ }\bibfield  {title} {\bibinfo {title} {{Neutrino-Nucleon Cross-Section Model Tuning in GENIE v3}},\ }\href {https://doi.org/10.1103/PhysRevD.104.072009} {\bibfield  {journal} {\bibinfo  {journal} {Phys. Rev. D}\ }\textbf {\bibinfo {volume} {104}},\ \bibinfo {pages} {072009} (\bibinfo {year} {2021})}\BibitemShut {NoStop}%
\bibitem [{\citenamefont {Rafique}(2024)}]{DUNEnuscale}%
  \BibitemOpen
  \bibfield  {author} {\bibinfo {author} {\bibfnamefont {A.}~\bibnamefont {Rafique}},\ }\bibfield  {title} {\bibinfo {title} {{Neutrino Energy Scale Measurements for Final State Interactions Using Advanced Computing in DUNE}}\ }(\bibinfo {year} {2024})\ \Eprint {https://arxiv.org/abs/2403.14858} {arXiv:2403.14858 [hep-ex]} \BibitemShut {NoStop}%
\bibitem [{\citenamefont {Abd~Alrahman}\ \emph {et~al.}(2025)\citenamefont {Abd~Alrahman} \emph {et~al.}}]{ICARUSexotics}%
  \BibitemOpen
  \bibfield  {author} {\bibinfo {author} {\bibfnamefont {F.}~\bibnamefont {Abd~Alrahman}} \emph {et~al.} (\bibinfo {collaboration} {ICARUS Collaboration}),\ }\bibfield  {title} {\bibinfo {title} {{Search for a Hidden Sector Scalar from Kaon Decay in the Dimuon Final State at ICARUS}},\ }\href {https://doi.org/10.1103/PhysRevLett.134.151801} {\bibfield  {journal} {\bibinfo  {journal} {Phys. Rev. Lett.}\ }\textbf {\bibinfo {volume} {134}},\ \bibinfo {pages} {151801} (\bibinfo {year} {2025})}\BibitemShut {NoStop}%
\bibitem [{\citenamefont {Abratenko}\ \emph {et~al.}(2025{\natexlab{f}})\citenamefont {Abratenko} \emph {et~al.}}]{SBNDPRISM}%
  \BibitemOpen
  \bibfield  {author} {\bibinfo {author} {\bibfnamefont {P.}~\bibnamefont {Abratenko}} \emph {et~al.} (\bibinfo {collaboration} {SBND Collaboration}),\ }\href@noop {} {\bibinfo {title} {{SBND-PRISM: Sampling Off-Axis Neutrino Fluxes with the Short-Baseline Near Detector}}} (\bibinfo {year} {2025}{\natexlab{f}}),\ \Eprint {https://arxiv.org/abs/2508.20239} {arXiv:2508.20239 [hep-ex]} \BibitemShut {NoStop}%
\bibitem [{\citenamefont {≈ªmuda}\ \emph {et~al.}(2015)\citenamefont {≈ªmuda}, \citenamefont {Graczyk}, \citenamefont {Juszczak},\ and\ \citenamefont {Sobczyk}}]{NuWro}%
  \BibitemOpen
  \bibfield  {author} {\bibinfo {author} {\bibfnamefont {J.}~\bibnamefont {≈ªmuda}}, \bibinfo {author} {\bibfnamefont {K.}~\bibnamefont {Graczyk}}, \bibinfo {author} {\bibfnamefont {C.}~\bibnamefont {Juszczak}},\ and\ \bibinfo {author} {\bibfnamefont {J.}~\bibnamefont {Sobczyk}},\ }\bibfield  {title} {\bibinfo {title} {{{\tt NuWro} Monte Carlo Generator of Neutrino Interactions --- First Electron Scattering Results}},\ }\href {https://doi.org/10.5506/aphyspolb.46.2329} {\bibfield  {journal} {\bibinfo  {journal} {{Acta Phys. Pol. B}}\ }\textbf {\bibinfo {volume} {46}},\ \bibinfo {pages} {2329} (\bibinfo {year} {2015})}\BibitemShut {NoStop}%
\bibitem [{\citenamefont {{Llewellyn Smith}}(1972)}]{LLEWELLYNSMITH1972}%
  \BibitemOpen
  \bibfield  {author} {\bibinfo {author} {\bibfnamefont {C.}~\bibnamefont {{Llewellyn Smith}}},\ }\bibfield  {title} {\bibinfo {title} {{Neutrino Reactions at Accelerator Energies}},\ }\href {https://doi.org/https://doi.org/10.1016/0370-1573(72)90010-5} {\bibfield  {journal} {\bibinfo  {journal} {Phys. Rep.}\ }\textbf {\bibinfo {volume} {3}},\ \bibinfo {pages} {261} (\bibinfo {year} {1972})}\BibitemShut {NoStop}%
\bibitem [{\citenamefont {Graczyk}\ \emph {et~al.}(2009)\citenamefont {Graczyk}, \citenamefont {Kielczewska}, \citenamefont {Przewlocki},\ and\ \citenamefont {Sobczyk}}]{AdlerRES}%
  \BibitemOpen
  \bibfield  {author} {\bibinfo {author} {\bibfnamefont {K.~M.}\ \bibnamefont {Graczyk}}, \bibinfo {author} {\bibfnamefont {D.}~\bibnamefont {Kielczewska}}, \bibinfo {author} {\bibfnamefont {P.}~\bibnamefont {Przewlocki}},\ and\ \bibinfo {author} {\bibfnamefont {J.~T.}\ \bibnamefont {Sobczyk}},\ }\bibfield  {title} {\bibinfo {title} {{${C}_{5}^{A}$ Axial Form Factor from Bubble Chamber Experiments}},\ }\href {https://doi.org/10.1103/PhysRevD.80.093001} {\bibfield  {journal} {\bibinfo  {journal} {Phys. Rev. D}\ }\textbf {\bibinfo {volume} {80}},\ \bibinfo {pages} {093001} (\bibinfo {year} {2009})}\BibitemShut {NoStop}%
\bibitem [{\citenamefont {Niewczas}\ and\ \citenamefont {Sobczyk}(2019)}]{NucTransp}%
  \BibitemOpen
  \bibfield  {author} {\bibinfo {author} {\bibfnamefont {K.}~\bibnamefont {Niewczas}}\ and\ \bibinfo {author} {\bibfnamefont {J.~T.}\ \bibnamefont {Sobczyk}},\ }\bibfield  {title} {\bibinfo {title} {{Nuclear Transparency in Monte Carlo Neutrino Event Generators}},\ }\href {https://doi.org/10.1103/PhysRevC.100.015505} {\bibfield  {journal} {\bibinfo  {journal} {Phys. Rev. C}\ }\textbf {\bibinfo {volume} {100}},\ \bibinfo {pages} {015505} (\bibinfo {year} {2019})}\BibitemShut {NoStop}%
\bibitem [{\citenamefont {Golan}\ \emph {et~al.}(2012)\citenamefont {Golan}, \citenamefont {Juszczak},\ and\ \citenamefont {Sobczyk}}]{FSIEffects}%
  \BibitemOpen
  \bibfield  {author} {\bibinfo {author} {\bibfnamefont {T.}~\bibnamefont {Golan}}, \bibinfo {author} {\bibfnamefont {C.}~\bibnamefont {Juszczak}},\ and\ \bibinfo {author} {\bibfnamefont {J.~T.}\ \bibnamefont {Sobczyk}},\ }\bibfield  {title} {\bibinfo {title} {{Effects of Final-State Interactions in Neutrino-Nucleus Interactions}},\ }\href {https://doi.org/10.1103/PhysRevC.86.015505} {\bibfield  {journal} {\bibinfo  {journal} {Phys. Rev. C}\ }\textbf {\bibinfo {volume} {86}},\ \bibinfo {pages} {015505} (\bibinfo {year} {2012})}\BibitemShut {NoStop}%
\bibitem [{\citenamefont {Hayato}\ and\ \citenamefont {Pickering}(2021)}]{NEUT}%
  \BibitemOpen
  \bibfield  {author} {\bibinfo {author} {\bibfnamefont {Y.}~\bibnamefont {Hayato}}\ and\ \bibinfo {author} {\bibfnamefont {L.}~\bibnamefont {Pickering}},\ }\bibfield  {title} {\bibinfo {title} {{The {NEUT} Neutrino Interaction Simulation Program Library}},\ }\href {https://doi.org/10.1140/epjs/s11734-021-00287-7} {\bibfield  {journal} {\bibinfo  {journal} {Eur. Phys. J. Spec. Top.}\ }\textbf {\bibinfo {volume} {230}},\ \bibinfo {pages} {4469} (\bibinfo {year} {2021})}\BibitemShut {NoStop}%
\bibitem [{\citenamefont {Auchincloss}\ \emph {et~al.}(1990)\citenamefont {Auchincloss} \emph {et~al.}}]{Auchincloss:1990tu}%
  \BibitemOpen
  \bibfield  {author} {\bibinfo {author} {\bibfnamefont {P.~S.}\ \bibnamefont {Auchincloss}} \emph {et~al.},\ }\bibfield  {title} {\bibinfo {title} {{Measurement of the Inclusive Charged Current Cross-section for Neutrino and Anti-neutrino Scattering on Isoscalar Nucleons}},\ }\href {https://doi.org/10.1007/BF01572022} {\bibfield  {journal} {\bibinfo  {journal} {Z. Phys. C}\ }\textbf {\bibinfo {volume} {48}},\ \bibinfo {pages} {411} (\bibinfo {year} {1990})}\BibitemShut {NoStop}%
\bibitem [{\citenamefont {Salcedo}\ \emph {et~al.}(1988)\citenamefont {Salcedo}, \citenamefont {Oset}, \citenamefont {Vicente-Vacas},\ and\ \citenamefont {Garcia-Recio}}]{SALCEDO1988557}%
  \BibitemOpen
  \bibfield  {author} {\bibinfo {author} {\bibfnamefont {L.}~\bibnamefont {Salcedo}}, \bibinfo {author} {\bibfnamefont {E.}~\bibnamefont {Oset}}, \bibinfo {author} {\bibfnamefont {M.}~\bibnamefont {Vicente-Vacas}},\ and\ \bibinfo {author} {\bibfnamefont {C.}~\bibnamefont {Garcia-Recio}},\ }\bibfield  {title} {\bibinfo {title} {{Computer Simulation of Inclusive Pion Nuclear Reactions}},\ }\href {https://doi.org/https://doi.org/10.1016/0375-9474(88)90310-7} {\bibfield  {journal} {\bibinfo  {journal} {Nuclear Physics A}\ }\textbf {\bibinfo {volume} {484}},\ \bibinfo {pages} {557} (\bibinfo {year} {1988})}\BibitemShut {NoStop}%
\bibitem [{\citenamefont {Gl√ºck}\ \emph {et~al.}(1998)\citenamefont {Gl√ºck}, \citenamefont {Reya},\ and\ \citenamefont {Vogt}}]{Gl_ck_1998}%
  \BibitemOpen
  \bibfield  {author} {\bibinfo {author} {\bibfnamefont {M.}~\bibnamefont {Gl√ºck}}, \bibinfo {author} {\bibfnamefont {E.}~\bibnamefont {Reya}},\ and\ \bibinfo {author} {\bibfnamefont {A.}~\bibnamefont {Vogt}},\ }\bibfield  {title} {\bibinfo {title} {Dynamical parton distributions revisited},\ }\href {https://doi.org/10.1007/s100529800978} {\bibfield  {journal} {\bibinfo  {journal} {The European Physical Journal C}\ }\textbf {\bibinfo {volume} {5}},\ \bibinfo {pages} {461‚Äì470} (\bibinfo {year} {1998})}\BibitemShut {NoStop}%
\bibitem [{\citenamefont {Bronner}(2016)}]{Bronner_2016}%
  \BibitemOpen
  \bibfield  {author} {\bibinfo {author} {\bibfnamefont {C.}~\bibnamefont {Bronner}},\ }\bibfield  {title} {\bibinfo {title} {{Generators for the SIS/DIS Region}},\ }in\ \href {http://dx.doi.org/10.7566/JPSCP.12.010025} {\emph {\bibinfo {booktitle} {Proceedings of the 10th International Workshop on Neutrino-Nucleus Interactions in Few-GeV Region (NuInt15)}}}\ (\bibinfo  {publisher} {J. Phys. Soc. Jpn.},\ \bibinfo {year} {2016})\BibitemShut {NoStop}%
\bibitem [{\citenamefont {Buss}\ \emph {et~al.}(2012)\citenamefont {Buss} \emph {et~al.}}]{GiBuu}%
  \BibitemOpen
  \bibfield  {author} {\bibinfo {author} {\bibfnamefont {O.}~\bibnamefont {Buss}} \emph {et~al.},\ }\bibfield  {title} {\bibinfo {title} {{Transport-Theoretical Description of Nuclear Reactions}},\ }\href {https://doi.org/10.1016/j.physrep.2011.12.001} {\bibfield  {journal} {\bibinfo  {journal} {Phys. Rep.}\ }\textbf {\bibinfo {volume} {512}},\ \bibinfo {pages} {1} (\bibinfo {year} {2012})}\BibitemShut {NoStop}%
\bibitem [{\citenamefont {Bogart}\ \emph {et~al.}(2024)\citenamefont {Bogart}, \citenamefont {Gallmeister},\ and\ \citenamefont {Mosel}}]{INMEDIUMGIBUU}%
  \BibitemOpen
  \bibfield  {author} {\bibinfo {author} {\bibfnamefont {B.}~\bibnamefont {Bogart}}, \bibinfo {author} {\bibfnamefont {K.}~\bibnamefont {Gallmeister}},\ and\ \bibinfo {author} {\bibfnamefont {U.}~\bibnamefont {Mosel}},\ }\bibfield  {title} {\bibinfo {title} {{In-Medium Changes of Nucleon Cross Sections Tested in Neutrino-Induced Reactions}},\ }\href {https://doi.org/10.1103/PhysRevC.110.044001} {\bibfield  {journal} {\bibinfo  {journal} {Phys. Rev. C}\ }\textbf {\bibinfo {volume} {110}},\ \bibinfo {pages} {044001} (\bibinfo {year} {2024})}\BibitemShut {NoStop}%
\bibitem [{\citenamefont {Mosel}(2019)}]{Mosel_2019}%
  \BibitemOpen
  \bibfield  {author} {\bibinfo {author} {\bibfnamefont {U.}~\bibnamefont {Mosel}},\ }\bibfield  {title} {\bibinfo {title} {{Neutrino Event Generators: Foundation, Status and Future}},\ }\href {https://doi.org/10.1088/1361-6471/ab3830} {\bibfield  {journal} {\bibinfo  {journal} {J. Phys. G}\ }\textbf {\bibinfo {volume} {46}},\ \bibinfo {pages} {113001} (\bibinfo {year} {2019})}\BibitemShut {NoStop}%
\bibitem [{\citenamefont {Leitner}\ \emph {et~al.}(2006)\citenamefont {Leitner}, \citenamefont {Alvarez-Ruso},\ and\ \citenamefont {Mosel}}]{GiBUUCCQE}%
  \BibitemOpen
  \bibfield  {author} {\bibinfo {author} {\bibfnamefont {T.}~\bibnamefont {Leitner}}, \bibinfo {author} {\bibfnamefont {L.}~\bibnamefont {Alvarez-Ruso}},\ and\ \bibinfo {author} {\bibfnamefont {U.}~\bibnamefont {Mosel}},\ }\bibfield  {title} {\bibinfo {title} {{Charged Current Neutrino-Nucleus Interactions at Intermediate Energies}},\ }\href {https://doi.org/10.1103/PhysRevC.73.065502} {\bibfield  {journal} {\bibinfo  {journal} {Phys. Rev. C}\ }\textbf {\bibinfo {volume} {73}},\ \bibinfo {pages} {065502} (\bibinfo {year} {2006})}\BibitemShut {NoStop}%
\bibitem [{\citenamefont {Stowell}\ \emph {et~al.}(2017)\citenamefont {Stowell} \emph {et~al.}}]{NUISANCE}%
  \BibitemOpen
  \bibfield  {author} {\bibinfo {author} {\bibfnamefont {P.}~\bibnamefont {Stowell}} \emph {et~al.},\ }\bibfield  {title} {\bibinfo {title} {{NUISANCE: a Neutrino Cross-Section Generator Tuning and Comparison Framework}},\ }\href {https://doi.org/10.1088/1748-0221/12/01/P01016} {\bibfield  {journal} {\bibinfo  {journal} {{J. Instrum.}}\ }\textbf {\bibinfo {volume} {12}},\ \bibinfo {pages} {P01016} (\bibinfo {year} {2017})}\BibitemShut {NoStop}%
\bibitem [{\citenamefont {Dytman}\ \emph {et~al.}(2021)\citenamefont {Dytman}, \citenamefont {Hayato}, \citenamefont {Raboanary}, \citenamefont {Sobczyk}, \citenamefont {Tena-Vidal},\ and\ \citenamefont {Vololoniaina}}]{CompHAHN}%
  \BibitemOpen
  \bibfield  {author} {\bibinfo {author} {\bibfnamefont {S.}~\bibnamefont {Dytman}}, \bibinfo {author} {\bibfnamefont {Y.}~\bibnamefont {Hayato}}, \bibinfo {author} {\bibfnamefont {R.}~\bibnamefont {Raboanary}}, \bibinfo {author} {\bibfnamefont {J.~T.}\ \bibnamefont {Sobczyk}}, \bibinfo {author} {\bibfnamefont {J.}~\bibnamefont {Tena-Vidal}},\ and\ \bibinfo {author} {\bibfnamefont {N.}~\bibnamefont {Vololoniaina}},\ }\bibfield  {title} {\bibinfo {title} {{Comparison of Validation Methods of Simulations for Final State Interactions in Hadron Production Experiments}},\ }\href {https://doi.org/10.1103/PhysRevD.104.053006} {\bibfield  {journal} {\bibinfo  {journal} {Phys. Rev. D}\ }\textbf {\bibinfo {volume} {104}},\ \bibinfo {pages} {053006} (\bibinfo {year} {2021})}\BibitemShut {NoStop}%
\bibitem [{\citenamefont {Wright}\ and\ \citenamefont {Kelsey}(2015)}]{BertiniG4}%
  \BibitemOpen
  \bibfield  {author} {\bibinfo {author} {\bibfnamefont {D.~H.}\ \bibnamefont {Wright}}\ and\ \bibinfo {author} {\bibfnamefont {M.~H.}\ \bibnamefont {Kelsey}},\ }\bibfield  {title} {\bibinfo {title} {{{The Geant4 Bertini Cascade}}},\ }\href {https://doi.org/10.1016/j.nima.2015.09.058} {\bibfield  {journal} {\bibinfo  {journal} {Nucl. Instrum. Meth. A}\ }\textbf {\bibinfo {volume} {804}},\ \bibinfo {pages} {175} (\bibinfo {year} {2015})}\BibitemShut {NoStop}%
\bibitem [{\citenamefont {Katori}(2015)}]{DytmanMECModel}%
  \BibitemOpen
  \bibfield  {author} {\bibinfo {author} {\bibfnamefont {T.}~\bibnamefont {Katori}},\ }\bibfield  {title} {\bibinfo {title} {{Meson Exchange Current ({MEC}) Models in Neutrino Interaction Generators}},\ }\href {https://doi.org/10.1063/1.4919465} {\bibfield  {journal} {\bibinfo  {journal} {AIP Conf. Proc.}\ }\textbf {\bibinfo {volume} {1663}},\ \bibinfo {pages} {030001} (\bibinfo {year} {2015})}\BibitemShut {NoStop}%
\bibitem [{\citenamefont {McKean}\ \emph {et~al.}(2025)\citenamefont {McKean}, \citenamefont {Gonz\'alez-Jim\'enez}, \citenamefont {Kabirnezhad}, \citenamefont {Ud\'{\i}as},\ and\ \citenamefont {Uchida}}]{NEUTEDRMF}%
  \BibitemOpen
  \bibfield  {author} {\bibinfo {author} {\bibfnamefont {J.}~\bibnamefont {McKean}}, \bibinfo {author} {\bibfnamefont {R.}~\bibnamefont {Gonz\'alez-Jim\'enez}}, \bibinfo {author} {\bibfnamefont {M.}~\bibnamefont {Kabirnezhad}}, \bibinfo {author} {\bibfnamefont {J.~M.}\ \bibnamefont {Ud\'{\i}as}},\ and\ \bibinfo {author} {\bibfnamefont {Y.}~\bibnamefont {Uchida}},\ }\bibfield  {title} {\bibinfo {title} {Implementation of a relativistic distorted wave impulse approximation model into the neut event generator},\ }\href {https://doi.org/10.1103/f7x5-snmz} {\bibfield  {journal} {\bibinfo  {journal} {Phys. Rev. D}\ }\textbf {\bibinfo {volume} {112}},\ \bibinfo {pages} {032009} (\bibinfo {year} {2025})}\BibitemShut {NoStop}%
\bibitem [{\citenamefont {Gonz\'alez-Jim\'enez}\ \emph {et~al.}(2019)\citenamefont {Gonz\'alez-Jim\'enez}, \citenamefont {Nikolakopoulos}, \citenamefont {Jachowicz},\ and\ \citenamefont {Ud\'{\i}as}}]{EDRMF}%
  \BibitemOpen
  \bibfield  {author} {\bibinfo {author} {\bibfnamefont {R.}~\bibnamefont {Gonz\'alez-Jim\'enez}}, \bibinfo {author} {\bibfnamefont {A.}~\bibnamefont {Nikolakopoulos}}, \bibinfo {author} {\bibfnamefont {N.}~\bibnamefont {Jachowicz}},\ and\ \bibinfo {author} {\bibfnamefont {J.~M.}\ \bibnamefont {Ud\'{\i}as}},\ }\bibfield  {title} {\bibinfo {title} {Nuclear effects in electron-nucleus and neutrino-nucleus scattering within a relativistic quantum mechanical framework},\ }\href {https://doi.org/10.1103/PhysRevC.100.045501} {\bibfield  {journal} {\bibinfo  {journal} {Phys. Rev. C}\ }\textbf {\bibinfo {volume} {100}},\ \bibinfo {pages} {045501} (\bibinfo {year} {2019})}\BibitemShut {NoStop}%
\bibitem [{\citenamefont {Sobczyk}\ and\ \citenamefont {Nieves}(2025)}]{MECValenciaUpdate}%
  \BibitemOpen
  \bibfield  {author} {\bibinfo {author} {\bibfnamefont {J.~E.}\ \bibnamefont {Sobczyk}}\ and\ \bibinfo {author} {\bibfnamefont {J.}~\bibnamefont {Nieves}},\ }\bibfield  {title} {\bibinfo {title} {Neutrino and antineutrino charged-current multinucleon cross sections reexamined},\ }\href {https://doi.org/10.1103/PhysRevC.111.025502} {\bibfield  {journal} {\bibinfo  {journal} {Phys. Rev. C}\ }\textbf {\bibinfo {volume} {111}},\ \bibinfo {pages} {025502} (\bibinfo {year} {2025})}\BibitemShut {NoStop}%
\bibitem [{\citenamefont {Boudard}\ \emph {et~al.}(2013)\citenamefont {Boudard}, \citenamefont {Cugnon}, \citenamefont {David}, \citenamefont {Leray},\ and\ \citenamefont {Mancusi}}]{INCL}%
  \BibitemOpen
  \bibfield  {author} {\bibinfo {author} {\bibfnamefont {A.}~\bibnamefont {Boudard}}, \bibinfo {author} {\bibfnamefont {J.}~\bibnamefont {Cugnon}}, \bibinfo {author} {\bibfnamefont {J.-C.}\ \bibnamefont {David}}, \bibinfo {author} {\bibfnamefont {S.}~\bibnamefont {Leray}},\ and\ \bibinfo {author} {\bibfnamefont {D.}~\bibnamefont {Mancusi}},\ }\bibfield  {title} {\bibinfo {title} {{New potentialities of the Li\`ege intranuclear cascade model for reactions induced by nucleons and light charged particles}},\ }\href {https://doi.org/10.1103/PhysRevC.87.014606} {\bibfield  {journal} {\bibinfo  {journal} {Phys. Rev. C}\ }\textbf {\bibinfo {volume} {87}},\ \bibinfo {pages} {014606} (\bibinfo {year} {2013})}\BibitemShut {NoStop}%
\bibitem [{\citenamefont {Mancusi}\ \emph {et~al.}(2014)\citenamefont {Mancusi}, \citenamefont {Boudard}, \citenamefont {Cugnon}, \citenamefont {David}, \citenamefont {Kaitaniemi},\ and\ \citenamefont {Leray}}]{INCL++}%
  \BibitemOpen
  \bibfield  {author} {\bibinfo {author} {\bibfnamefont {D.}~\bibnamefont {Mancusi}}, \bibinfo {author} {\bibfnamefont {A.}~\bibnamefont {Boudard}}, \bibinfo {author} {\bibfnamefont {J.}~\bibnamefont {Cugnon}}, \bibinfo {author} {\bibfnamefont {J.-C.}\ \bibnamefont {David}}, \bibinfo {author} {\bibfnamefont {P.}~\bibnamefont {Kaitaniemi}},\ and\ \bibinfo {author} {\bibfnamefont {S.}~\bibnamefont {Leray}},\ }\bibfield  {title} {\bibinfo {title} {{Extension of the Li\`ege intranuclear-cascade model to reactions induced by light nuclei}},\ }\href {https://doi.org/10.1103/PhysRevC.90.054602} {\bibfield  {journal} {\bibinfo  {journal} {Phys. Rev. C}\ }\textbf {\bibinfo {volume} {90}},\ \bibinfo {pages} {054602} (\bibinfo {year} {2014})}\BibitemShut {NoStop}%
\bibitem [{\citenamefont {Acciarri}\ \emph {et~al.}(2015)\citenamefont {Acciarri} \emph {et~al.}}]{SBN}%
  \BibitemOpen
  \bibfield  {author} {\bibinfo {author} {\bibfnamefont {R.}~\bibnamefont {Acciarri}} \emph {et~al.},\ }\href@noop {} {\bibinfo {title} {{A Proposal for a Three Detector Short-Baseline Neutrino Oscillation Program in the Fermilab Booster Neutrino Beam}}} (\bibinfo {year} {2015}),\ \Eprint {https://arxiv.org/abs/1503.01520} {arXiv:1503.01520 [physics.ins-det]} \BibitemShut {NoStop}%
\end{thebibliography}
\end{document}